\documentclass[12pt]{article}

\usepackage{amsfonts,amssymb,amsmath,epsfig,fullpage}

\newcommand{\R}{{\mathbb R}}
\newcommand{\Z}{{\mathbb Z}}

\newcommand{\cB}{{\cal B}}
\newcommand{\tr}{{\rm tr}}
\newcommand{\fix}{{\rm fix}}

\newtheorem{definition}{Definition}
\newtheorem{theorem}{Theorem}[section]
\newtheorem{lemma}{Lemma}[section]
\newtheorem{corollary}{Corollary}[section]
\newcommand{\proof}{\noindent {\bf Proof: }}
\newcommand{\loc}{\mathrm{loc}}
\newcommand{\qed}{\hfill{\bf QED}

\vspace{5mm}}

\begin{document}

\title{On local attraction properties and a stability index for heteroclinic connections}

\author{Olga Podvigina\\
Observatoire de la C\^ote d'Azur, CNRS UMR~6529\\
BP~4229, 06304 Nice Cedex 4, France, and\\
International Institute of Earthquake Prediction Theory\\
and Mathematical Geophysics\\
84/32 Profsoyuznaya St, 117997 Moscow, Russian Federation
\and
Peter Ashwin\\
Mathematics Research Institute\\
Harrison Building\\
University of Exeter, Exeter EX4 4QF, UK
}

\maketitle

\begin{abstract}
Some invariant sets may attract a nearby set of initial conditions but
nonetheless repel a complementary nearby set of initial conditions. For a
given invariant set $X\subset\R^n$ with a basin of attraction $N$, we define
a {\em stability index} $\sigma(x)$ of a point
$x\in X$ that characterizes the local extent of the basin. Let $B_{\epsilon}$
denote a ball of radius $\epsilon$ about $x$. If $\sigma(x)>0$, then
the measure of $B_{\epsilon}\setminus N$ relative the measure of the ball
is $O(\epsilon^{|\sigma(x)|})$, while if $\sigma(x)<0$, then the measure of
$B_{\epsilon}\cap N$ relative the measure of the ball is of this order.
We show that this index is constant
along trajectories, and we relate this orbit invariant to other notions
of stability such as Milnor attraction, essential asymptotic stability and
asymptotic stability relative to a positive measure set. We adapt
the definition to local basins of attraction (i.e. where $N$ is defined as
the set of initial conditions that are in the basin and whose trajectories
remain local to $X$).

This stability index is particularly useful for discussing the stability of
robust heteroclinic cycles, where several authors have studied the appearance of cusps
of instability near cycles that are Milnor attractors. We study simple
(robust heteroclinic) cycles in $\R^4$ and show that the
local stability indices (and hence local stability properties) can be
calculated in terms of the eigenvalues of the linearization of the vector field
at steady states on the cycle. In doing this, we extend previous results of
Krupa and Melbourne (1995,2004) and give criteria for simple heteroclinic cycles in $\R^4$ to be Milnor attractors.

\end{abstract}

\section{Introduction}\label{sec_intro}

For many choices of smooth vector field $f:\R^n\to\R^n$, the system on $x\in\R^n$
\begin{equation}\label{eq_ode}
\dot{x}=f(x)
\end{equation}
has a small subset of $\R^n$ (an attractor) that attracts a large set of initial conditions; these attractors are important for understanding the long term behaviour of trajectories of the system. In this paper we explore the local attraction structure of invariant sets, while in the latter part we focus on a particular class of examples - attracting heteroclinic cycles. More precisely, an invariant set is {\em asymptotically stable} if it attracts all nearby points; many systems are found to possess invariant sets that are not asymptotically stable, but that are attractors in a weaker sense (e.g. in the sense of Milnor \cite{Mil85}).

Now consider $\xi_1,\ldots,\xi_m$ to be hyperbolic equilibria of (\ref{eq_ode}). A set of connecting trajectories $W^u(\xi_j)\cap W^s(\xi_{j+1})\ne\emptyset$, $j=1,\ldots,m$, $\xi_{m+1}=\xi_1$, is called a {\em heteroclinic cycle} between these equilibria. It has been shown that heteroclinic cycles can be robust (persistent to small perturbations) if $f$ is constrained to be symmetric with respect to certain group representations \cite{Kru97,AshMon02,Sot03}, or if $f$ is constrained to preserve certain invariant subspaces \cite{Kru97}. Heteroclinic cycles that are not asymptotically stable may often be observed to be apparently stable in computations. To explain this, weaker notions of stability for heteroclinic cycles sets were introduced in \cite{Mel91,KruMel95b,KirSil94} - they do not require attraction in a full neighbourhood of the invariant set; they may even be repelling in a region that is typically cusp-shaped in Poincar\'{e} sections to the cycle. The papers \cite{Mel91,KruMel95b} define a heteroclinic cycle to be {\em essentially asymptotically stable} (e.a.s.) if it attracts almost all nearby trajectories, and they define it to be {\em almost completely unstable} (a.c.u.), if it attracts almost no nearby trajectories. However, as shown in \cite{KirSil94} these definitions are not mutually exclusive. (Brannath \cite{Bra94} similarly discusses e.a.s.\ using the notion of {\em relative asymptotic stability} from Ura \cite{Ura64}.)

The paper is organized as follows: in Section~\ref{sec_defs} we discuss various definitions of stability, and we relate them to the notion of Milnor attractor and the local geometry of the basin of attraction. We introduce a {\em stability index} that characterizes the local geometry of the basin of attraction. After proving some basic properties about this invariant of the dynamics, we generalize to a {\em local stability index} that is the limit of stability indices of local basins of attraction. In Section~\ref{sec_hetc} we discuss the structure of heteroclinic cycles and describe the geometry of local basins of attraction by way of the local stability index and (Poincar\'{e}) surfaces of section. We show, under certain assumptions, that the stability index of a connecting trajectory is the stability index on a surface of section.

Section~\ref{sec_ABC} computes the stability indices for robust heteroclinic cycles in $\R^4$; we employ the classification of simple cycles in $\R^4$ into Types A-C by Krupa and Melbourne in a series of papers \cite{KruMel95a,KruMel95b,KruMel04} and calculate the stability indices of the connections in terms of eigenvalues of the linearization at equilibria in the cycle. Finally we discuss some of the limitations and possible further uses of stability indices and related concepts in Section~\ref{sec_conc}.

\section{Attractors and the stability index}\label{sec_defs}

Various definitions of attraction of invariant sets have been introduced \cite{Bra94,KruMel95b,Mel91,Ura64} to describe sets that are not asymptotically stable but that are nevertheless attracting in some sense. We review these notions and relate them to Milnor's notion of a measure attractor \cite{Mil85}.

\subsection{Notions of attraction for invariant sets}

In this section we consider a smooth flow $\Phi_t(x)$ on $\R^n$. Two very general notions of attraction are the Milnor and weak attractors discussed in \cite{Mil85} and \cite{AshTer00} respectively. For an invariant set $X\subset \R^n$ we define the (global) {\em 

basin of attraction} of $X$ to be
$$
{\cal B}(X)= \{x\in \R^n~:~ \omega(x)\subset X\}
$$
where $\omega(x)=\bigcap_{T>0}\overline{\{\Phi_t(x)~:~t>T\}}$ is the $\omega$-limit of $x$. The following defines attraction properties of $X$ in terms of this basin. We use $\ell(\cdot)$ to denote Lebesgue measure on $\R^n$.

\begin{definition}\cite{AshTer00}
We say a compact invariant set $X$ is a {\em weak attractor} if $\ell(\cB(X))>0$. We say a compact invariant set $X$ is a {\em Milnor attractor} if it is a weak attractor such that for any proper subset $Y\subset X$ that is compact and invariant we have
$$
\ell(\cB(X)\setminus\cB(Y))>0.
$$
\end{definition}

We do not assume transitivity of $X$ (a dense orbit); indeed, the main examples we will consider later on are heteroclinic cycles that are not transitive. Note that any weak attractor contains a Milnor attractor \cite[Lemma~3.2]{AshTer00}. There are various examples of robust heteroclinic cycles (e.g. \cite{Mel91,Bra94,KirSil94}) that are Milnor attractors, even though they are not asymptotically stable. Let $d(\cdot,\cdot)$ denote the Hausdorff distance between two sets, let
$$
B_{\epsilon}(X) := \{ x \in \R^n ~:~ d(x, X) < \epsilon \}
$$
denote the $\epsilon$-parallel body of $X$, and let $D^c$ denote the complement of $D$ in $\R^n$.

\begin{definition}\label{def2}\cite{Mel91}
We say a compact invariant set $X$ is {\em essentially asymptotically stable} (e.a.s.), if there is a set $D$ such that for any open neighbourhood $U$ of $X$ and any $\epsilon>0$ there exists an open neighbourhood $V\subset U$ of $X$ such that:
\begin{itemize}
\item[(a)] If $x_0\in V\cap D^c$ then $\Phi_t(x_0)\in U$ for all $t>0$ and
$\lim_{t\to\infty}d(\Phi_t(x_0),X)=0$,
\item[(b)] $\ell(V\cap D^c)/\ell(V)>1-\epsilon$.
\end{itemize}
\end{definition}

Intuitively, if $X$ is e.a.s.\ one might expect that it attracts ``almost all'' nearby trajectories, while \cite{KruMel95b} says $X$ is almost completely unstable\footnote{A flow-invariant set $X$ is called {\it almost completely unstable (a.c.u)}, if there is a set $D$ and an open neighbourhood $U$ of $X$ such that for some $\epsilon>0$ there exists an open neighbourhood $V$ of $X$, $V\subset U$, such that (a) for $x_0\in V\setminus D$ there exists a $t>0$ with $\Phi_t(x_0)\notin U$; and (b) $\ell(V\cap D^c)/\ell(V)>1-\epsilon$.} if it attracts ``almost none'' of them. However, these definitions do not formalise these intuitive categories very well; as highlighted in \cite{KirSil94}, they are not mutually exclusive and so may yield classifications that are not intuitively helpful. Another useful definition is that of \cite{Ura64} which is used in \cite{Bra94}: for this we consider a set $N\subset \R^n$.

\begin{definition}\cite{Ura64}
We say a compact invariant set $X$ with $X\subset \overline{N}$ is {\em asymptotically stable, relative to} (a.s.r.t.) $N$ if for every neighbourhood $U$ of $X$ there is a neighbourhood $V$ of $X$ such that for all initial $x\in V\cap N$ we have $\Phi_t(x)\in U$ for $t>0$, and $\omega(x)\subset X$.
\end{definition}

In fact, Brannath \cite{Bra94} interestingly suggests that the authors of \cite{Mel91,KruMel95b} had the following definition in mind for e.a.s., but we name it differently to distinguish from the original definition in \cite{Mel91}.

\begin{definition}\label{def4}(Adapted from \cite{Bra94})
We say a compact invariant set $X$ is {\em predominantly asymptotically stable} (p.a.s.) if there is an $N$ such that $X$ is asymptotically stable relative to $N$ and
$$
\lim_{\epsilon\rightarrow 0} \frac{\ell(B_{\epsilon}(X)\cap N)}{\ell(B_{\epsilon}(X))}=1.
$$
\end{definition}

We now give a result that relates these concepts of attraction.

\begin{theorem}\label{thm_main}
Suppose that $X$ is a compact invariant set for a continuous flow $\Phi_t$.
\begin{itemize}
\item[(a)] $X$ is p.a.s.\ $\Rightarrow$ $X$ is e.a.s.
\item[(b)] $X$ is e.a.s.\ $\Rightarrow$ $X$ contains a Milnor attractor.
\item[(c)] $X$ is e.a.s.\ $\Leftrightarrow$ there is an $N$ with $\ell(N\cap A)>0$ for any neighbourhood $A$ of $X$, such that $X$ is a.s.r.t.\ $N$.
\end{itemize}
\end{theorem}

Before proving this theorem, we give a useful lemma that will be used in the proof. For any measurable set $N$ we define the density of $N$ at $x$ to be
$$
F(x)=\lim_{\epsilon\rightarrow 0} \frac{\ell(B_{\epsilon}(x)\cap N)}{\ell(B_{\epsilon}(x))}
$$
and recall that the Lebesgue Density Theorem \cite{Fal86} states that for
$\ell$-almost all $x\in N$ we have $F(x)=1$. In such a case we say that $x$ is a {\em point of Lebesgue density} for $N$.

\begin{lemma}\label{lem_density}
Suppose that $N$ has positive measure and $Y$ be any closed and bounded subset of $N$ with zero measure. Then for any $\epsilon>0$ one can find an open set $V$ containing $Y$ with
$$
\ell(V\cap N)/\ell(V)>1-\epsilon.
$$
\end{lemma}

\proof
Although $Y$ need not contain any points of Lebesgue density for $N$, there is at least one point $x\in N$ of Lebesgue density, and so we choose $\delta>0$ such that
$$
\ell(B_{\delta}(x)\cap N)/\ell(B_{\delta}(x))>1-\frac{\epsilon}{2}.
$$
Now let $V=B_{\delta}(x)\cup B_{\eta}(Y)$. Because of outer regularity of $\ell$, $\eta$ can be chosen small enough to ensure that $\ell(B_{\eta}(Y))$ is as small as desired, and hence the result holds.
\qed

With a slight modification of the argument, one can assume that $V$ is connected and open in the statement of the above Lemma; however it may be very far from being a ball in terms of the relationship between diameter and volume of the set.

~

\proof[of Theorem~\ref{thm_main}]
For (a) suppose that $X$ is p.a.s.\ and let $N$ be a set for which $X$ is a.s.r.t.. By Definition~\ref{def4}, for any $\epsilon>0$ there exists a
$\delta_0>0$ such that
$$
\frac{\ell(B_{\delta}(X)\cap N)}{\ell(B_{\delta}(X))}>1-\epsilon
$$
for all $\delta<\delta_0$. In Definition~\ref{def2} we set $N=D^c$ and
$V=B_{\delta}(X)$ (where $\delta$ is sufficiently small so that
$B_{\delta}(X)\subset U$), and we prove that $X$ is e.a.s.. For (b), note that this
follows because $N=D^c$ is a subset of the basin of attraction of $X$ and has
positive measure as $\ell(V\cap N)>0$ for some set $V$. Hence it is a weak attractor, and contains a Milnor attractor \cite{AshTer00}. Finally, for case (c) suppose
firstly that $X$ is e.a.s., then it is stable relative to the set $N=D^c$ and
$\ell(N\cap A)>0$ for any neighbourhood $A$ of $X$. The converse for (c)
follows similarly, on applying Lemma~\ref{lem_density}.
\qed

There are examples that show that, in general, converses of (a,b) do not hold; for a counterexample to the converse of (a) we refer to \cite{KirSil94} who present heteroclinic cycles that are, in our terminology, e.a.s.\ but not p.a.s.. For a counterexample to the converse of (b), there are ``unstable attractors'' \cite{AshTim05}, though only for a weaker assumption - that $\Phi_t$ is a semiflow. These ``unstable attractors'' are Milnor attractors that have zero basin measures within a small enough neighbourhood of the attractor. It is not clear whether the converse of (b) is true for flows (possibly subject to some smoothness assumptions). Note that Theorem~\ref{thm_main} is a generalization of comments already made in \cite[p1369]{Bra94} which assume $N$ to be an open set. There are examples of heteroclinic cycles that are not asymptotically stable relative to any open set, but that do seem to be asymptotically stable relative to a positive measure ``riddled'' set \cite{AshCho00}.

\subsection{Geometry of global basins: the stability index}

We suggest that it is useful to distinguish between different local geometries for e.a.s.\ sets. To this end, consider $X$ an invariant set in $\R^n$ and let $N=\cB(X)$ denote its (global) basin of attraction; we assume that the flow $\Phi_t$ is smooth. Pick a point $x\in X$, define
\begin{equation}\label{eq:sigma_eps}
\Sigma_{\epsilon}(x)= \frac{\ell(B_{\epsilon}(x)\cap N)} {\ell(B_{\epsilon}(x))}
\end{equation}
and note that $0\leq \Sigma_{\epsilon}(x)\leq 1$.

\begin{definition}\label{def_stab_index}
For a point $x\in X$ we define the {\em stability index} of $X$ at $x$ to be
$$
\sigma(x):=\sigma_+(x)-\sigma_-(x)
$$
which exists when the following converge
$$
\sigma_-(x):=\lim_{\epsilon\rightarrow 0} \left[\frac{\ln(\Sigma_{\epsilon}(x))
}{\ln(\epsilon)}\right],~~~~
\sigma_+(x):=\lim_{\epsilon\rightarrow 0} \left[\frac{\ln(1-\Sigma_{\epsilon}(x))}{\ln(\epsilon)}\right].
$$
We use the convention that $\sigma_-(x)=\infty$ if there is an $\epsilon_0>0$ such that $\Sigma_{\epsilon}(x)=0$  for all $\epsilon<\epsilon_0$, and $\sigma_+(x)=\infty$ if $\Sigma_{\epsilon}(x)=1$ for all $\epsilon<\epsilon_0$. Note that $\sigma_{\pm}(x)\geq 0$ and so we can assume that $\sigma(x)\in[-\infty,\infty]$.
\end{definition}

\begin{figure}
\centerline{\epsfig{file=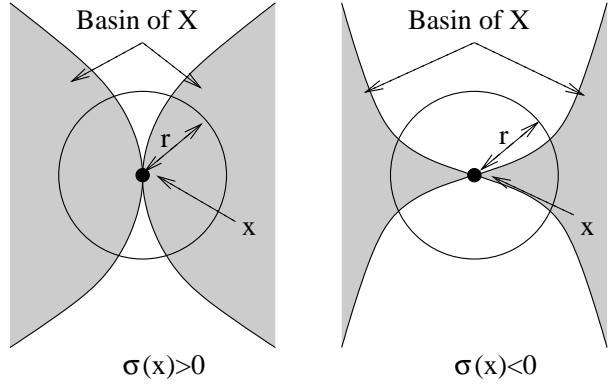,width=8cm}}
\caption{
Schematic diagram illustrating how the stability index $\sigma(x)$ of
a point $x\in X$ relates to the local geometry of the basin of
attraction of $X$ (shaded region).
For $\sigma(x)>0$, the measure of points in a ball of radius $r$ that are in
the complement of the basin goes to zero, relative the measure of the ball, as
$r^{|\sigma(x)|}$. For $\sigma(x)<0$, this estimate applies to the basin itself.
}
\label{fig_schem_index}
\end{figure}

The stability index may not exist at certain points in $X$ (an example is given in Section~\ref{sec_conc}), and may vary throughout $X$ when it does exist. Figure~\ref{fig_schem_index} illustrates how the local geometry of the basin relates to the sign of $\sigma(x)$ for a point $x\in X$. Note that $\sigma(x)=+\infty$ is the ``strongest'' form of local stability while $-\infty$ is the ``weakest''.  The following Lemma characterizes some basic properties of the index:

\begin{lemma}\label{lemma2}
Suppose that $\sigma(x)$ is defined for some $x\in X\subset \R^n$; then the following hold:
\begin{itemize}
\item[(a)] If one of $\sigma_{\pm}(x)$ converges to a positive value then the other converges to zero (i.e. only one of $\sigma_+(x)$ and $\sigma_-(x)$ can be non-zero).
\item[(b)] If $\sigma(x)=c>0$ then $1-\Sigma_{\epsilon}(x)=O(\epsilon^c)$
(and in particular $\Sigma_{\epsilon}(x)\rightarrow 1$ as $\epsilon\rightarrow 0$).
\item[(c)] If $\sigma(x)=-c<0$ then $\Sigma_{\epsilon}(x)=O(\epsilon^c)$
(and in particular $\Sigma_{\epsilon}(x)\rightarrow 0$ as $\epsilon\rightarrow 0$).
\end{itemize}
\end{lemma}

\proof
For (a) note that if $\sigma_-(x)>0$ then $\lim_{\epsilon\rightarrow 0} \Sigma_{\epsilon}(x)=0$; this implies that $1-\Sigma_{\epsilon}$ converges to $1$ as $\epsilon\rightarrow 0$ and so $\sigma_+(x)=0$. The other case is argued in a similar way. (b) follows on noting by (a) that $c=\sigma(x)=\sigma_+(x)>0$ and $\sigma_-(x)=0$. Hence we have from the definition of $\sigma_+(x)$ that $1-\Sigma_{\epsilon}(x)=O(\epsilon^c)$. A similar argument gives (c).
\qed

The main result in this section is the following; this can be generalized to cases where $\sigma(x)$ is measured relative to any measurable invariant set $N$.

\begin{theorem}\label{thm_sigma_const}
Suppose that $N=\cB(X)$ is the basin of a compact invariant set X
for a $C^1$-smooth flow $\Phi_t(x)$. Then for any $x$ the index $\sigma(x)$ is constant on trajectories, whenever it is defined.
\end{theorem}

\proof
Fix $x\in X$ such that $\sigma(x)$ is defined, and pick any $t>0$. Let $\phi(x)=\Phi_t(x)$. Because $\phi$ is a $C^1$ diffeomorphism, one can find an $\eta>0$ such that there is an $L>1$ and an $M>1$ with
\begin{equation}\label{eq:bounds}
\frac{1}{L}< \det(D\phi(y))< L,~~\frac{1}{M}< \|D\phi(y)\|< M
\end{equation}
for all $y\in B_{\eta}(x)$, where $D$ denotes the derivative (Jacobian) of the map. We assume that $L,M$ are chosen so that the same inequalities are satisfied by $D\phi^{-1}$ for $z\in B_{\eta}(\phi(x))$. As a consequence of this, one can find $\eta'$ with $0<\eta'<\eta$ such that
\begin{equation}\label{eq:contains}
B_{\epsilon/M}(x)\subset \phi^{-1}(B_{\epsilon}(\phi(x)))\subset B_{M\epsilon}(x)
\end{equation}
for any $\epsilon<\eta'$. Writing $\chi_N(y)$ as the indicator function for $N$ and $y\in B_{\eta}(x)$ we have
\begin{eqnarray*}
\Lambda=\ell\left(B_{\epsilon}(\phi(x))\cap N\right)
&=& \int_{y\in B_{\epsilon}(\phi(x))} \chi_N(y)\,d\ell(y)\\
&=& \int_{z\in \phi^{-1}(B_{\epsilon}(\phi(x)))} \chi_N(z)\det(D\phi^{-1}(z)) \,d\ell(z)
\end{eqnarray*}
where in the last line we have substituted $y=\phi(z)$ and we have used the fact that $N$ is invariant, so that
\begin{equation}
\chi_N(y)=1~~~\Leftrightarrow~~~\chi_N(\phi(y))=1.
\label{eq_N_invart}
\end{equation}
Hence, using (\ref{eq:bounds},\ref{eq:contains}),
\begin{eqnarray*}
\frac{1}{L}\int_{z\in B_{\epsilon/M}(x)} \chi_N(z) \,d\ell(z)
&\leq & \frac{1}{L}\int_{z\in \phi^{-1}(B_{\epsilon}(\phi(x))} \chi_N(z) \,d\ell(z)\\
&<& \int_{z\in \phi^{-1}(B_{\epsilon}(\phi(x)))} \chi_N(z)\det(D\phi^{-1}(z)) \,d\ell(z)=\Lambda\\
&<& L\int_{z\in \phi^{-1}(B_{\epsilon}(\phi(x)))} \chi_N(z) \,d\ell(z)\\
&\leq & L\int_{z\in B_{\epsilon M}(x)} \chi_N(z) \,d\ell(z)
\end{eqnarray*}
meaning that for all $\epsilon<\eta'$ we have
\begin{equation}
\frac{1}{L}\ell(B_{\epsilon/M}(x)\cap N)
< \ell(B_{\epsilon}(\phi(x))\cap N)
< L\ell(B_{\epsilon M}(x)\cap N).
\end{equation}
This means that from (\ref{eq:sigma_eps}), there is a $K=LM^n$ such that for all small enough $\epsilon$ we have
$$
\frac{1}{K} \Sigma_{\epsilon/M}(x)<\Sigma_{\epsilon}(\phi(x))< K \Sigma_{\epsilon M}(x)
$$
(we have used the property that $\ell(B_{M\epsilon}(x))=M^n\ell(B_{\epsilon}(x))$ for any $\epsilon>0$ and $x$). Hence we have
\begin{eqnarray*}
\left[\frac{\ln \epsilon-\ln M}{\ln \epsilon}\right]
\frac{\ln(\Sigma_{\epsilon/M}(x))}{\ln (\epsilon/M)}-\frac{\ln K}{\ln \epsilon}
&=& \frac{\ln(\Sigma_{\epsilon/M}(x))}{\ln \epsilon}-\frac{\ln K}{\ln \epsilon}
= \frac{\ln \left(\frac{1}{K} \Sigma_{\epsilon/M}(x)\right)}{\ln \epsilon} \\
&<&\frac{\ln(\Sigma_{\epsilon}(\phi(x)))}{\ln \epsilon}\\
&<& \frac{\ln \left(K \Sigma_{\epsilon M}(x)\right)}{\ln \epsilon} = \frac{\ln(\Sigma_{\epsilon M}(x))}{\ln \epsilon} +\frac{\ln K}{\ln \epsilon}\\
&=&\left[\frac{\ln \epsilon+\ln M}{\ln \epsilon}\right]
\frac{\ln(\Sigma_{\epsilon M}(x))}{\ln (\epsilon M)}+\frac{\ln K}{\ln \epsilon}
\end{eqnarray*}
and taking the limits as $\epsilon\rightarrow 0$ we have
\begin{equation}
\sigma_-(x) \leq \sigma_-(\phi(x))\leq \sigma_-(x).
\end{equation}
A similar argument on substituting $N$ by its complement gives $\sigma_+(x)=\sigma_+(\phi(x))$ and hence the value of $\sigma(x)$ is constant along trajectories of $\Phi_t$.
\qed

Note that this argument works for any $C^1$-diffeomorphism $\phi$ for which $N$ is invariant, meaning the result can be used to show that $\sigma(x)$ is invariant under $C^1$-conjugation - it is an invariant of the dynamics. Note also that although $\sigma(x)$ is constant on a given trajectory, it may depend on which trajectory is chosen.

The stability index can be used to determine e.a.s.\ and p.a.s.\ by the following theorem. However, converses of the following theorem are not expected to be true in general as $\sigma(x)$ may be negative on a ``lower dimensional'' set of trajectories within $X$, or may not converge.

\begin{theorem}\label{thm_milnor}
Suppose that for all $x\in X$ the stability index $\sigma(x)\in[-\infty,\infty]$ is defined.
\begin{itemize}
\item If there is a point $x\in X$ such that $-\infty<\sigma(x)$ then $X$ is essentially asymptotically stable (e.a.s.), and contains a Milnor attractor.
\item If there is a $c>0$ such that $c<\sigma(x)$ for all $x\in X$ then $X$ is predominantly asymptotically stable (p.a.s.).
\end{itemize}
\end{theorem}

\proof
(a) The fact that $-\infty<\sigma(x)$ implies in particular that $N=\cB(X)$ contains a set of positive measure, and so by Theorem~\ref{thm_main}(c) it is e.a.s.. By the Theorem~\ref{thm_main}(b), $X$ contains a Milnor attractor.
(b) Note that the basin of attraction $N$ of $X$ is such that for any $\delta$, $\Sigma_{\epsilon}(x) > 1-\delta$  for all $x$, and some $\epsilon$ depending on $x$. By compactness of $X$ one can choose an $\epsilon$ small enough that $\ell(B_{\epsilon}(X)\cap N)\geq (1-\delta)\ell(B_{\epsilon}(X))$, implying p.a.s.\ of $X$.
\qed

\subsection{The local stability index}\label{seclsind}

While Definition~\ref{def_stab_index} considers the global basin of attraction, the stability index $\sigma(x)$ can be adapted to provide a useful concept from purely local properties of the attractor. We define the $\delta$-local basin of attraction to be the basin of attraction of $X$ relative to $B_{\delta}(X)$; that is,
\begin{equation}\label{eq:Sigma2}
\cB_{\delta}(X):= \{ x~:~ \omega(x)\subset X \mbox{ and }\Phi_t(x)\in B_{\delta}(X) \mbox{ for all }t>0\}.
\end{equation}
Note that $\cB_{\delta}(X)$ is forwards, but not necessarily backwards invariant under the flow. The limit of the stability index for points relative to the $\delta$-local basin as $\delta\rightarrow 0$ is called the {\em local stability index} $\sigma_{\loc}(x)$ for $X$. More precisely, we define
\begin{equation}\label{eq:sigma_eps2}
\Sigma_{\epsilon,\delta}(x)= \frac{\ell\left(B_{\epsilon}(x)\cap \cB_{\delta}(X)\right)} {\ell(B_{\epsilon}(x))}
\end{equation}
and for a point $x\in X$ we define the local stability index of $X$ at $x$ to be
$$
\sigma_{\loc}(x):=\sigma_{\loc,+}(x)-\sigma_{\loc,-}(x)
$$
which exists when the following converge
$$
\sigma_{\loc,-}(x):=\lim_{\delta\rightarrow 0}\lim_{\epsilon\rightarrow 0} \left[\frac{\ln(\Sigma_{\epsilon,\delta}(x))
}{\ln(\epsilon)}\right],~~~~
\sigma_{\loc,+}(x):=\lim_{\delta\rightarrow 0}\lim_{\epsilon\rightarrow 0} \left[\frac{\ln(1-\Sigma_{\epsilon,\delta}(x))
}{\ln(\epsilon)}\right]
$$
with the same conventions as before. The definition works for discrete time systems as well as continuous time, without further modification. Note that the local stability index is computed for $\delta$ small
and fixed before taking the limit as $\delta\rightarrow 0$.

One can weaken the assumptions in  Theorem~\ref{thm_sigma_const} to give the same conclusion; the critical step is that if $\eta<\delta$ and we assume that $\Phi_s(y)\in B_{\eta}(X)$ for $0<s<t$ then (\ref{eq_N_invart}) still holds- and by continuity of $\phi$ one can choose a small enough $\epsilon$ that $x$ and $\phi(x)$ are guaranteed to be in such a $B_{\eta}(X)$.

\subsection{Stability indices for sections to the flow}

Suppose that $\Phi_t(x)$ has an attractor $X\subset \R^n$ and pick a point $x\in X$. Let $S\subset \R^n$ be a smooth $n-1$-dimensional subspace containing $x$ that is transverse to the flow at $x$. One can relate the stability index $\sigma(x)$ or $\sigma_{\loc}(x)$ to the stability index for the dynamics defined by the return map $F$ on $S$ as follows.

\begin{theorem}\label{thm_flow_to_map}
Suppose that $X$ is invariant for a $C^1$-smooth flow $\Phi_t(x)$ and that $N$ is a (local) basin for $X$. Suppose that $S$ is a codimension one surface that is transverse to the flow at $x$; then $\sigma(x)$ can be computed relative to the intersection of $N$ with $S$ on substituting $\Sigma_{\epsilon}(x)$ by
$$
\Sigma_{\epsilon,S}(x)= \frac{\ell_S(B_{\epsilon}(x)\cap N\cap S)} {\ell_S(B_{\epsilon}(x)\cap S)}.
$$
\end{theorem}

\proof
Let $N=\cB(X)$; the argument for local basins will be similar. Note that $N$
is invariant implies that it is a union of trajectories. We consider local
coordinates in $\R^n$ near $x$ that are the coordinates in $S$ and time. Pick any small
$\epsilon>0$; by simple geometric arguments (i.e. you can always put a cylinder in a larger sphere, and a sphere in a
larger cylinder) there is a constant $K>1$ such that
$$
B_{\epsilon/K}(x) \subset (B_{\epsilon}(x)\cap S) \times [-\epsilon,\epsilon] \subset B_{\epsilon K}(x).
$$
Using the product structure of Lebesgue measure we have
$$
\ell(B_{\epsilon/K}(x)) < 2\epsilon \times \ell_S(B_{\epsilon}(x)\cap S) < \ell(B_{\epsilon K}(x))
$$
with a similar inequality for $\ell_S(B_{\epsilon}(x)\cap N\cap S)$. Hence
$$
\frac{\ell(B_{\epsilon/K}(x)\cap N )}{\ell(B_{\epsilon K}(x))} <
\frac{\ell_S(B_{\epsilon}(x)\cap N\cap S)}{\ell_S(B_{\epsilon}(x)\cap S)} <
\frac{\ell(B_{\epsilon K}(x)\cap N )}{\ell(B_{\epsilon/K}(x))}
$$
meaning that
$$
\Sigma_{\epsilon,S}(x)= \frac{\ell_S(B_{\epsilon}(x)\cap N\cap S)} {\ell_S(B_{\epsilon}(x)\cap S)}
$$
and as before (for the flow) $\Sigma_{\epsilon}(x)$ satisfies the inequalities
$$
\frac{1}{(2K)^n} \Sigma_{\epsilon/K}(x) <\Sigma_{\epsilon,S}(x)< (2K)^n\Sigma_{\epsilon K}(x),
$$
where we have used the fact that $\ell(B_{\epsilon K}(x))=(2K)^n \ell(B_{\epsilon/K}(x))$. In particular, the scalings of these quantities are the same as $\epsilon\rightarrow 0$.
\qed

Theorem~\ref{thm_flow_to_map} implies, for example, that if there is a return map for the flow on $S$ then the stability index of trajectories for a flow can be computed by examining the stability index for the intersection of the basin with a suitable surface of section.

\section{Robust heteroclinic cycles}\label{sec_hetc}

Suppose that $\Gamma$ is a finite group acting orthogonally on $\R^n$, and that $f:\R^n\to\R^n$ is a $\Gamma$-equivariant vector field, i.e.
$$
f(\gamma x)=\gamma f(x),\quad \mbox{ for all }\gamma\in\Gamma.
$$
Let $\xi_j$, $j=1,\ldots,m$, be hyperbolic equilibria for
$$
\dot{x}=f(x)
$$
with stable and unstable manifolds $W^s(\xi_j)$ and $W^u(\xi_j)$
respectively, and let $s_j=W^u(\xi_j)\cap W^s(\xi_{j+1})\ne\emptyset$ be
connections between $\xi_j$ and $\xi_{j+1}$, where $\xi_{m+1}=\xi_1$; then the group orbit $X$ of the equilibria and the connections
$$
X=\mbox{clos}\left(\{\gamma s_j~:~\ j=1,\ldots,m,\ \gamma\in\Gamma\}\right)
$$
is called a {\it heteroclinic cycle}. Recall that for a group $\Gamma$ acting on $\R^n$, the isotropy of the point $x\in\R^n$ is the subgroup
$$
\Sigma_x=\{\gamma\in\Gamma\ ~:~\ \gamma x=x\}
$$
while for a subgroup $\Sigma\subset\Gamma$, a {\em fixed-point subspace} of $\Sigma$ is the linear subspace
$$
{\rm Fix}(\Sigma)=\{x\in\R^n\ ~: \ \sigma x=x\mbox{ for all } \sigma\in\Sigma\}.
$$

In the absence of symmetry or other constraints, a vector field with a
heteroclinic cycle is structurally unstable, i.e. there are arbitrarily
small perturbations of $f$ to $g$, such that the heteroclinic cycle does not
exist for the vector field $g$. For symmetric vector fields, heteroclinic cycles
may be robust, as long as each connection is robust within some invariant subspace and only symmetric perturbations are allowed \cite{Kru97}.

\subsection{Local structure: eigenspaces and simple cycles}

A sufficient condition for a cycle $X$ to be {\em structurally stable}
(or {\em robust}), is that for all $j$ there exists a subspace
$P_j$ such that $P_j={\rm Fix}(\Sigma_j)$  for some $\Sigma_j\subset\Gamma$,
$s_j\subset P_j$, $\xi_{j+1}$ is a sink in $P_j$. Denote $L_j=P_j\cap P_{j-1}$.
We denote the isotropy subgroup of points in $L_j\setminus\{0\}$ by $T_j$. Note that $X\ominus Y$, where $Y$ is a linear subspaces of the inner product space $X$, denotes the orthogonal complement to $Y$ in $X$.

If $X$ is a structurally stable heteroclinic cycle then the eigenvalues of
$(df)_{\xi_j}$ can be divided into four classes:
\begin{itemize}
\item Eigenvalues with associated eigenvectors in $L_j$ are called {\it radial},
the maximal real part of radial eigenvalues being $-r_j$.
\item Eigenvalues with associated
eigenvectors in $P_{j-1}\ominus L_j$ are called {\it contracting},
the maximal real part of contracting eigenvalues being $-c_j$.
\item Eigenvalues
with associated eigenvectors in $P_j\ominus L_j$ are called {\it expanding},
the maximal real part of expanding eigenvalues being $e_j$.
\item The remaining eigenvalues are called {\it transverse}, the maximal real part of
transverse eigenvalues being $t_j$.
\end{itemize}

The heteroclinic cycle $X\in\R^4\setminus \{0\}$ is called a {\it simple robust heteroclinic cycle} (in $\R^4$) \cite{KruMel04} if for all $j$:
\begin{itemize}
\item All eigenvalues of $(df)_{\xi_j}$ are distinct, $\dim P_j=2$ and $X$
intersects with each connected component of $L_j\setminus\{0\}$ in at most one point.
\end{itemize}

For simple cycles, $L_j$ and the three remaining subspaces are one-dimensional,
hence there is a unique real eigenvalue of each type. Moreover, for simple
cycles either $T_j\cong\Z^2_2$ and $\Sigma_j\cong\Z_2$ for all $j$ or
$T_j\cong\Z^3_2$ and $\Sigma_j\cong\Z^2_2$ for all $j$ (see Proposition 3.1 in
\cite{KruMel04}), and each simple cycle is of one of three types discussed by \cite{KruMel04}.

\begin{definition}
Suppose that $X$ is a simple heteroclinic cycle that is robust for a vector
field in $\R^4$ with a finite symmetry group. We say
\label{def_ABC}
\begin{itemize}
\item $X$ is of {\it Type A} if $\Sigma_j\cong\Z_2$ for all $j$.
\item $X$ is of {\it Type B} if there is a subspace $Q$ of $\R^4$ with $\dim(Q)=3$ such that $Q=\fix(\tilde{\Sigma})$ for some $\tilde{\Sigma}\subset\Gamma$ and $X\subset Q$.
\item $X$ is of {\it Type C} if it is neither of Type A nor of Type B.
\end{itemize}
\end{definition}

The work of \cite{KruMel04} goes on to differentiate between four varieties of
Type B cycles (denoted by $B^+_1$, $B^+_2$, $B^-_1$ and $B^-_3$) and three
varieties of Type C cycles (denoted by $C^-_1$, $C^-_2$ and $C^-_4$), depending
on the number of equilibria involved in the cycle and action of the group $\Gamma$.
In Section~\ref{sec_ABC} we examine the stability of cycles in $\R^4$ using Poincar\'e maps, where the structure of the maps depend on the type of cycle.

\subsection{Local stability for heteroclinic cycles}\label{sec_stab}

For a heteroclinic cycle $X$ comprised of one-dimensional connections $s_j$,
$j=1,\ldots,m$ ($s_j$ is the connection from $\xi_{j-1}$ to $\xi_j$),
its local attraction properties are described by the set of stability
indices of the trajectories
$$
{\mathbf\sigma}=(\sigma_1,\ldots,\sigma_m).
$$
where
$$
\sigma_j=\sigma(x)
$$
for $x$ an arbitrary point on $s_j$. The following lemma will be useful later on:

\begin{lemma}\label{lem_con}
Let a simple heteroclinic cycle $X$ be comprised of one-dimensional connections
and suppose that $-\infty<\sigma_j$ for some $j$. Then $X$ is a Milnor attractor.
\end{lemma}

\proof
This follows from the fact that $-\infty<\sigma_j(x)$ implies that
$\ell(\cB(X))>0$ and so $X$ is a weak attractor. Since no invariant subset of the cycle
can be a Milnor attractor, $X$ must itself be a Milnor attractor.
\qed

In what follows, we calculate local stability indices for different types of
simple robust heteroclinic cycles in $\R^4$. Following \cite{KirSil94,KruMel04},
to examine stability we construct a Poincar\'e map in the vicinity of the cycle.

\subsection{Stability indices for return maps}\label{secindmap}

Section~\ref{sec_hetc} gave definitions for radial, contracting,
expanding and transverse eigenvalues of the linearization $(df)_{\xi_j}$.
Simple cycles in $\R^4$ will possess a single eigenvalue of each type.
Let $(\tilde u,\tilde v,\tilde w,\tilde z)$ be local coordinates near $\xi_j$
in the basis of the four associated eigenvectors, and
$\tilde B_{\tilde\delta}(\xi_j)$ be a neighbourhood of $\xi_j$ defined as
$$
\tilde B_{\tilde\delta}(\xi_j)=\{(\tilde u,\tilde v,\tilde w,\tilde z)\ |\
\max(|\tilde u|,|\tilde v|,|\tilde w|,|\tilde z|)<\tilde\delta\}
$$
where $\tilde\delta$ is small, denote by $(u,v,w,z)$ the scaled coordinates
$(u,v,w,z)=(\tilde u,\tilde v,\tilde w,\tilde z)/\tilde\delta$.
In the $(u,v)$ plane we will also employ plane polar coordinates $(r,\theta)$,
$u=r\cos\theta$ and $v=r\sin\theta$. For a small $\tilde\delta$, the vector field $f$ can be linearly approximated in $\tilde B_{\tilde\delta}(\xi_j)$. We assume that $\delta$ in (\ref{eq:Sigma2}) is sufficiently small, so that
$$
B_{\delta}(\xi_j)\subset\tilde B_{\tilde\delta}(\xi_j)~\mbox{for all}~~ j.
$$
In $\tilde B_{\tilde\delta}(\xi_j)$ we consider the linearised system (\ref{eq_ode})
\begin{equation}\label{lmap}
\begin{array}{l}
\dot u=-r_j u\\
\dot v=-c_j v\\
\dot w=e_j w\\
\dot z=t_j z
\end{array}
\end{equation}
This gives an accurate approximation of the nonlinear flow as long
as the linear system has no low order resonances.

The connection $s_j$ is tangent to the subspace $u=v=z=0$. The heteroclinic
connection to $\xi_j$ lies in $P_{j-1}$ where local coordinates are
$u$ and $v$. For
$$
H^{(out)}_j=\{(u,v,w,z)~:~|u|,|v|,|z|\le1, w=1\},
$$
$$
H^{(in)}_j=\{(r,\theta,w,z)~:~r=1,|w|,|z|\le1\}.
$$
the {\em first return map} $\phi_j:H^{(in)}_j\to H^{(out)}_j$ is defined
near each equilibrium in $\tilde H^{(in)}_j=Q_j^{(in)}\cap H^{(in)}_j$, where
$Q_j^{(in)}=\{(z,w)|\ |z|<|w|^{t_j/e_j}\}$\footnote{Strictly speaking,
the maps are defined for $|z|<K(1-\delta)|w|^{t_j/e_j}$, where $K$ is a constant
and $\delta$ is small \cite{KirSil94}. For simplicity, we ignore $K$ and $\delta$,
because for small $z$ and $w$ they do not enter into asymptotically significant terms.}.
For each connection we define {\em connecting diffeomorphisms} $\psi_j:H^{(out)}_j\to H^{(in)}_{j+1}$
and their compositions $\tilde g_j=\psi_j\circ\phi_j:\tilde H^{(in)}_j\to H^{(in)}_{j+1}$.
The Poincar\'e map is the composition $\tilde g=\tilde g_m\circ\ldots\circ \tilde g_1:H^{(in)}_1\to H^{(in)}_1$.

As shown in \cite{KirSil94,KruMel95a}, at leading order the maps have the
form\footnote{The maps for negative $w$ are defined as follows. Consider a
neighbourhood of a point $\xi_j$. Two heteroclinic connections enter this
neighbourhood, $\xi_{j-1}\to\xi_j$ and $\gamma^{(in)}\xi_{j-1}\to\xi_j$,
where $\gamma^{(in)}$ is any symmetry, satisfying $\gamma^{(in)}\in T_j$ and
$\gamma^{(in)}\not\in\Sigma_{j-1}$. Two heteroclinic connections exit the
neighbourhood: $\xi_j\to\xi_{j+1}$ and $\xi_j\to\gamma^{(out)}\xi_{j+1}$, where
$\gamma^{(out)}$ is a symmetry, satisfying $\gamma^{(out)}\in T_j$ and
$\gamma^{(out)}\not\in\Sigma_j$. The local map
$\phi_j:H^{(in)}_j\to H^{(out)}_j$ is defined only for $w$ and $v$ of
particular signs, say $w>0$ and $v>0$. For $w<0$, the local map acts to
$\tilde H^{(out)}_j$, where $\tilde H^{(out)}_j=\gamma^{(out)}H^{(out)}_j$;
for $v<0$, it is defined in $\gamma^{(in)}H^{(in)}_j$. Due to existence
of the symmetries $\gamma^{(in)}$ and $\gamma^{(out)}$, we can consider
$\phi_j$ for arbitrary $w$ and $v$: by applying these symmetries the local map
can be defined for $w$ and $v$ of arbitrary signs.}
\begin{equation}\label{fhit}
\phi_j(u,v,w,z)=(u|w|^{r_j/e_j},vw^{c_j/e_j},1,z|w|^{-t_j/e_j})
\end{equation}
and
\begin{equation}\label{cdif}
\psi_j(u,v,w,z)=(1,\theta_0+\beta_ju,\alpha_{j_1}v+\alpha_{j_2}z,
\alpha_{j_3}v+\alpha_{j_4}z).
\end{equation}
For these maps only $(w,z)$ coordinates are important \cite{KirSil94,KruMel95a},
and restricting to these coordinates the map $g_j:\R^2\to\R^2$ we have
\begin{equation}\label{eq_mapg0a}
g_j(w,z)=(\alpha_{j_1}w^{c_j/e_j}+\alpha_{j_2}z|w|^{-t_j/e_j},
\alpha_{j_3}w^{c_j/e_j}+\alpha_{j_4}z|w|^{-t_j/e_j}).
\end{equation}
For cycles of Type A, generically $\alpha_{jk}\ne0$ for all $j=1,\ldots,m$,
$k=1,2,3,4$. For cycles of Type B, $\alpha_{j_2}=\alpha_{j_3}=0$ and
$\alpha_{j_1}\alpha_{j_4}\ne0$. For cycles of Type C, $\alpha_{j_1}=\alpha_{j_4}=0$
and $\alpha_{j_2}\alpha_{j_3}\ne 0$.

Thus, for a point $x= X\cap H^{(in)}_1$ on the cycle $X$ we have associated
the map
\begin{equation}\label{mapg}
g=g_m\circ g_{m-1}\circ\ldots\circ g_1:\R^2\to \R^2.
\end{equation}
We will also use the notation $g_{l,k}=g_l\circ g_{l-1}\circ\ldots\circ g_1\circ g^k$.

Similarly to the local stability index for a point $x\in X$
(see Section~\ref{seclsind}), we define a stability index $\sigma^g$ for the map (\ref{mapg}).
Note that $B_{\epsilon}$ is the ball of radius $\epsilon$ in $\R^n$ centered at $0$, and we define
\begin{equation}%\label{eq:Sigma_g}
\cB_{\delta}^g:= \{ x~:~ x\in\R^n,\
|g_{l,k}x|<\delta \mbox{ for all }0\le l\le m-1,\ k\ge0\},
\end{equation}
to be the $\delta$-local basin of attraction of $0$ in $\R^n$ for the map $g$ (\ref{mapg}). The local stability index is defined to be
$$
\sigma^g:=\sigma_+^g-\sigma_-^g
$$
where
$$
\sigma_-^g:=\lim_{\delta\rightarrow 0}\lim_{\epsilon\rightarrow 0}
\left[\frac{\ln(\Sigma_{\epsilon,\delta})}{\ln(\epsilon)}\right],~~~~
\sigma_+^g:=\lim_{\delta\rightarrow 0}\lim_{\epsilon\rightarrow 0}
\left[\frac{\ln(1-\Sigma_{\epsilon,\delta})}{\ln(\epsilon)}\right],
$$
with
\begin{equation}
\label{eq:Sigma_g}
\Sigma_{\epsilon,\delta}= \frac{\ell(B_{\epsilon}(0)\cap \cB_{\delta}^g)} {\ell(B_{\epsilon}(0))}.
\end{equation}

Because of the asymptotic independence of the return map on two of the
coordinates, we can effectively reduce the computation of the stability index
for heteroclinic cycles in $\R^4$ to a calculation on a section to the cycle in $\R^2$.

\begin{theorem}\label{th_xg}
Let $g:\R^2\to\R^2$ (\ref{mapg}) be the map associated with a point $x=X\cap H_1^{(in)}$, where $X$ is a simple heteroclinic cycle in $\R^4$. Then
$$
\sigma_{\loc}(x)=\sigma^g.
$$
\end{theorem}

\proof
First, we prove that $\sigma_{\loc}(x)\ge\sigma^g$. Denote
$Q=[-\epsilon,\epsilon]^2\times\cB_{\delta}^g$. If
$y\equiv(u,v,w,z)\in B_{\epsilon}(x)\setminus Q$, then there exist
$j$ and $k$, such that $|g_{j,k}(y)|>\delta$. For small $\delta$, a trajectory
near the heteroclinic cycle is approximated by the maps $\phi_j$
(\ref{fhit}) and $\psi_j$ (\ref{cdif}),
where the coordinates $(w,z)$ are independent of $(u,v)$.
Hence, for the point $(\tilde u,\tilde v,\tilde w,\tilde z)$, which is the $(k+1)$st intersection of $\Phi_t(y)$ with $H^{(in)}_j$, we have $|(\tilde w,\tilde z)|>\delta/2$. Hence $y\notin\cB_{\delta/2}(X)$, and therefore
$$
\ell(B_{\epsilon}(x)\cap \cB_{\delta/2}(X))<
4\epsilon^2\ell(B_{\epsilon}\cap \cB_{\delta}^g),
$$
implying that $\sigma_{\loc}(x)\ge\sigma^g$.

Second, we prove that $\sigma_{\loc}(x)\le\sigma^g$. If
$y\equiv(u,v,w,z)\in B_{\epsilon}(x)\cap Q$, then for any intersection
$(1,\theta_1,\tilde w,\tilde z)$ (here we are using polar coordinates
in the $(u,v)$ plane, $\theta_1$ is the difference between $\theta$ and
$\theta_0$, the distance of the point from the cycle is denoted
$|(\theta_1,\tilde w,\tilde z)|$) of $\Phi_t(y)$ with $H^{(in)}_j$,
$|(\tilde w,\tilde z)|<2\delta$ holds true.
Together with (\ref{fhit}) and (\ref{cdif}), it implies that in any intersection
$\theta_1<\beta\delta^r$, where $\beta=\max(\max_j(\beta_j),1)$ and
$r=\min(\min_j(r_j/e_j),1)$. Thus, we have proved that
for any intersection $\tilde y$ of $\Phi_t(y)$ with $H^{(in)}_j$ for any $j$ we have
\begin{equation}\label{ints}
d(\tilde y,X)=|(\theta_1,\tilde w,\tilde z)|<2\beta\delta^r.
\end{equation}
If a trajectory is close to the heteroclinic cycle, then there exist
constants $K_j$ such that
$d(s_j,\Phi_t(x))< K_j d(s_j,\Phi_{T_{j,k}}(x))$
in the interval $\tilde T_{j-1,k}<t<T_{j,k}$.
(Here $T_{j,k}$ is the time when $\Phi_t(x)$ crosses $H^{(in)}_j$ and $\tilde T_{j-1,k}$ is the time when it crosses
$H^{(out)}_{j-1}$.) Denote $K=\beta\max_j(K_j)$. In the vicinity of $\xi_j$ we
can consider the system (\ref{lmap}) using the approximated map $f$, hence
(\ref{ints}) is satisfied at the points of intersection, which implies that
$$
d(\Phi_t(y),X)<2K\delta^r\hbox{ for all }t>0.
$$
Hence $y\in\cB_{2K\delta^r}(X)$, and therefore
$$\ell(B_{\epsilon}(x)\cap \cB_{2K\delta^r}(X))>
\epsilon^2\ell(B_{\epsilon}\cap \cB_{\delta}^g).$$
Since $K>0$ and $r>0$ do not depend on $\delta$, this implies $\sigma_{\loc}(x)\le\sigma^g$ and therefore $\sigma_{\loc}(x)=\sigma^g$.
\qed

\section{Stability indices for heteroclinic cycles in $\R^4$}\label{sec_ABC}

In this section the stability indices for the connections of simple robust
heteroclinic cycles in $\R^4$ are calculated in terms of ratios of eigenvalues
$a_j=c_j/e_j$ and $b_j=-t_j/e_j$, $1\le j\le m$. We do this relative to the
classification of simple heteroclinic cycles in $\R^4$ of
Definition~\ref{def_ABC} and \cite{KruMel04}. Here only statements of the main
theorems and a sketch of the proof of the main theorem for type A cycles are presented.
The complete proofs are given in the Appendices.

Using the maps $\{g_j~,~j=1,\ldots,m\}$ from the previous section, we define return maps on sections
to any of the connections
$$
g^{(j)}:H_j^{(in)}\rightarrow H_j^{(in)}
$$
to be $g^{(j)}=g_{j-1}\circ\ldots\circ g_1\circ g_m\circ\ldots\circ g_{j}$, let
$$
\sigma_j=\sigma_{j,+}-\sigma_{j,-}
$$
and
$$
\sigma_{j,\pm}=\sigma_{\pm}^{g^{(j)}},
$$
where $\sigma_{\pm}^g$ are defined in Section~\ref{secindmap}. Note that in the previous section we used
$$
g=g^{(1)}
$$
and that $\sigma_j$ will effectively give the stability index for the connection that intersects $H_j^{(in)}$.

\subsection{Type A cycles}
%\label{typeA}

Consider the map $g=g_m\circ\ldots\circ g_1:\R^2\to\R^2$ where
\begin{equation}\label{eq_mapg}
g_j(w,z)\equiv(g_j^w(w,z),g_j^z(w,z))=
(A_jw^{a_j}+B_j|w|^{b_j}z,C_jw^{a_j}+D_j|w|^{b_j}z).
\end{equation}
For generic cycles of Type A we can assume $A_jB_jC_jD_j\neq 0$ and
$A_jD_j\neq B_jC_j$ for all $j$. Recall that
$g_{l,k}=g_l\circ\ldots\circ g_1 \circ g^k$.

Let us define
\begin{eqnarray*}
Q(r,\epsilon)&=&\{(w,z)~:~|(w,z)|<\epsilon,\ |z|>|w|^{1+r},\ |w|>|z|^{1+r}\},\\
Q_z(f_1(z),f_2(z),\epsilon)&=&\{(w,z)~:~|(w,z)|<\epsilon,\ f_1(z)<w<f_2(z)\},\\
Q_w(f_1(w),f_2(w),\epsilon)&=&\{(w,z)~:~|(w,z)|<\epsilon,\ f_1(w)<z<f_2(w)\},
\end{eqnarray*}
where we assume $f_1(z)={\rm O}(z^{1+r})$, $f_1(w)={\rm O}(w^{1+r})$,
$f_1(z)-f_2(z)={\rm O}(z^{1+r_1})$, $f_1(w)-f_2(w)={\rm O}(w^{1+r_1})$ for some
$r>0$ and $r_1\ge r$. In line with the definitions, the areas of the sets
$Q_z(f_1(z),f_2(z),\epsilon)$ and $Q_w(f_1(w),f_2(w),\epsilon)$ are
${\rm O}(\epsilon^{r_1+2})$.
Examples of the sets $Q(r,\epsilon)$, $Q_z(f_1(z),f_2(z),\epsilon)$ and
$Q_w(f_1(w),f_2(w),\epsilon)$ are shown in Figure \ref{fig_example1}.
By Theorem~\ref{th_A} below, if the stability
index satisfies $\sigma_j>-\infty$, then either the complement to
$\cB_{\delta}^g$ in $B_{\epsilon}(0)\subset H^{(in)}_j$ is empty, or it is the
union of the sets $Q_z(f_1(z),f_2(z),\epsilon)$, or the union of
$Q_w(f_1(w),f_2(w),\epsilon)$. Whether it is the union of the sets $Q_z$ or $Q_w$,
depends on the difference between the contracting and the transverse eigenvalues.

\begin{figure}
\vspace*{2mm}

\centerline{
\epsfig{file=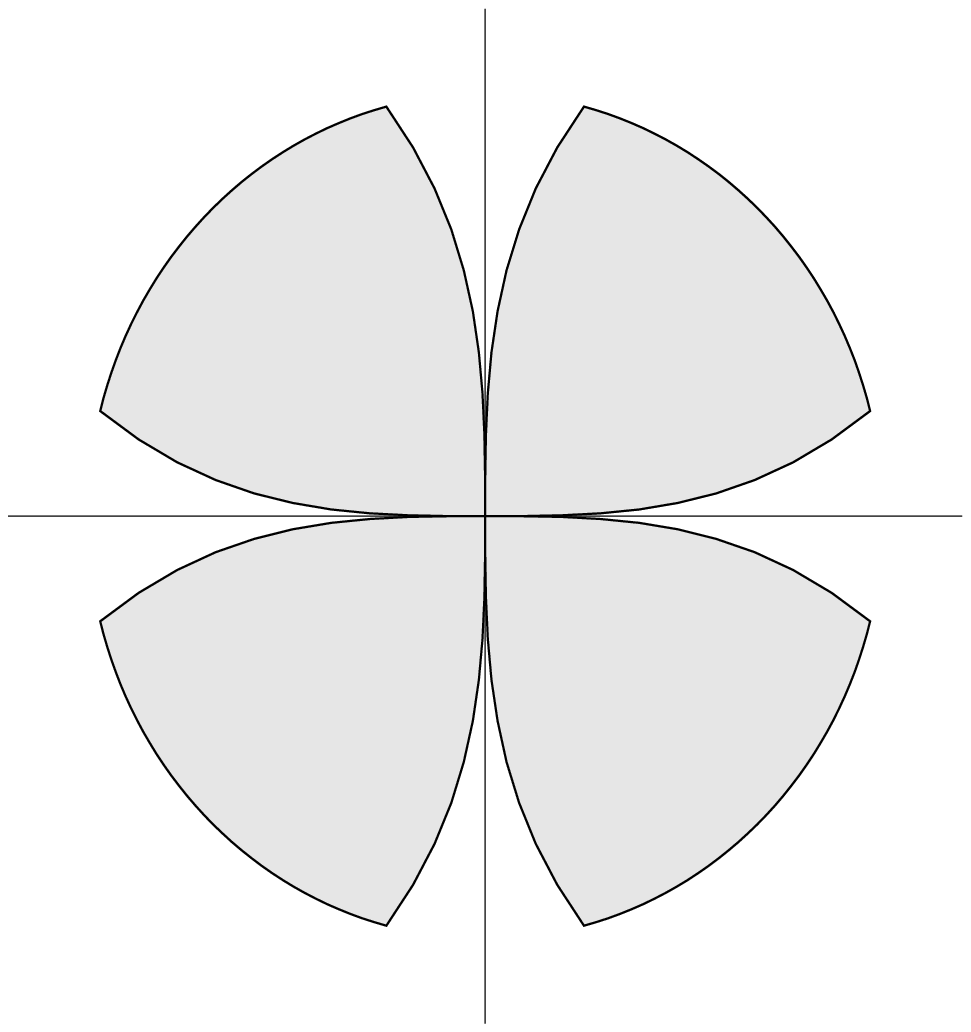,width=6cm}~\hspace{-8mm}
\epsfig{file=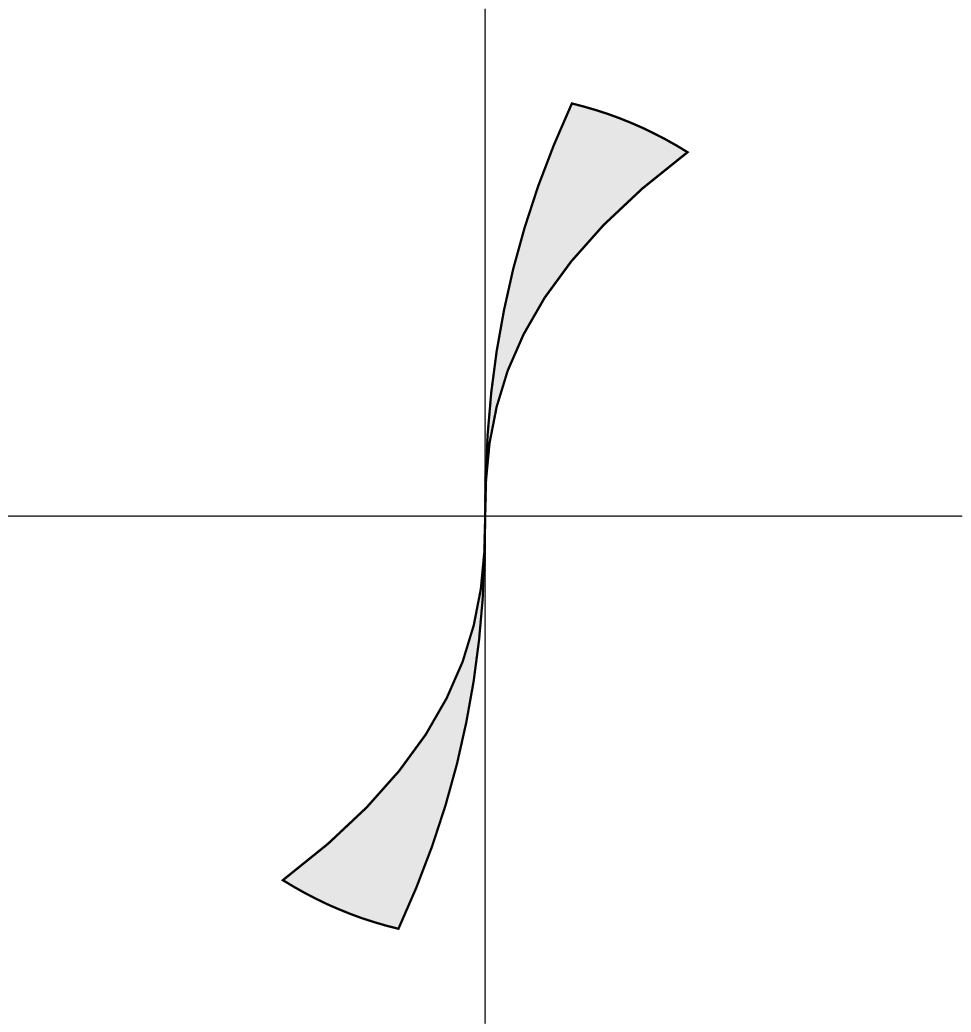,width=6cm}~\hspace{-8mm}
\epsfig{file=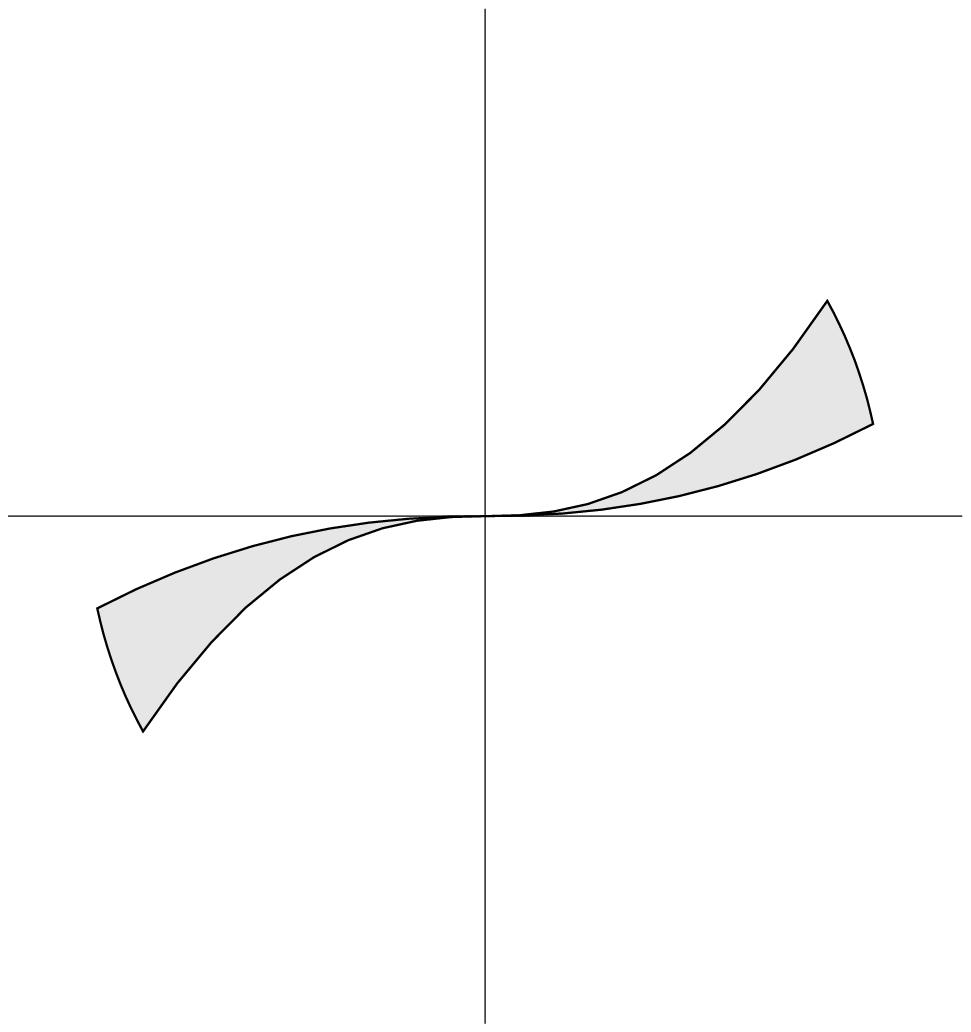,width=6cm}
}

\vspace*{-56mm}
{~}
\hspace*{20mm}
{\large $z$}

\vspace*{-.5cm}
\hspace*{80mm}
{\large $z$}

\vspace*{-.5cm}
\hspace*{135mm}
{\large $z$}

\vspace*{1.8cm}
\hspace*{46mm}
{\large $w$}

\vspace*{-.5cm}
\hspace*{96mm}
{\large $w$}

\vspace*{-.5cm}
\hspace*{151mm}
{\large $w$}

\vspace{12mm}
\hspace{40mm}(a)\hspace{50mm}(b)\hspace{50mm}(c)
\vspace{10mm}

\caption{
Examples of sets (a) $Q(r,\epsilon)$, (b) $Q_z(f_1(z),f_2(z),\epsilon)$ and
(c) and $Q_w(f_1(w),f_2(w),\epsilon)$ used in the discussion of stability of Type A cycles. The sets are shaded.
}
\label{fig_example1}
\end{figure}

\begin{figure}
\vspace*{2mm}

\centerline{
\epsfig{file=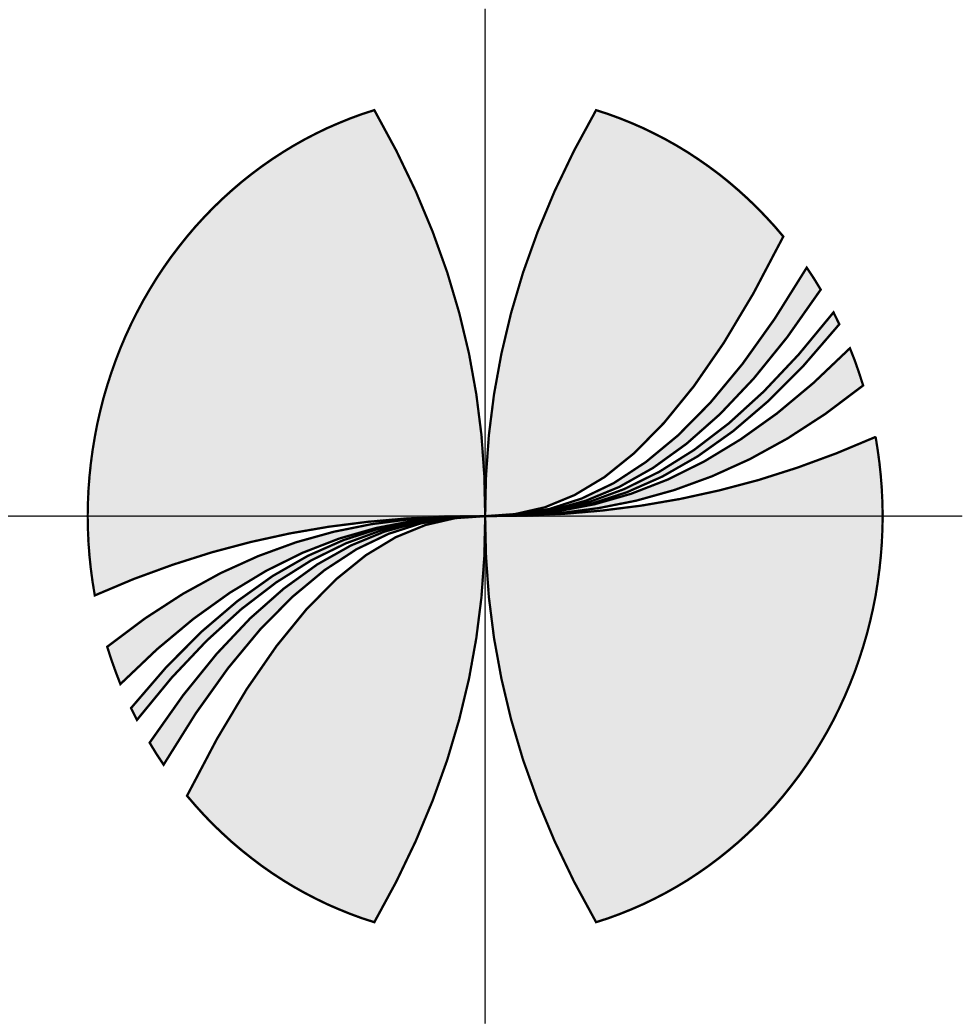,width=6cm}~
\epsfig{file=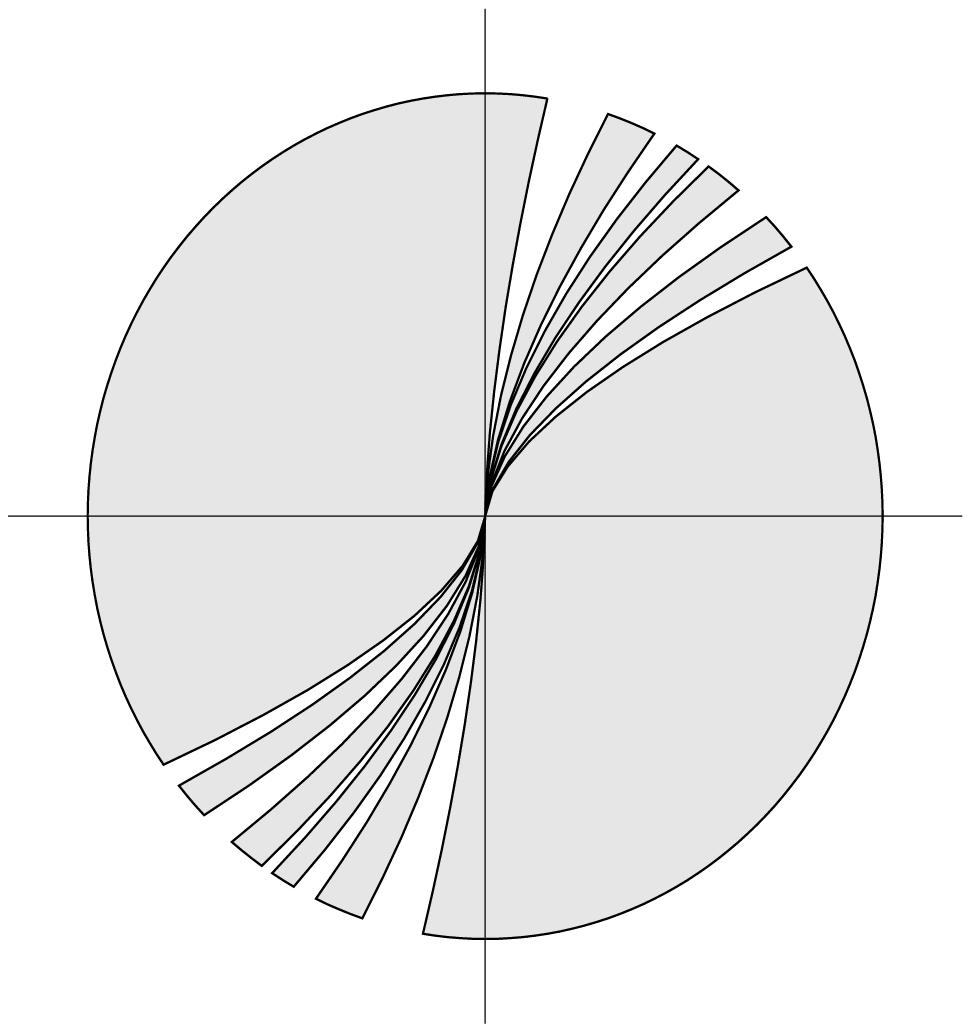,width=6cm}
}

\vspace*{-56mm}
{~}
\hspace*{45mm}
{\large $z$}

\vspace*{-.5cm}
\hspace*{110mm}
{\large $z$}

\vspace*{1.8cm}
\hspace*{72mm}
{\large $w$}

\vspace*{-.5cm}
\hspace*{134mm}
{\large $w$}

\vspace{12mm}
\hspace{40mm}(a)\hspace{60mm}(b)
\vspace{10mm}

\caption{
Examples of sets $\cB^g_{\delta}$ (shaded) for Type A cycles
for the cases (a) $b_1<0$, $a_1-b_1<1$ and (b) $b_1>0$, $a_1-b_1>1$.
}
\label{fig_example11}
\end{figure}

Examples of the sets $\cB_{\delta}^g$ are shown in Figure \ref{fig_example11}
for different signs of $b_j$ and $a_j-b_j-1$. It can be observed that the
sets are invariant with respect to the symmetry $(w,z)\to(-w,-z)$. This is so,
because the linearised systems (\ref{lmap}) evidently possess the symmetry,
and the global maps $\phi_j$ are symmetric (being linear).

For calculation of stability indices, we introduce
the collection of functions $\{h_{l,j}(y)\}$ for $1\le j\le m$, $l\le j$,
which are defined as follows\footnote{If an index takes values $1,\ldots,m$,
then the index value modulo $m$ is understood here and below.
Note that $l$ in $h_{l,j}$ can take negative values.}:
\begin{eqnarray*}
h_{j,j}(y)&=&y,\\
h_{l,j}(y)&=&
\left\{
\renewcommand{\arraystretch}{1.2}
\begin{array}{rl}
\infty & \mbox{ if } a_l-b_l<0\\
\displaystyle \frac{a_lh_{l+1,j}(y)-a_l+1}{ a_l-b_l} & \mbox{ if } 0<a_l-b_l<1\\
a_lh_{l+1,j}(y)-b_l & \mbox{ if } a_l-b_l>1
\end{array}\right.
\end{eqnarray*}

The next theorem is the main result for Type A cycles,
namely it gives the stability indices $\sigma_j$ for the
collection of maps $g_j$ related to Type A cycles. Recall that the
coefficients $a_j$ and $b_j$ of the
map $g_j$ are related to the eigenvalues of linearisation
of (\ref{eq_ode}) near $\xi_j$ as $a_j=c_j/e_j$ and $b_j=-t_j/e_j$.
Recall, that $c_j>0$ and $e_j>0$ for all $j$, therefore $a_j>0$.
Following \cite{KruMel95a}, we denote
$$
\rho_j=\min(a_j,1+b_j),
$$
$\rho=\rho_1\cdots\rho_m$, and note that generically the non-degeneracy
conditions
\begin{equation}\label{eq_typea_nondeg}
a_j\ne 1+b_j,~b_j\ne -1,~\rho\ne 1
\end{equation}
are satisfied. The Theorem below is stated and proved more precisely as Theorem~\ref{th_A_repeat}.

\begin{theorem}\label{th_A}
For the collection of maps $g_j$ associated with a Type A cycle, the stability indices are:
\begin{itemize}
\item[(a)] If $\rho>1$ and $b_j>0$ for all $j$ then
$\sigma_{j,+}=\infty$ and $\sigma_{j,-}=0$ for any $j$.
\item[(b)] If $\rho>1$, $b_j>-1$ for all $j$ and $b_j<0$ for
$j=J_1,\ldots,J_L$ then $\sigma_{j,-}=0$ and $\sigma_{j,+}$ are:
$$
\sigma_{j,+}=\min_{s=J_1,\ldots,J_L}h_{\tilde j,s}\left(-\frac{1}{ b_s}\right)-1,
\hbox{ where }\tilde j=
\left\{
\begin{array}{rl}
j, & j\le s\\
j-m, & j>s
\end{array}\right. .
$$
\item[(c)] If $\rho<1$ or there exists $j$ such that $b_j<-1$ then
$\sigma_{j,+}=0$, $\sigma_{j,-}=\infty$
and the cycle is not an attractor.
\end{itemize}
\end{theorem}

\proof
(a) Since $\rho>1$, there exists a $q>0$ such that
\begin{equation}\label{A_ea0}
\tilde\rho=\prod_{j=1}^m(\rho_j-q)>1.
\end{equation}
By Lemma \ref{lem_411}, for sufficiently small $\epsilon$
\begin{equation}\label{A_ea1}
|g_j(w,z)|<|(w,z)|^{\rho_j-q}\hbox{ for any }(w,z),\ |(w,z)|<\epsilon.
\end{equation}
Consider $(w,z)\in H^{(in)}_1$. Inequality (\ref{A_ea1}) implies that
for a given $\delta$, we can find an $\epsilon>0$ such that
$$|g_{j,k}(w,z)|=|g_{j,1}(w,z)||g^k(w,z)|<|g_{j,1}(w,z)||(w,z)|^{\tilde\rho^k}<\delta$$
$$\hbox{ for all }1\le j\le m,\ k\ge0,\hbox{ and }|(w,z)|<\epsilon.$$
Therefore $\sigma_{1,+}=\infty$ and $\sigma_{1,-}=0$. The proof for $j>1$ is similar.
\vspace{3mm}

(b) Consider $s=J_l$ for some $l$, whereby $-1<b_s<0$.
For any small $q>0$ and $\delta>0$ we can find $\epsilon$, such that
$$(w,z)\in Q^{s,s}(\epsilon)\stackrel{\rm def}{=}
Q_z(-|z|^{-1/b_s+q},|z|^{-1/b_s+q},\epsilon)$$
implies
$$|g_s(w,z)|>\delta$$
and
$$(w,z)\in H^{(in)}_s\setminus\tilde Q^{s,s}(\epsilon),\hbox{ where }
\tilde Q^{s,s}(\epsilon)\stackrel{\rm def}{=}
Q_z(-|z|^{-1/b_s-q},|z|^{-1/b_s-q},\epsilon)$$
implies
$$|g_s(w,z)|<\delta.$$

For simplicity, we ignore small $q$ and say that if
$$(w,z)\in \hat Q^{s,s}(\epsilon)\stackrel{\rm def}{=}
Q_z(-|z|^{-1/b_s},|z|^{-1/b_s},\epsilon),$$
then
$$|g_s(w,z)|>\delta,$$
and if
$$(w,z)\in H^{(in)}_s\setminus\hat Q^{s,s}(\epsilon),$$
then
$$|g_s(w,z)|<\delta.$$
We also assume that $\epsilon$ is sufficiently small and all estimates for
$(w,z)$ required in all the applied lemmas are satisfied.

Denote by $\hat Q^{s-1,s}(\epsilon)\subset H^{(in)}_{s-1}$ the
preimage of $\hat Q^{s,s}(\epsilon)$ under the map $g_{s-1}$,
and by $\hat Q^{j,s}(\epsilon)\subset H^{(in)}_j$ the preimage
of $\hat Q^{s,s}(\epsilon)$ under $g_s\circ g_{s-1}\circ\ldots\circ g_j$.
By construction of the sets $\hat Q^{j,s}(\epsilon)$,
$1\le j\le s$, for any $(w,z)\in\hat Q^{j,s}(\epsilon)$ the inequality
$$|g_s\circ g_{s-1}\circ\ldots\circ g_j(w,z)|>\delta$$
is valid.

The measure (area) of the set $\hat Q^{s,s}(\epsilon)$ can be estimated as
$$\ell(\hat Q^{s,s}(\epsilon))={\rm O}(\epsilon^{-1/b_s+1})={\rm O}(\epsilon^{h_{s,s}(-1/b_s)+1}).$$
By virtue of the definition of functions $h_{l,j}$ and due to Lemmas \ref{lem_43}
and \ref{lem_44}, the measure of the set $\hat Q^{1,s}(\epsilon)$ is
$$\ell(\hat Q^{1,s}(\epsilon))={\rm O}(\epsilon^{h_{1,s}(-1/b_s)+1}).$$
(Here and below, the measure of an empty set $\hat Q^{j,s}(\epsilon)$ is supposed to
be $\epsilon^{\infty}$.)

Denote by $\hat Q^{1-m,s}(\epsilon)$ the preimage of the set $\hat Q^{1,s}(\epsilon)$
under a complete iteration $g$ along the cycle, and by $\hat Q^{1-km,s}(\epsilon)$
the preimage under $k$ iterations.
The measure of the set $\hat Q^{1-km,s}(\epsilon)$ is
$$\ell(\hat Q^{1-km,s}(\epsilon))={\rm O}(\epsilon^{h_{1-km,s}(-1/b_s)+1}).$$

Since $\rho>1$, by the same arguments as used in the proof of part (a),
if $(w,z)\in H^{(in)}_1\cap B_{\epsilon}$ does not belong to any
$\hat Q^{1-km,s}(\epsilon)$ for all $s=J_1,\ldots,J_L$ and $k\ge0$, then
$$|g_{j,k}(w,z)|<\delta\hbox{ for all }1\le j\le m\hbox{ and }k\ge0.$$
By construction of the sets
$\hat Q^{1-km,s}(\epsilon)$, if $(w,z)\in\hat Q^{1-km,s}(\epsilon)$, then
$$|g_{j,k}(w,z)|>\delta\hbox{ for some }j\hbox{ and }k.$$
By properties of the functions $h_{l,j}$,
the measure of the set $\hat Q^{1,s}(\epsilon)$ is larger than that of any
other set $\hat Q^{1-km,s}(\epsilon)$ for $k>0$.
Since
$$\ell(\cup_{1\le s\le J_l}\hat Q^{1,s}(\epsilon))={\rm O}(\epsilon^{\min_s h_{1,s}(-1/b_s)+1}),$$
by definition of the stability index, for $j=1$
the statement of the theorem, part (b), holds true. For other $j$ the proof is similar.

\vspace{3mm}

(c) Below we assume that $\delta$ and $\epsilon$ are sufficiently small, so that for
$|(w,z)|<\max(\delta,\epsilon)$ the conditions of all lemmas to be applied hold true.

We consider three following cases, which cover exhaustively all possibilities:
\begin{itemize}
\item
Suppose $\rho<1$ and $b_j>-1$ for all $j$.
Since $\rho<1$, there exists a $q>0$ such that
\begin{equation}\label{A_ea00}
\tilde\rho=\prod_{j=1}^m(\rho_j+q)<1.
\end{equation}

By Corollary \ref{cor_lem_41}(b), there exists $r$ such that if
$(w,z)\in Q(r,\delta)$ then
\begin{equation}\label{A_eaX}
|g_{l,k}(w,z)|>\delta
\end{equation}
for some $l$ and $k$. Consequently, if $(w,z)\in(g^K)^{-1}Q(r,\delta)$ for some $K>0$,
then the inequality (\ref{A_eaX}) is satisfied for $k\equiv k+K$.
The complement to $Q(r,\delta)$ in $B_{\delta}$ is the union of the sets
$\tilde Q_z=Q_z(|z|^{1+r},-|z|^{1+r},\delta)$ and
$\tilde Q_w=Q_w(|w|^{1+r},-|w|^{1+r},\delta)$.
Corollaries~\ref{cor_lem_43} and \ref{cor_lem_44} imply existence of the limit sets
$$Q^{\rm lim}_z=\lim_{k\to \infty}(g^k)^{-1}\tilde Q_z\ne\emptyset\mbox{ and }
Q^{\rm lim}_w=\lim_{k\to \infty}(g^k)^{-1}\tilde Q_w\ne\emptyset,$$
and by Lemmas \ref{lem_43} and \ref{lem_44},
(\ref{A_eaX}) is satisfied for $(w,z)\in Q^{\rm lim}_z\cup Q^{\rm lim}_w$ for some
$l$ and $k$. Therefore, $\sigma_{1,-}=\infty$.

\item
Suppose that $b_s<-1$ for some $s$, and also at least one of the inequalities,
$\rho>1$, or $a_t-b_t<0$ for some $t$, is satisfied.
Let the set $\hat Q^{s,s}(\epsilon)$ be defined as in the proof of part (b).
Denote by $R^{s,s}(\epsilon)$
the complement to the set $\hat Q^{s,s}(\epsilon)$ in $B_{\epsilon}$,
and by $R^{1-km,s}(\epsilon)$ the preimage of $R^{s,s}(\epsilon)$ under
$g_s\circ g_{s-1}\circ\ldots\circ g_1g^k$. By the arguments of the
proof of part (b), the measure of the set $R^{1-km,s}(\epsilon)$ is
$$\ell(R^{1-km,s}(\epsilon))={\rm O}(\epsilon^{h_{1-km,s}(-b_s)+1}).$$
By properties of the functions $h_{l,j}$, $\lim_{k\to\infty}h_{1-km,s}(-b_s)=\infty$,
and therefore $\sigma_{1,-}=\infty$.

\item
Suppose $\rho<1$, $a_t-b_t>0$ for all $t$,  and $b_s<-1$ for some $s$.
Let the sets $R^{1-km,s}(\epsilon)$ be defined as in the previous paragraph.
Since $\rho<1$ and $a_t-b_t>0$ for all $t$, the sets do not vanish in the limit $k\to\infty$.
By Corollaries~\ref{cor_lem_43} and \ref{cor_lem_44}, the limit set
$R^{\rm lim}=\lim_{k\to\infty}R^{1-km,s}(\epsilon)$ does exist, and by
Lemmas \ref{lem_43} and \ref{lem_44}, inequality
(\ref{A_eaX}) is satisfied for all $(w,z)\in R^{\rm lim}$
for some $l$ and $k$. Therefore, $\sigma_{1,-}=\infty$.
\end{itemize}

The proof for $\sigma_{j,-}$, $j>1$, is similar.
\qed

\subsection{Type B and C cycles}\label{typeBC}

\begin{figure}
\vspace*{2mm}

\centerline{
\epsfig{file=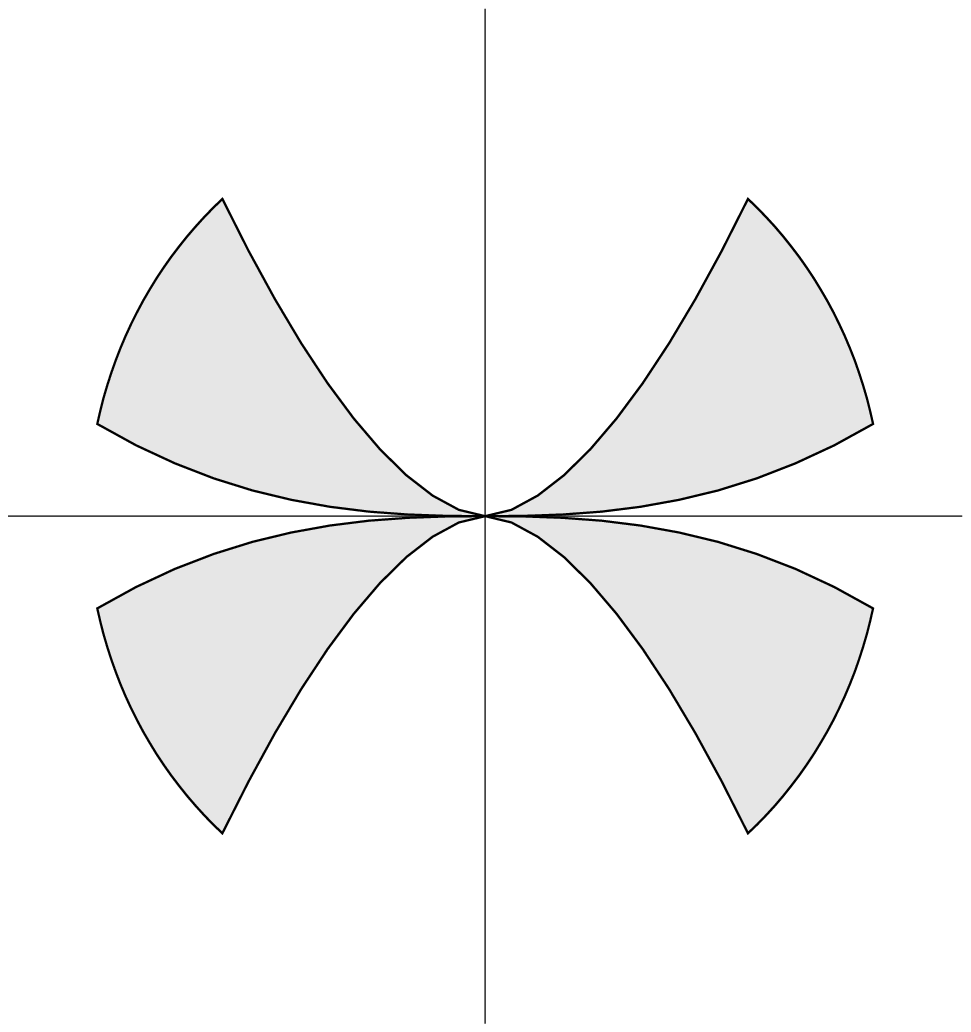,width=6cm}~
\epsfig{file=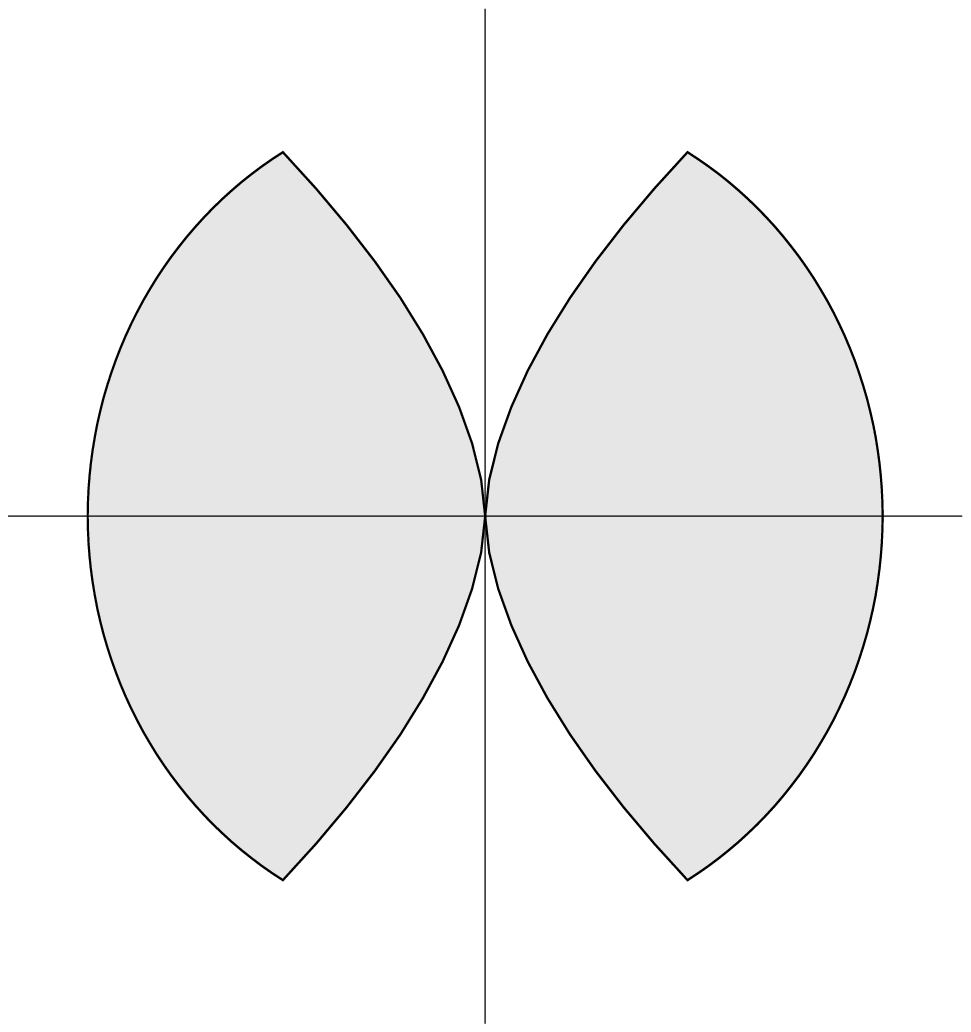,width=6cm}
}

\vspace*{-56mm}
{~}
\hspace*{45mm}
{\large $z$}

\vspace*{-.5cm}
\hspace*{110mm}
{\large $z$}

\vspace*{1.8cm}
\hspace*{72mm}
{\large $w$}

\vspace*{-.5cm}
\hspace*{134mm}
{\large $w$}

\vspace{12mm}
\hspace{40mm}(a)\hspace{60mm}(b)
\vspace{10mm}

\caption{
Examples of sets $\cB^g_{\delta}$ (shaded) for Type B or C cycles.}
\label{fig_example12}
\end{figure}

Examples of sets $\cB_{\delta}^g$ for Type B and C cycles are shown in Figure
\ref{fig_example12}. The sets are invariant with respect to the symmetries
$(w,z)\to(-w,z)$ and $(w,z)\to(w,-z)$, because the linearised systems
(\ref{lmap}) evidently possess these symmetries, and the global maps $\phi_j$
are symmetric due to linearity and invariance of the subspaces $(w,0)$ and
$(0,z)$. Therefore, we consider in this subsection only positive values of $w$
and $z$; components of $\cB_{\delta}^g$ in other three quadrants are obtained
on applying the symmetries.

As noted in \cite{KruMel04}, the maps $g_j:H^{(in)}_j\to H^{(in)}_{j+1}$
related to the cycles of Types B and C asymptotically to the lowest order have
the form $(Ew^{a_j},Fw^{b_j}z)$ and $(Ew^{b_j}z,Fw^{a_j})$, respectively.
In the new coordinates $(\zeta,\eta)$, $\zeta=\ln z$ and $\eta=\ln w$,
the maps $g_j$ are linear:
$$g_j(\zeta,\eta)=M_j\left(
\begin{array}{c}
\zeta\\
\eta
\end{array}\right)+\left(
\begin{array}{c}
\ln E\\
\ln F
\end{array}\right),$$
where the {\em transition matrices} of the maps are
$$
M_j=\left(
\begin{array}{cc}
a_j&0\\
b_j&1
\end{array}
\right)\hbox{ and }
M_j=\left(
\begin{array}{cc}
b_j&1\\
a_j&0
\end{array}\right)
$$
for cycles of Types B and C, respectively.
In the definition of stability indices, asymptotically small $z$ and $w$
(and therefore asymptotically large negative $\zeta$ and $\eta$)
are assumed. Hence, we ignore finite $\ln E$ and $\ln F$.

Recall that the coefficients $a_j$ and $b_j$ of the matrices are related
to the eigenvalues of linearisation of (\ref{eq_ode}) near
$\xi_j$ as $a_j=c_j/e_j$ and $b_j=-t_j/e_j$. For the map
$g=g_m\circ\ldots\circ g_1$ the transition matrix is $M=M(g)=M_m\cdots M_1$.
We introduce the notation: $M_{j,k}$ and $M^{(j)}$ denote transition matrices
for the maps $g_{j,k}$ and $g^{(j)}$, respectively; $M^{(l,j)}=M_l\cdots M_j$;
$\lambda^j_1$, $\lambda^j_2$, ${\bf v}^j_1=(v^j_{11},v^j_{12})$
and ${\bf v}^j_2=(v^j_{21},v^j_{22})$ denote eigenvalues and associated
eigenvectors of the matrix $M^{(j)}$, respectively. If the eigenvalues are real,
$\lambda^j_1\ge\lambda^j_2$ is assumed. (Generically the eigenvalues are different.)

A necessary condition for $(w,z)$ to belong to $\cB_{\delta}^g$ (see
Subsection~\ref{secindmap}) is that $g^k(w,z)$ is bounded for all $k$. To
leading order, the map $g:(\zeta,\eta)\to(\zeta,\eta)$ is described by the
transition matrix $M(g)$. Due to linearity of the map, in the new coordinates
the condition that the iterates $(\zeta_k,\eta_k)^t=M^k(\zeta,\eta)^t$ are
bounded by an $S<0$ (i.e., $\zeta_n<S$ and $\eta_n<S$, or, in the original
coordinates, $w<{\rm e}^S$ and $z<{\rm e}^S$) generically is equivalent to
$\lim_{k\to\infty}M^k(\zeta,\eta)^t=(-\infty,-\infty)^t$.

We denote
$$
U^{-\infty}(M)=
\{(x,y):\ x\le0,\ y\le0,\ \lim_{n\to\infty}M^n(x,y)^t=(-\infty,-\infty)^t\}.
$$
Evidently, $U^{-\infty}(M)=\emptyset$ implies $\cB_{\delta}^g=\emptyset$. The conditions
for $U^{-\infty}(M)\not=\emptyset$ are given in Lemma \ref{lem_Uinf} in terms
of eigenvalues and eigenvectors of matrix $M$. They are:
\begin{itemize}
\item[(i)] the eigenvalues are real;
\item[(ii)] $\lambda_1>1$;
\item[(iii)] $\lambda_1>|\lambda_2|$;
\item[(iv)] $v_{11}v_{12}>0$.
\end{itemize}

In terms of entries of a $2\times2$ matrix $M=(a_{ij})$ (where we assume $a_{11}>a_{22}$) the conditions are (Lemma \ref{lem_eig}):
\begin{eqnarray}
\mbox{(i)}&&~~~~ \frac{(a_{11}-a_{22})^2}{ 4}+a_{12}a_{21}\ge 0
\\
\label{condii}
\mbox{(ii)}&&~~~~\max\left(\frac{a_{11}+a_{22}}{ 2},~a_{11}+a_{22}-a_{11}a_{22}+a_{12}a_{21}\right)>1
\\
\mbox{(iii)}&&~~~~\frac{a_{11}+a_{22}}{ 2}>0
\\
\mbox{(iv)}&&~~~~a_{21}>0.
\end{eqnarray}

For calculation of stability indices we introduce the sets
$U_R=\{(\zeta,\eta)~|~\max(\zeta,\eta)<R\}$ and
$$
\begin{array}{c}
U_R(\alpha_1,\beta_1,q_1;\alpha_2,\beta_2,q_2)=\\
\{(\zeta,\eta)\in U_R~:~(\alpha_1+q_1)\zeta+(\beta_1+q_1)\eta<0,\
(\alpha_2+q_2)\zeta+(\beta_2+q_2)\eta<0\},
\end{array}
$$
where $R<0$.
If all entries of the matrices $M_j$, $1\le j\le m$, are non-negative,
the stability indices of the related map can be calculated using the following Theorem which is stated and proved as Theorem~\ref{th_BC1_repeat}.

\begin{theorem}\label{th_BC1}
Let $g$ be a map related to simple heteroclinic cycle of Types B or C
and $M_j$, $1\le j\le m$, its transition matrices. Suppose that for all $j$,
$1\le j\le m$, all entries of the matrices are non-negative. Then:
\begin{itemize}
\item[(a)] If the transition
matrix $M=M_m\cdots M_1$ satisfies condition (ii) (see (\ref{condii})),
then $\sigma_{j,+}=\infty$ and $\sigma_{j,-}=0$ for all $j$ and moreover
the cycle is asymptotically stable.
\item[(b)] Otherwise, $\sigma_{j,+}=0$ and $\sigma_{j,-}=\infty$ for all $j$ and
the cycle is not an attractor.
\end{itemize}
\end{theorem}

For calculation of stability indices of matrices with negative entries
we introduce the following functions
$$
f^+(\alpha,\beta)=\left\{
\renewcommand{\arraystretch}{2.0}
\begin{array}{rl}
\infty, &\alpha\ge0,\ \beta\ge0,\\
0,&\alpha\le0,\ \beta\le0,\\
-\displaystyle\frac{\beta}{\alpha}-1, &\alpha<0,\ \beta>0,
\ \displaystyle\frac{\beta}{\alpha}<-1,\\
0, &\alpha<0,\ \beta>0,\ \displaystyle\frac{\beta}{\alpha}>-1,\\
-\displaystyle\frac{\alpha}{\beta}-1, &\alpha>0,\ \beta<0,
\ \displaystyle\frac{\alpha}{\beta}<-1,\\
0, &\alpha>0,\ \beta<0,\ \displaystyle\frac{\alpha}{\beta}>-1
\end{array}\right.,
$$
$f^-(\alpha,\beta)=f^+(-\alpha,-\beta)$ and
$$f^{\rm index}(\alpha,\beta)=f^+(\alpha,\beta)-f^-(\alpha,\beta)$$
and prove in the following theorem that the set $\cB_{\delta}(g)$ in
$\hat H^{(in)}_1$ (and similarly for $\hat H^{(in)}_j$ with $j>1$)
in the $(\zeta,\eta)$ coordinates
is $U_R(\alpha_1,\beta_1,0;\alpha_2,\beta_2,0)$.
We denote the latter set in original coordinates $(w,z)$ as
$$
\tilde U_R(\alpha_1,\beta_1,0;\alpha_2,\beta_2,0)
$$
and then by Definition~\ref{def_stab_index}
\begin{eqnarray*}
 \sigma &=&
\lim_{R\to-\infty}
\frac{
\ln(\ell(\tilde U_R\setminus\tilde U_R(\alpha_1,\beta_1,0;\alpha_2,\beta_2,0)))-
\ln(\ell(\tilde U_R(\alpha_1,\beta_1,0;\alpha_2,\beta_2,0)))
} {\ln(\ell(\tilde U_R))}\\
&=& \min(f^{\rm index}(\alpha_1,\beta_1),f^{\rm index}(\alpha_2,\beta_2)).
\end{eqnarray*}
The Theorem below is stated and proved as Theorem~\ref{th_BC2_repeat}.

\begin{theorem}\label{th_BC2}
Let $X$ be a simple heteroclinic cycle of Type B or C
and $M_j$, $1\le j\le m$ the associated transition matrices. We denote by
$j=j_1,\ldots j_L$ the indices, for which some of the entries of $M_j$
are negative; they are all non-negative for all remaining $j$.
\begin{itemize}
\item[(a)] If for at least one of $j=j_l+1$
the matrix $M^{(j)}$ does not satisfy conditions (i)-(iv)
of Lemma \ref{lem_eig}, then the cycle is repelling and
$\sigma_j=-\infty$ for all $j$.
\item[(b)] If the matrices $M^{(j)}$ satisfy conditions (i)-(iv) of Lemma
\ref{lem_eig} for all $j=j_l+1$, then there exist numbers
$(\alpha_1^j,\beta_1^j,\alpha_2^j,\beta_2^j)$, $1\le j\le m$, such that
\begin{itemize}
\item[(i)] $U_0(\alpha_1^j,\beta_1^j,0;\alpha_2^j,\beta_2^j,0)\ne\emptyset$, $1\le j\le m$.
\item[(ii)] For any $S<0$ and $q>0$ there exists $R<0$ such that
$$M^{(l,j)}(M^{(j)})^k
(U_R(\alpha_1^j,\beta_1^j,-q;\alpha_2^j,\beta_2^j,-q))
\subset U_S\ \mbox{ for all }\,l,\ \ 1\le l\le m,\ k\ge0,$$
\item[(iii)] $$
\lim_{k\to\infty}(M^{(l,j)})(M^{(j)})^k(\zeta,\eta)^t=(-\infty,-\infty)
\mbox{ for all }\ (\zeta,\eta)\in
U_0(\alpha_1^j,\beta_1^j,0;\alpha_2^j,\beta_2^j,0).
$$
\item[(iv)]
$$
U_0(\alpha_1^j,\beta_1^j,0;\alpha_2^j,\beta_2^j,0)=U^{-\infty}(M^{(j)})
\cap \bigcap_{1\le l\le L}(M^{(j_l,j)})^{-1}U_0
\cap \bigcap_{1\le l\le L}(M^{(j_l+m,j)})^{-1}U_0.
$$
\item[(v)] If $\lambda_2\ge0$ then
$$
U_0(\alpha_1^j,\beta_1^j,0;\alpha_2^j,\beta_2^j,0)=U^{-\infty}(M^{(j)})
\cap \bigcap_{1\le l\le L}(M^{(j_l,j)})^{-1}U_0.
$$
\end{itemize}
and the cycle is a Milnor attractor.
\end{itemize}
\end{theorem}

Note that
$$
(M^{(j,s)})^{-1}U_0\cap U_0=U_0(\alpha^{(j,s)}_{11},\alpha^{(j,s)}_{12},0;
\alpha^{(j,s)}_{21},\alpha^{(j,s)}_{22},0),
$$
where $\alpha^{(j,s)}$ are entries of the matrix $M^{(j,s)}$.

Theorems~\ref{th_A}-\ref{th_BC2} imply the following Corollary:

\begin{corollary}\label{cor_BC}
For simple heteroclinic cycles in $\R^4$, $\sigma_j=-\infty$ for some $j$ if and only if $\sigma_j=-\infty$ for all $j$.
\end{corollary}

\subsubsection{Calculation of stability indices for Type B cycles}\label{typeB}

\paragraph{Types $B^+_1$ and $B^-_1$}

The cycles of Types $B^+_1$ and $B^-_1$ have transition matrices
$$
M=\left(
\begin{array}{cc}
a&0\\
b&1
\end{array}
\right).
$$
Corollary~\ref{cor_lem_eig} implies that if $a<1$ or $b<0$, then the cycles are
not attractors and the stability index is $-\infty$, otherwise they are
attracting and the stability index is $\infty$.

\paragraph{Types $B^+_2$}

For cycles of Type $B^+_2$ the product of transition matrices is
$$
M_1M_2=\left(
\begin{array}{cc}
a_1a_2&0\\
b_1a_2+b_2&1
\end{array}
\right)
$$
with eigenvalues $a_1a_2$ and 1, and the associated eigenvectors
$(a_1a_2-1,b_1a_2+b_2)$ and $(0,1)$, respectively
(for $M_2M_1$, simply swap the indices 1 and 2 in the expressions to obtain
the corresponding eigenvectors).

Theorems \ref{th_BC1} and \ref{th_BC2} imply to obtain the following
classification:
\begin{itemize}
\item
If $b_1<0$ and $b_2<0$, then the cycle is not an attractor and all stability
indices are $-\infty$.
\item
Suppose $b_1>0$ and $b_2>0$.
\begin{itemize}
\item
If $a_1a_2<1$, then the cycle is not an attractor
and the stability indices are $-\infty$.
\item
If $a_1a_2>1$, then
the cycle is locally attracting and the stability indices are $\infty$.
\end{itemize}
\item
Suppose $b_1<0$ and $b_2>0$.
\begin{itemize}
\item
If $a_1a_2<1$ or $b_1a_2+b_2<0$, then the cycle
is not an attractor and the stability indices are $-\infty$.
\item
If $a_1a_2>1$
and $b_1a_2+b_2>0$, then the stability indices are
$\sigma_1=f^{\rm index}(b_1,1)$ and $\sigma_2=\infty$.
\end{itemize}
\end{itemize}

\paragraph{Type $B^-_3$}

For cycles of Type $B^-_3$ the product of transition matrices is
$$
M_3M_2M_1=\left(
\begin{array}{cc}
a_1a_2a_3&0\\
b_3a_1a_2+b_2a_1+b_1&1
\end{array}
\right).
$$
Its eigenvalues are $a_1a_2a_3$ and 1 with associated eigenvectors
$$(a_1a_2a_3-1,b_3a_1a_2+b_2a_1+b_1), ~~(0,1).$$
(For $M_2M_1M_3$ and $M_1M_3M_2$, the quantities
are obtained by cyclic permutation of the indices.) Theorems \ref{th_BC1} and \ref{th_BC2} imply obtain the following classification:
\begin{itemize}
\item
If $b_1<0$, $b_2<0$ and $b_3<0$, then the cycle is not an attractor
and the stability indices are all $-\infty$.
\item
Suppose $b_1>0$, $b_2>0$ and $b_3>0$.
\begin{itemize}
\item
If $a_1a_2a_3<1$, then the cycle is
not an attractor and the stability indices are $-\infty$.
\item
If $a_1a_2a_3>1$,
then the cycle is locally attracting and the stability indices are $\infty$.
\end{itemize}
\item
Suppose $b_1<0$, $b_2>0$ and $b_3>0$.
\begin{itemize}
\item
If $a_1a_2a_3<1$ or $b_1a_2a_3+b_3a_2+b_2<0$, then the cycle not an attractor
and the stability indices are $-\infty$.
\item
If $a_1a_2a_3>1$
and $b_1a_2a_3+b_3a_2+b_2>0$, then the stability indices are
$\sigma_1=f^{\rm index}(b_1,1)$, $\sigma_2=\infty$ and
$\sigma_3=f^{\rm index}(b_3+b_1a_3,1)$.
\end{itemize}
\item
Suppose $b_1<0$, $b_2<0$ and $b_3>0$.
\begin{itemize}
\item
If $a_1a_2a_3<1$ or
$b_2a_1a_3+b_1a_3+b_3<0$ or $b_1a_2a_3+a_2b_3+b_2<0$, then the cycle is not an
attractor and the stability indices are $-\infty$.
\item
If $a_1a_2a_3>1$,
$b_2a_1a_3+b_1a_3+b_3>0$ and $b_1a_2a_3+a_2b_3+b_2>0$, then the stability
indices are $\sigma_1=\min(f^{\rm index}(b_1,1),f^{\rm index}(b_1+b_2a_1,1))$,
$\sigma_2=f^{\rm index}(b_2,1)$ and
$\sigma_3=f^{\rm index}(b_3+b_1a_3,1)$.
\end{itemize}
\end{itemize}

\subsubsection{Calculation of stability indices for Type C cycles}\label{typeC}

\paragraph{Type $C^-_1$}

Cycles of Type $C^-_1$ have a transition matrix of the form
$$
M=\left(
\begin{array}{cc}
b&1\\
a&0
\end{array}\right).
$$
Corollary~\ref{cor_lem_eig} implies that the cycle is attracting and the
stability index is $\infty$ whenever $b\ge0$ and $a+b>1$;
otherwise it is not an attractor and the stability index is $-\infty$.

\paragraph{Type $C^-_2$}

The product of transition matrices for cycles of Type $C^-_2$ is
$$
M_1M_2=\left(
\begin{array}{cc}
b_1b_2+a_2&b_1\\
a_1b_2&a_1
\end{array}\right).
$$
Denote by $\lambda_1$ and $\lambda_2$ eigenvalues of the matrix $M_1M_2$
(which will be the same as those for $M_2M_1$).

Theorems~\ref{th_BC1} and \ref{th_BC2} imply the following:
\begin{itemize}
\item
If $b_1<0$ and $b_2<0$, then the cycle is not an attractor and the stability
indices are $-\infty$.
\item
Suppose that $b_1>0$ and $b_2>0$.
\begin{itemize}
\item
If
$$
\max(b_1b_2+a_2+a_1,2(b_1b_2+a_2+a_1-a_1a_2))<2
$$
then the cycle is not an attractor and the stability indices are $-\infty$.
\item
Otherwise the cycle is locally attracting and the stability indices are $\infty$.
\end{itemize}
\item
Suppose $b_1<0$ and $b_2>0$.
\begin{itemize}
\item
If
$$
(b_1b_2+a_2-a_1)^2+4a_1b_1b_2<0,
$$
or
$$
\max(b_1b_2+a_2+a_1,2(b_1b_2+a_2+a_1-a_1a_2))<2,
$$
or
$$
b_1b_2-a_1+a_2<0,
$$
then the cycle is not an attractor and the stability indices are $-\infty$.
\item
If none of the listed conditions are satisfied, the stability indices are
\begin{equation}\label{c4_c1-}
\sigma_1=f^{\rm index}(\frac{b_1b_2+a_1-\lambda_2}{b_2},1)~\mbox{ and }~
\sigma_2=f^{\rm index}(\frac{\lambda_2-b_1b_2-a_2}{b_1},-1).
\end{equation}
\end{itemize}
\end{itemize}

\paragraph{Type $C^-_4$}

The transition matrix for cycles of Type $C^-_4$ is
\begin{eqnarray*}
&M\equiv M^{4,1}\equiv M_4M_3M_2M_1=&\\
&
\left(\begin{array}{cc}
(b_1b_2+a_1)(b_3b_4+a_3)+b_1a_2b_4 & (b_3b_4+a_3)b_2+b_4a_2\\
a_4b_3(b_1b_2+a_1)+a_2a_4b_1       & a_4b_2b_3+a_2a_4
\end{array}\right)&
\end{eqnarray*}
Denote by $\lambda_1$ and $\lambda_2$ eigenvalues and by ${\bf v}_1^{4,1}$
and ${\bf v}_2^{4,1}$ the associated eigenvectors of the matrix. The trace and
determinant are:
$$
\tr(M)=b_1b_2b_3b_4+b_1b_2a_3+b_3b_4a_1+b_1b_4a_2+a_4b_2b_3+a_1a_3+a_2a_4,\
\det(M)=a_1a_2a_3a_4.
$$

Theorems~\ref{th_BC1} and \ref{th_BC2} imply the following:
\begin{itemize}
\item
If $b_j<0$ for all $j$, then the cycle is not an attractor and the stability
indices are $-\infty$.
\item
Suppose that $b_j>0$ for all $j$.
\begin{itemize}
\item
If
\begin{equation}\label{cond1}
\max\left(\tr M,2\,\tr M-2\det M\right)<2,
\end{equation}
then the cycle is not an attractor and the stability indices are $-\infty$.
\item
Otherwise the cycle is locally attracting and the stability indices are $\infty$.
\end{itemize}
\item
Suppose that $b_1<0$ and $b_j>0$ for $2\le j\le4$.
\begin{itemize}
\item
If (\ref{cond1}) holds, or
\begin{equation}\label{cond2}
\left(\tr M\right)^2-4\det M<0
\end{equation}
or
\begin{equation}\label{cond3}
v_{11}^{1,2}v_{12}^{1,2}<0
\end{equation}
then the cycle is not an attractor and the stability indices are $-\infty$.
\item
If none of the listed conditions are satisfied, the stability indices are
$$
\sigma_j=f^{\rm index}(v_{22}^{j+3,j}/h^{j+3,j},-v_{21}^{j+3,j}/h^{j+3,j})
~\mbox{ where }
$$
$$
h^{j+3,j}=v_{11}^{j+3,j}v_{22}^{j+3,j}-v_{12}^{j+3,j}v_{21}^{j+3,j}
~\mbox{ and }~v_{11}^{j+3,j}>0,\ v_{12}^{j+3,j}>0.
$$
($v_{11}^{j+3,j}>0$ and $v_{12}^{j+3,j}>0$ can be assumed, because
$v_{11}^{1,2}v_{12}^{1,2}>0$ implies that
$v_{11}^{j+3,j}v_{12}^{j+3,j}>0$ for all $j$).
\end{itemize}
\item
Suppose that $b_1<0$, $b_2<0$, $b_3>0$ and $b_4>0$.
\begin{itemize}
\item
If (\ref{cond1}), or (\ref{cond2}), or (\ref{cond3}), or
\begin{equation}\label{cond4}
v_{11}^{2,3}v_{12}^{2,3}<0,
\end{equation}
holds then the cycle is not an attractor and the stability indices are $-\infty$.
\item
If none of the listed conditions are satisfied then the stability indices are
$$\sigma_1=\min(f^{\rm index}(v_{22}^{4,1}/h^{4,1},-v_{21}^{4,1}/h^{4,1}),
f^{\rm index}(b_1b_2+a_1,b_2),f^{\rm index}(b_1,1)),$$
$$\sigma_2=\min(f^{\rm index}(v_{22}^{1,2}/h^{1,2},-v_{21}^{1,2}/h^{1,2}),
f^{\rm index}(b_2,1)),$$
$$\sigma_3=\min(f^{\rm index}(v_{22}^{2,3}/h^{2,3},-v_{21}^{2,3}/h^{2,3}),
f^{\rm index}(b_1b_3b_4+b_1a_3+b_3a_4,b_1b_4+a_4)),$$
$$\sigma_4=\min(f^{\rm index}(v_{22}^{3,4}/h^{3,4},-v_{21}^{3,4}/h^{3,4}),$$
$$f^{\rm index}(b_1b_4+a_4,b_1),
f^{\rm index}(b_1b_2b_4+b_2a_4+a_1b_4,b_1b_2+a_1)).$$
\end{itemize}
\item
Suppose that $b_1<0$, $b_2>0$, $b_3<0$ and $b_4>0$.
\begin{itemize}
\item
If (\ref{cond1}), or (\ref{cond2}), or (\ref{cond3}), or
\begin{equation}\label{cond5}
v_{11}^{3,4}v_{12}^{3,4}<0,
\end{equation}
holds then the cycle is not an attractor and the stability indices are $-\infty$.
\item
If none of the listed conditions are satisfied then the stability indices are
$$
\sigma_j=f^{\rm index}(v_{22}^{j+3,j}/h^{j+3,j},-v_{21}^{j+3,j}/h^{j+3,j}).
$$
\end{itemize}
\item
Suppose that $b_1<0$, $b_2<0$, $b_3<0$ and $b_4>0$.
\begin{itemize}
\item
If at least one of (\ref{cond1})-(\ref{cond5}) is satisfied,
then the cycle is not an attractor and the stability indices are $-\infty$.
\item
If none of the listed conditions are satisfied, the stability indices are
$$\sigma_1=\min(f^{\rm index}(v_{22}^{4,1}/h^{4,1},-v_{21}^{4,1}/h^{4,1}),
f^{\rm index}(b_1,1),f^{\rm index}(b_1b_2+a_1,b_2),$$
$$f^{\rm index}(b_1b_2b_3+a_1b_3+b_1a_2,b_2b_3+a_2)),$$
$$\sigma_2=\min(f^{\rm index}(v_{22}^{1,2}/h^{1,2},-v_{21}^{1,2}/h^{1,2}),
f^{\rm index}(b_2b_3+a_2,b_3),f^{\rm index}(b_2,1)),$$
$$\sigma_3=\min(f^{\rm index}(v_{22}^{2,3}/h^{2,3},-v_{21}^{2,3}/h^{2,3}),
f^{\rm index}(b_1b_3b_4+b_1a_3+b_3a_4,b_1b_4+a_4),f^{\rm index}(b_3,1)),$$
$$\sigma_4=\min(f^{\rm index}(v_{22}^{3,4}/h^{3,4},-v_{21}^{3,4}/h^{3,4}),
f^{\rm index}(b_1b_4+a_4,b_1),
f^{\rm index}(b_1b_4a_2+a_2a_4,b_1a_2)).$$
\end{itemize}
\end{itemize}

\subsection{Comparison with earlier results}\label{comp}

Asymptotic stability of heteroclinic cycles has been previously  examined in a number of papers.
Type A heteroclinic cycles were considered by Krupa and Melbourne \cite{KruMel95a,KruMel95b}.
In the first paper \cite{KruMel95a}, cycles with negative transverse eigenvalues were investigated
and the condition $\rho>1$ was found to be necessary and sufficient for asymptotic stability of cycles.
In the second paper \cite{KruMel95b} it was shown that cycles with some positive transverse eigenvalues
are essentially asymptotically stable, if and only if $\rho>1$ and the condition $t_j>-1$ is
satisfied for all $j$. This result is a special cases of our Theorem~\ref{th_A}.
The stability of Types B and C cycles with positive transverse eigenvalues was studied in
\cite{KruMel04}. The conditions for asymptotic stability presented in this paper are equivalent
to ours as given in subsections \ref{typeB} and \ref{typeC} for the special cases $b_j>0$.

A cycle of Type $C^-_2$ with one positive and one negative transverse eigenvalues
was considered in \cite{Pos10}. Conditions (12)-(14) in \cite{Pos10},
under which the set of points satisfying $\lim_{k\to\infty}g^k(w,z)=(0,0)$ is
non-empty, are equivalent to our conditions that $\sigma_1>-\infty$
for $b_1<0$ and $b_2>0$. Existence of the set
$U_0(\alpha_1^j,\beta_1^j,0;\alpha_2^j,\beta_2^j,0)$ was also noted
in \cite{Pos10}. It was termed $\hat\Sigma_{\beta_+}$ and was defined
by the condition $z>w^{\beta_+}$, which we express as $\beta_+\zeta-\eta<0$.
This inequality implies that the stability index $\sigma_1$ is equal to
$f^{\rm index}(\beta_+,-1)$. Substituting in it $\beta_+$ defined
in the beginning of \cite[Theorem 3.2]{Pos10}, we obtain the value $\sigma_2$
given in (\ref{c4_c1-}) (subject to the appropriate change of indices of $t_j$ and $c_j$).

\section{Discussion}\label{sec_conc}

Although it is natural to investigate cusp-like basins of
attraction for heteroclinic cycles, to our knowledge this is the first paper to identify the algebraic
order of the cusp as being an invariant of the dynamics --- we use the stability index $\sigma(x)$ to characterize the local geometry of basins of attraction near invariant sets $X$ in general, and heteroclinic cycles in particular.
This quantity might be especially useful in describing the local structure
of a range of invariant sets, for example, for riddled and intermingled basins of attraction \cite{Ale&al92,AshTer00}.

In the latter part of the paper we have calculated how
the stability indices depend on the cycle structure and eigenvalues for simple (robust heteroclinic)
cycles in $\R^4$. Clearly, transition matrices can be used to study
the stability of simple cycles in higher-dimensional systems or for more
complex cycles; however, we expect such a classification to be so complex that
the results can hardly be enlightening --- and so we have not attempted this.

Our approach should give some insight into the structure of heteroclinic
networks \cite{AshFie99,AshPod10} and heteroclinic switching
\cite{AshOroWorTow,DriHom09,HomKno10,KirSil94,KirLanRucSil10}. For some cycles (of Type A
where at least one difference between transverse and contracting eigenvalues
%, $t_j+c_j$,
is negative, or of Type B where at most one transverse eigenvalue is
positive) we find one of the stability indices to be $\infty$. Consequently,
if a heteroclinic network involves such a cycle, no switching is
possible for perturbations on that particular connecting trajectory (almost all
trajectories that are near a connection where $\sigma=\infty$ will stay near
the cycle for all $t>0$). Some of the results in this paper may be
extended to general robust heteroclinic cycles; it seems plausible that Corollary~\ref{cor_BC}
holds for simple heteroclinic cycles in $\R^n$ where all connections are
one-dimensional manifolds, because for such cycles the Poincar\'e map along the
cycle becomes linear after the same change of coordinates. It should also be possible to extend to cases of compact but not finite symmetry, though again this is likely to be quite involved owing to the complexity of heteroclinic cycles between relative equilibria.

Finally, we emphasise that there is no {\em a priori} reason why the limits in the
definition of the stability index exist. We give below an example where the
stability index can be shown not to converge. This is probably not a generic
situation though; the example below is highly degenerate, and the generic
conditions for the cycles in $\R^4$, as detailed in the previous section, all result in computable stability indices.

\paragraph{An example where $\sigma_+$ and $\sigma_-$ do not exist.}

We consider a (non-invertible) map $M:[0,\infty)\rightarrow [0,\infty)$
with $X=\{0\}$ and consider its basin of attraction $N=\cB(X)$. Define a
sequence $\epsilon_k=\exp(-2^k)$ and
$$
M(y)=\left\{
\begin{array}{rl}
(\epsilon_{2k-1}(y-\epsilon_{2k+2})-\epsilon_{2k}(y-\epsilon_{2k+1}))/(\epsilon_{2k+1}-
\epsilon_{2k+2}), & \mbox{ if }\epsilon_{2k+2}<|y|\leq \epsilon_{2k+1}\\
0, & \mbox{ if }\epsilon_{2k+1}<|y|\leq \epsilon_{2k}\\
0, & \mbox{ if }y=0
\end{array}\right..
$$
Then
$$
\Sigma_{\epsilon_{2k}}>
\frac{\epsilon_{2k}-\epsilon_{2k+1}}{\epsilon_{2k}}=1-\epsilon_{2k},
$$
whereas
$$
\Sigma_{\epsilon_{2k+1}}<
\frac{\epsilon_{2k+2}}{\epsilon_{2k+1}}=\epsilon_{2k+1}.
$$
Hence,
$\ln(\Sigma_{\epsilon_{2k}})/\ln(\epsilon_{2k})<\ln(1-\epsilon_{2k})/\ln(\epsilon_{2k})$ and
$\ln(\Sigma_{\epsilon_{2k+1}})/\ln(\epsilon_{2k+1})>1$, and therefore the limit defining
$\sigma_-(0)$ does not exist; it can similarly be shown that $\sigma_+(0)$ does not exist.
This example can clearly be extended to a continuous map $M$. Although it is not easy to see how to extend to a map that is differentiable at $y=0$ with the same properties, it may well be possible to produce a smooth example in dimension two or more.

\paragraph{Other global measures of stability}

The stability index $\sigma(x)$ describes the local geometry of the basin of
attraction of an invariant set $X$. One can define global and local {\em stability numbers} of a flow invariant set $X$ as follows:
$$
n_{\rm glob}(X)=\lim_{\epsilon\to 0}
\frac{\ell(B_{\epsilon}(X)\cap \cB(X))} {\ell (B_\epsilon(X))},
$$
$$
n_{\rm loc}(X)=\lim_{\delta\to 0}\lim_{\epsilon\to 0}
\frac{\ell(B_{\epsilon}(X)\cap \cB_{\delta}(X))} {\ell (B_\epsilon(X))}
$$
where $\cB_{\delta}$ is defined as in (\ref{eq:Sigma2}). Clearly, from the definition one can verify that $0\leq n_{\rm glob}(X)\leq 1$, $0\leq n_{\rm loc}(X)\leq 1$, and $n_{\rm glob}(X)=1$ if and only if $X$ is p.a.s.\ for a local basin of attraction. Note however that stability number is not an invariant of the dynamics; we believe that only the classification into whether $n(X)=0$, $0<n(X)<1$ or $n(X)=1$ and scaling properties will be invariant under smooth conjugation. For a heteroclinic cycle comprised of one-dimensional connections, the stability number can be related to its stability indices by the following:
$$
n_{\rm glob}(X)=\frac{\sum_{\sigma_j>0}\ell^1(s_j)}{\sum \ell^1(s_j)}.
$$
where $\ell^1(s_j)$ is the length of the connection $s_j$.

\subsection*{Acknowledgements}

Part of the research of OP was carried out during visit to the University of Exeter during January to April 2008. We are grateful to the Royal Society for the support of the visit. OP was also
financed by the grants ANR-07-BLAN-0235 OTARIE from Agence Nationale de
la Recherche, France, and 07-01-92217-CNRSL{\_}a from the Russian foundation
for basic research.

\appendix
\setcounter{theorem}{5}

\section{Type A cycles}\label{typeA}

In this Appendix we present a proof of the main theorem for calculation
of stability indices for Type A cycles. Consider the map
$g=g_m\circ\ldots\circ g_1:\R^2\to\R^2$, where\footnote{By virtue of
(\ref{eq_mapg0a}), $a_{2j}=0$ and $b_{2j}=1$.
%, or $a_{2j}=1$ and $b_{2j}=0$.
However, the first two lemmas in this Appendix are proved for arbitrary
$a_{ij}$ and $b_{ij}$.}
\begin{equation}\label{eq_mapg2}
g_j(w,z)\equiv(g_j^w(w,z),g_j^z(w,z))=
(A_jw^{a_{1j}}|z|^{a_{2j}}+B_j|w|^{b_{1j}}z^{b_{2j}},
C_jw^{a_{1j}}|z|^{a_{2j}}+D_j|w|^{b_{1j}}z^{b_{2j}}).
\end{equation}
For generic cycles of Type A, $A_jB_jC_jD_j\neq 0$ and $A_jD_j\neq B_jC_j$ holds
true for all $j$. In the first two lemmas in this appendix we assume that
$a_{1j}+a_{2j}<b_{1j}+b_{2j}$. No generality is lost, because the expression
(\ref{eq_mapg}) for the map $g_j(w,z)$ is invariant under the transformation
$(A_j,B_j,C_j,D_j;a_{1j},a_{2j},b_{1j},b_{2j};w,z)\to
(B_j,A_j,D_j,C_j;b_{2j},b_{1j},a_{2j},a_{1j};z,w).$
Recall that $g_{l,k}=g_l\circ\ldots\circ g_1 \circ g^k$.

Let us define
\begin{eqnarray*}
Q(r,\epsilon)&=&\{(w,z):|(w,z)|<\epsilon,\ |z|>|w|^{1+r},\ |w|>|z|^{1+r}\},\\
Q_z(f_1(z),f_2(z),\epsilon)&=&\{(w,z):|(w,z)|<\epsilon,\ f_1(z)<w<f_2(z)\},\\
Q_w(f_1(w),f_2(w),\epsilon)&=&\{(w,z):|(w,z)|<\epsilon,\ f_1(w)<z<f_2(w)\},
\end{eqnarray*}
where we assume $f_1(z)={\rm O}(z^{1+r})$, $f_1(w)={\rm O}(w^{1+r})$,
$f_1(z)-f_2(z)={\rm O}(z^{1+r_1})$, $f_1(w)-f_2(w)={\rm O}(w^{1+r_1})$ for some
$r>0$ and $r_1\ge r$.

We start the study of stability by proving a few lemmas about properties of
the maps $g_j$.

\begin{lemma}\label{lem_41}
Suppose $a_{1j}+a_{2j}>0$. For any $q>0$ and $r>0$ satisfying
$$(1-r)(b_{1j}-a_{1j})+b_{2j}-a_{2j}>0,\ b_{1j}-a_{1j}+(1-r)(b_{2j}-a_{2j})>0,\ a_{1j}+a_{2j}-q>0$$
and
\begin{equation}\label{cond41_2}
r<\min\left(\left|\frac{q}{2a_{1j}}\right|,\left|\frac{q}{2a_{2j}}\right|\right)
\end{equation}
there exists an $\epsilon_j>0$, such that
\begin{equation}\label{st41_1}
g_j(Q(r,\epsilon_j))\subset Q(r,\epsilon_j^{a_{1j}+a_{2j}-q}),
\end{equation}
\begin{equation}\label{st41_2}
|g_j(w,z)|<|(w,z)|^{a_{1j}+a_{2j}-q}\hbox{ for any }(w,z)\in Q(r,\epsilon_j)
\end{equation}
and
\begin{equation}\label{st41_3}
|g_j(w,z)|>|(w,z)|^{a_{1j}+a_{2j}+q}\hbox{ for any }(w,z)\in Q(r,\epsilon_j).
\end{equation}
\end{lemma}

\proof
At least one of the coefficients, $a_{1j}$ or $a_{2j}$, is positive. Assume
$a_{1j}>0$. The inequality $a_{1j}+a_{2j}<b_{1j}-b_{2j}$ implies
$b_{1j}-a_{1j}+b_{2j}-a_{2j}>0$ and hence at least one of two differences,
$b_{1j}-a_{1j}$ or $b_{2j}-a_{2j}$, is positive. For the sake of definiteness
we assume without loss of generality that
$b_{1j}-a_{1j}>0$. Set $\epsilon_j<1$ satisfying the following inequalities:
\begin{equation}\label{es41_1}
\epsilon_j^{(1-r)(b_{1j}-a_{1j})+b_{2j}-a_{2j}}<
\min\left(\left|\frac{A_j}{ 2B_j}\right|,\left|\frac{C_j}{2D_j}\right|\right)
\end{equation}
\begin{equation}\label{es41_2}
\epsilon_j^{q/2}|(3A_j/2,3C_j/2)|<1,
\end{equation}
\begin{equation}\label{es41_3}
\epsilon_j^{r(a_{1j}+a_{2j}-q)}<\left|\frac{A_j}{3C_j}\right|~~\mbox{ and }
\end{equation}
\begin{equation}\label{es41_4}
\epsilon_j^{q/2}K<|(A_j/2,C_j/2)|,
\end{equation}
where\footnote{Here, and below, we use the norm
$|(w,z)|=(w^2+z^2)^{1/2}$. If a different norm is employed, the proofs remain similar but some constants will be modified.} $K=2^{a+b+q+1}$.

Assume that
\begin{equation}\label{as41_1}
(w,z)\in Q(r,\epsilon_j)
\end{equation}
and re-write (\ref{eq_mapg}) as
$$g_j(w,z)=w^{a_{1j}}|z|^{a_{2j}}(A_j+B_jw^{b_{1j}-a_{1j}}z^{b_{2j}-a_{2j}},
C_j+D_jw^{b_{1j}-a_{1j}}z^{b_{2j}-a_{2j}}).$$
Due to (\ref{es41_1}),
\begin{equation}\label{es41_5}
|w^{b_{1j}-a_{1j}}z^{b_{2j}-a_{2j}}|<|z|^{(b_{1j}-a_{1j})/(1+r)+b_{2j}-a_{2j}}<
\epsilon_j^{(1-r)(b_{1j}-a_{1j})+b_{2j}-a_{2j}}<
\min\left(\left|\frac{A_j}{2B_j}\right|,\left|\frac{C_j}{2D_j}\right|\right).
\end{equation}
Therefore, due to (\ref{cond41_2}), (\ref{es41_2}) and (\ref{as41_1})
\begin{equation}\label{es41_6}
\begin{split}
|g_j(w,z)|<|w^{a_{1j}}z^{a_{2j}}(3A_j/2,3C_j/2)|<|z|^{a_{1j}/(1+r)+a_{2j}}|
(3A_j/2,3C_j/2)|\\
<|(w,z)|^{a_{1j}(1-r)+a_{2j}-q/2}\epsilon_j^{q/2}|(3A_j/2,3C_j/2)|<
|(w,z)|^{a_{1j}+a_{2j}-q}.
\end{split}
\end{equation}
Thus, (\ref{st41_2}) is proved.

The inequalities (\ref{es41_5}), (\ref{es41_3}) and (\ref{es41_6}) imply that for any $q>0$ and $r>0$ the $\epsilon_j$ can be chosen so that
$$
\left|\frac{g_j^w(w,z)}{g_j^z(w,z)}\right|>\left|\frac{A_j}{ 3C_j}\right|>\epsilon_j^{r(a_{1j}+a_{2j}-q)}
>|g_j^z(w,z)|^r.
$$
Hence
$$
|g_j^w(w,z)|>|g_j^z(w,z)|^{1+r},
$$
and similarly
$$
|g_j^z(w,z)|>|g_j^w(w,z)|^{1+r}.
$$
Therefore (\ref{st41_1}) holds because of (\ref{es41_6}).

Combining (\ref{es41_5}), (\ref{as41_1}), (\ref{cond41_2}) and (\ref{es41_4}) we obtain
\begin{equation}\label{es41_7}
\begin{split}
|g_j(w,z)|>|w^{a_{1j}}z^{a_{2j}}(A_j/2,C_j/2)|>|z|^{a_{1j}(1+r)+a_{2j}}|(A_j/2,C_j/2)|\\
>|z|^{a_{1j}(1+r)+a_{2j}+q/2}\epsilon_j^{-q/2}|(A_j/2,C_j/2)|>K|z|^{a_{1j}+a_{2j}+q}.
\end{split}
\end{equation}
Assume that $a_{2j}<0$. Estimates (\ref{es41_5}), (\ref{as41_1}), (\ref{cond41_2})
and (\ref{es41_4}) imply that
\begin{equation}
\begin{split}
|g_j(w,z)|>|w^{a_{1j}}z^{a_{2j}}(A_j/2,C_j/2)|>
|w|^{a_{1j}+a_{2j}/(1+r)}|(A_j/2,C_j/2)|\\
>|w|^{a_{1j}+a_{2j}(1-r)}|(A_j/2,C_j/2)|>
|w|^{a_{1j}+a_{2j}(1-r)+q/2}\epsilon_j^{-q/2}|(A_j/2,C_j/2)|\\
>K|w|^{a_{1j}+a_{2j}+q}.
\end{split}
\end{equation}
If $a_{2j}>0$ then
$$
|g_j(w,z)|>|w|^{a_{1j}+a_{2j}(1+r)}|(A_j/2,C_j/2)|>K|w|^{a_{1j}+a_{2j}+q}.
$$
Thus,
$$
|g_j(w,z)|>\frac{1}{2}K(|z|^{a_{1j}+a_{2j}+q}+|w|^{a_{1j}+a_{2j}+q})>|(w,z)|^{a_{1j}+a_{2j}+q}.
$$
\qed

\begin{corollary}\label{cor_lem_41}
Suppose that the conditions of Lemma~\ref{lem_41} are satisfied for $r>0$, $q>0$ and for all $1\le j\le m$
\begin{itemize}
\item[(a)]
If $\prod_{1\le j\le m}(a_{1j}+a_{2j}-q)>1$ then for any $\delta>0$ there
exists an $\epsilon>0$ such that $|g_{l,k}(w,z)|<\delta$ for all
$(w,z)\in Q(r,\epsilon)$ and for any $1\le l\le m$ and $k\ge0$.
\item[(b)]
Denote $\delta_0=\min_{1\le j\le m}\epsilon_j$.
If $\prod_{1\le j\le m}(a_{1j}+a_{2j}+q)<1$ then for any $\delta<\delta_0$ and
any $(w,z)\in Q(r,\delta)$ there exist
$l$ and $k$ such that $|g_{l,k}(w,z)|>\delta$.
\end{itemize}
\end{corollary}

\begin{lemma}\label{lem_411}
Suppose $a_{1j},a_{2j},b_{1j},b_{2j}\ge0$. For any $q>0$ there exists an $\epsilon>0$ such that
\begin{equation}\label{st411_1}
|g_j(w,z)|<|(w,z)|^{a_{1j}+a_{2j}-q}\mbox{ for any }(w,z)\mbox{ with }|(w,z)|<\epsilon.
\end{equation}
\end{lemma}

\proof
Let $\epsilon<\min((|A_j|+|C_j|+|B_j|+|D_j|)^{-1/q},1)$. Thus, for such $(w,z)$ we have
$$
\begin{array}{rcl} |g_j(w,z)| &<& (|A_j|+|C_j|)|w^{a_{1j}}z^{a_{2j}}|+(|B_j|+|D_j|)|w^{b_{1j}}z^{b_{2j}}|\\
&<& (|A_j|+|C_j|+|B_j|+|D_j|)|(w,z)|^{a_{1j}+a_{2j}}<|(w,z)|^{a_{1j}+a_{2j}-q}.
\end{array}
$$
\qed

The following lemmas and the main theorem of this subsection consider the
special case relevant to Type A cycles, where the maps $g_j(w,z)$ have
$a_{2j}=0$ and $b_{2j}=1$, i.e. they simplify to
\begin{equation}\label{eq_mapg0}
g_j(w,z)=(A_jw^{a_j}+B_j|w|^{b_j}z,C_jw^{a_j}+D_j|w|^{b_j}z)
\end{equation}
(we set $a_{1j}\equiv a_j$, $b_{1j}\equiv b_j$ and
do not assume necessarily that $a_j<b_j+1$).

\begin{lemma}\label{lem_43}
Suppose $a_j-b_j>1$.
\begin{itemize}
\item[(a)]
For any $q>0$, $f_1(z)$, $f_2(z)$, $\tilde{p}$ and $p$, such that
$\tilde{p}>p>1$ and
\begin{equation}\label{cond43_1}
f'_1(z)={\rm O}(z^{p-1}),\ f_1(z)-f_2(z)={\rm O}(z^{\tilde p}),
\hbox{ as }z\to 0
\end{equation}
there exist $\tilde f_1(w)$, $\tilde f_2(w)$ and $\epsilon_j>0$ such that
$$
\tilde f'_1(w)={\rm O}(w^{a_j-b_j-1}),\
\tilde f_1(w)-\tilde f_2(w)={\rm O}(w^{a_j\tilde p-b_j})\hbox{ for }w\to 0,
$$
and
$$
g_j(Q_w(\tilde f_1(w),\tilde f_2(w),\epsilon_j))\subset
Q_z(f_1(z),f_2(z),\epsilon_j^{a_j-q}),
$$
$$
g_j(Q_w(\tilde f_1(w),\tilde f_2(w),\epsilon_j))\supset
Q_z(f_1(z),f_2(z),\epsilon_j^{a_j+q}),
$$
$$
|g_j(w,z)|<|(w,z)|^{a_j-q},\ |g_j(w,z)|>|(w,z)|^{a_j+q}
$$
for all $(w,z)\in Q_w(\tilde f_1(w),\tilde f_2(w),\epsilon_j)$.
\item[(b)]
For any $q>0$, $f_1(w)$, $f_2(w)$, $\tilde{p}$ and $p$, such that
$\tilde{p}>p>1$ and
\begin{equation}\label{cond43_5}
f'_1(w)={\rm O}(w^{p-1}),\ f_1(w)-f_2(w)={\rm O}(w^{\tilde p}),
\hbox{ as }w\to 0,
\end{equation}
there exist $\tilde f_1(w)$, $\tilde f_2(w)$ and $\epsilon_j>0$ such that
$$
\tilde f'_1(w)={\rm O}(w^{a_j-b_j-1}),\
\tilde f_1(w)-\tilde f_2(w)={\rm O}(w^{a_j\tilde p-b_j})\hbox{ as }w\to 0,
$$
and
$$
g_j(Q_w(\tilde f_1(w),\tilde f_2(w),\epsilon_j))\subset
Q_w(f_1(w),f_2(w),\epsilon_j^{a_j-q}),
$$
$$
g_j(Q_w(\tilde f_1(w),\tilde f_2(w),\epsilon_j))\supset
Q_w(f_1(w),f_2(w),\epsilon_j^{a_j+q}),
$$
$$
|g_j(w,z)|<|(w,z)|^{a_j-q},\ |g_j(w,z)|>|(w,z)|^{a_j+q}
$$
for all $(w,z)\in Q_w(\tilde f_1(w),\tilde f_2(w),\epsilon_j)$.
\end{itemize}
\end{lemma}

\proof
(a) Let $\tilde f_{1,2}(w)$ be respectively the solutions of
\begin{equation}\label{es43_1}
A_jw^{a_j}+B_j|w|^{b_j}\tilde f_1(w)=f_1(C_jw^{a_j}+D_j|w|^{b_j}\tilde f_1(w))
\end{equation}
and
\begin{equation}\label{es43_2}
A_jw^{a_j}+B_j|w|^{b_j}\tilde f_2(w)=f_2(C_jw^{a_j}+D_j|w|^{b_j}\tilde f_2(w)).
\end{equation}
The functions $\tilde f_{1,2}(w)$ are defined for any small $w$ and due
to (\ref{cond43_1})
\begin{equation}\label{es43_3}
\tilde f_{1,2}(w)=-\frac{A_j}{B_j}w^{a_j-b_j}+{\rm O}(w^{a_jp-b_j}).
\end{equation}
Substitution of $w\equiv w+\delta w$ into (\ref{es43_1}) implies
\begin{equation}\label{es43_5}
A_j(w+\delta w)^{a_j}+B_j|(w+\delta w)|^{b_j}\tilde f_1(w+\delta w)=
f_1(C_j(w+\delta w)^{a_j}+D_j|(w+\delta w)|^{b_j}\tilde f_1(w+\delta w)).
\end{equation}
Subtracting now (\ref{es43_1}) from (\ref{es43_5}), dividing the result by
$\delta w$ and taking the limit $\delta w\to0$, we obtain that
$$\tilde f'_1(w)=-\frac{A_j(a_j-b_j)}{B_j}w^{a_j-b_j-1}+{\rm O}(w^{a_jp-b_j-1}).$$

Subtraction of (\ref{es43_2}) from (\ref{es43_1}) yields
\begin{equation}\label{eq_sub}
\begin{split}
B_j|w|^{b_j}(\tilde f_1(w)-\tilde f_2(w))=f_1(C_jw^{a_j}+D_j|w|^{b_j}
\tilde f_1(w))-f_1(C_jw^{a_j}+D_j|w|^{b_j}\tilde f_2(w))+\\
f_1(C_jw^{a_j}+D_j|w|^{b_j}\tilde f_2(w))-
f_2(C_jw^{a_j}+D_j|w|^{b_j}\tilde f_2(w)).
\end{split}
\end{equation}
Because of (\ref{cond43_1}) and (\ref{es43_3})
\begin{equation}
\begin{split}
f_1(C_jw^{a_j}+D_j|w|^{b_j}\tilde f_1(w))-
f_1(C_jw^{a_j}+D_j|w|^{b_j}\tilde f_2(w))\\
\approx f_1'((C_j-\frac{D_jA_j}{B_j})w^{a_j})D_j|w|^{b_j}(\tilde f_1(w)-\tilde f_2(w))=
{\rm O}(w^{a_j(p-1)+b_j})(\tilde f_1(w)-\tilde f_2(w))
\end{split}
\end{equation}
and
$$
f_1(C_jw^{a_j}+D_j|w|^{b_j}\tilde f_2(w))-
f_2(C_jw^{a_j}+D_j|w|^{b_j}\tilde f_2(w))={\rm O}(w^{a_j\tilde p}).
$$
Hence (\ref{eq_sub}) takes the form
$$B_jw^{b_j}(\tilde f_1(w)-\tilde f_2(w))=
{\rm O}(w^{a_j(p-1)+b_j})(\tilde f_1(w)-\tilde f_2(w))+{\rm O}(w^{a_j\tilde p}).$$
Since $p>1$ and $a_j>1$, the first term in the r.h.s. of this expression is
asymptotically smaller than the l.h.s. and it can be ignored. Thus,
$$
\tilde f_1(w)-\tilde f_2(w)={\rm O}(w^{a_j\tilde p-b_j}).
$$

The asymptotic relation (\ref{es43_3}) implies that
there exist $\tilde\epsilon$ and $K$ such that
$$
\left|z+\frac{A_j}{B_j}w^{a_j-b_j}\right|<K|w^{a_jp-b_j}|
$$
for $(w,z)\in Q_w(\tilde f_1(w),\tilde f_2(w),\tilde\epsilon)$. Suppose that $\epsilon_j$ satisfies
$$
\epsilon_j<\min(\tilde\epsilon,1),\
\epsilon_j^q\left(|C_j|+\left|\frac{A_jD_j}{B_j}\right|+(|B_j|+|D_j|)K\right)<1,
$$
$$
K\epsilon_j^{a_j(p-1)}<\min\left(\frac{1}{2}\left|\frac{C_j}{D_j}-\frac{A_j}{B_j}\right|,\left|\frac{A_j}{B_j}\right|\right),
$$
$$
\epsilon_j^{-q}\left|C_j-\frac{D_jA_j}{B_j}\right|>4,~~
\epsilon_j^{a_j-b_j-1}<\left|\frac{B_j}{A_j}\right|.
$$
Then
$$
\begin{array}{rcl}
|g_j(w,z)|&=&|w^{a_j}(A_j+B_jw^{b_j-a_j}z,C_j+D_jw^{b_j-a_j}z)|\\
&<&|w|^{a_j}(|C_j|+|\frac{A_jD_j}{B_j}|+(|B_j|+|D_j|)K)<|(w,z)|^{a_j-q}
\end{array}
$$
and
$$
|g_j(w,z)|>|w^{a_j}(C_j+D_jw^{b_j-a_j}z)|>
|w|^{a_j}\frac{1}{2}\left|C_j-\frac{D_jA_j}{B_j}\right|>2|w|^{a_j+q}>
|(w,z)|^{a_j+q}.
$$
From (\ref{es43_1}) and (\ref{es43_2}), the proof of part (a) is complete.

(b) Set $\tilde f_{1,2}(w)$ to be respectively
the solutions of
$$
C_jw^{a_j}+D_j|w|^{b_j}\tilde f_1(w)=f_1(A_jw^{a_j}+B_j|w|^{b_j}\tilde f_1(w))
$$
and
$$
C_jw^{a_j}+D_j|w|^{b_j}\tilde f_2(w)=f_2(A_jw^{a_j}+B_j|w|^{b_j}\tilde f_2(w)).
$$
The remainder of the proof is similar to case (a).
\qed

\begin{corollary}\label{cor_lem_43}
Lemma \ref{lem_43} implies that the maps $f_l\to\tilde f_l$ are
monotonic in the following sense:
\begin{itemize}
\item[(a)]
Let $f_0$ be such that $f_1(z)<f_0(z)<f_2(z)$ for all $|z|<\epsilon_j^{a_j-q}$.
Define $\tilde f_0(w)$ by analogy with $\tilde f_{1,2}(w)$ in part (a) of that
lemma.
$$
\hbox{If }\tilde f_1(w)>\tilde f_2(w),\hbox{ then }
\tilde f_1(w)>\tilde f_0(w)>\tilde f_2(w),
$$
$$
\hbox{If }\tilde f_1(w)<\tilde f_2(w),\hbox{ then }
\tilde f_1(w)<\tilde f_0(w)<\tilde f_2(w)
$$
for all $|w|<\epsilon_j$.
\item[(b)]
Let $f_0$ be such that $f_1(w)<f_0(w)<f_2(w)$ for all $|w|<\epsilon_j^{a_j-q}$.
Define $\tilde f_0(w)$ by analogy with $\tilde f_{1,2}(w)$ in part (b) of that
lemma.
$$
\hbox{If }\tilde f_1(w)>\tilde f_2(w),\hbox{ then }
\tilde f_1(w)>\tilde f_0(w)>\tilde f_2(w),
$$
$$
\hbox{If }\tilde f_1(w)<\tilde f_2(w),\hbox{ then }
\tilde f_1(w)<\tilde f_0(w)<\tilde f_2(w)
$$
for all $|w|<\epsilon_j$.
\end{itemize}
\end{corollary}

\begin{lemma}\label{lem_44}
Suppose that $0<a_j-b_j<1$.
\begin{itemize}
\item[(a)]
For any $q>0$, $f_1(z)$, $f_2(z)$, $\tilde{p}$ and $p$, such that
$\tilde{p}>p>1$ and
\begin{equation}\label{cond44_1}
f'_1(z)={\rm O}(z^{p-1}),\ f_1(z)-f_2(z)={\rm O}(z^{\tilde p}),
\hbox{ as }z\to 0,
\end{equation}
there exist $\tilde f_1(z)$, $\tilde f_2(z)$ and $\epsilon_j>0$ such that
$$
\tilde f'_1(z)={\rm O}(z^{1/(a_j-b_j)-1}),\
\tilde f_1(z)-\tilde f_2(z)={\rm O}(w^{(a_j\tilde p-a_j+1)/(a_j-b_j)})
\hbox{ for }z\to 0,
$$
and
$$
g_j(Q_z(\tilde f_1(z),\tilde f_2(z),\epsilon_j))\subset
Q_z(f_1(z),f_2(z),\epsilon_j^{a_j/(a_j-b_j)-q}),
$$
$$
g_j(Q_z(\tilde f_1(z),\tilde f_2(z),\epsilon_j))\supset
Q_z(f_1(z),f_2(z),\epsilon_j^{a_j/(a_j-b_j)+q}),
$$
$$
|g_j(w,z)|<|(w,z)|^{a_j/(a_j-b_j)-q},\ |g_j(w,z)|>|(w,z)|^{a_j/(a_j-b_j)+q},
$$
for all $(w,z)\in Q_z(\tilde f_1(z),\tilde f_2(z),\epsilon_j)$.
\item[(b)]
For any $q>0$, $f_1(w)$, $f_2(w)$, $\tilde{p}$ and $p$, such that
$\tilde{p}>p>1$ and
\begin{equation}%\label{cond43_5b}
f'_1(w)={\rm O}(w^{p-1}),\ f_1(w)-f_2(w)={\rm O}(w^{\tilde p}),
\hbox{ as }w\to 0,
\end{equation}
there exist $\tilde f_1(z)$, $\tilde f_2(z)$ and $\epsilon_j>0$ such that
$$
\tilde f'_1(z)={\rm O}(z^{1/(a_j-b_j)-1}),\
\tilde f_1(z)-\tilde f_2(z)={\rm O}(z^{(a_j\tilde p-a_j+1)/(a_j-b_j)})
\hbox{ for }z\to 0,
$$
and
$$
g_j(Q_z(\tilde f_1(z),\tilde f_2(z),\epsilon_j)\subset
Q_w(f_1(w),f_2(w),\epsilon_j^{a_j/(a_j-b_j)-q}),
$$
$$
g_j(Q_z(\tilde f_1(z),\tilde f_2(z),\epsilon_j)\supset
Q_w(f_1(w),f_2(w),\epsilon_j^{a_j/(a_j-b_j)+q}),
$$
$$
|g_j(w,z)|<|(w,z)|^{a_j/(a_j-b_j)-q},\ |g_j(w,z)|>|(w,z)|^{a_j/(a_j-b_j)+q},
$$
$$
\hbox{ for all }(w,z)\in Q_z(\tilde f_1(z),\tilde f_2(z),\epsilon_j).
$$
\end{itemize}
\end{lemma}

\proof
(a) Let $\tilde f_{1,2}(z)$ be solutions of
\begin{equation}\label{es44_1}
A_j\tilde f_1^{a_j}(z)+B_j|\tilde f_1|^{b_j}(z)z=
f_1(C_j\tilde f_1^{a_j}(z)+D_j|\tilde f_1|^{b_j}(z)z)
\end{equation}
and
\begin{equation}\label{es44_2}
A_j\tilde f_2^{a_j}(z)+B_j|\tilde f_2|^{b_j}(z)z=
f_2(C_j\tilde f_2^{a_j}(z)+D_j|\tilde f_2|^{b_j}(z)z).
\end{equation}
The functions $\tilde f_{1,2}(z)$ are defined for any small $z$. Note that
\begin{equation}\label{es44_3}
\tilde f_{1,2}(z)=-\left(\frac{B_j}{A_j}\right)^{1/(a_j-b_j)}z^{1/(a_j-b_j)}+
{\rm O}(z^{(a_jp-a_j+1)/(a_j-b_j)}).
\end{equation}
We substitute $z\to z+\delta z$ into (\ref{es44_1}),
subtract (\ref{es44_1}) from the obtained equation, divide the result by
$\delta z$ and take the limit $\delta z\to0$. Since
$$f_1^{a_j}(z+\delta z)-f_1^{a_j}(z)=
f_1^{a_j}(z)\left(1+\frac{f_1(z+\delta z)-f_1(z)}{ f_1(z)}\right)^{a_j}-f_1^{a_j}(z)\approx
a_jf_1^{a_j-1}(f_1(z+\delta z)-f_1(z)),$$
and similar estimate holds true for $b_j$, we obtain that
\begin{equation}\label{es44_3_2}
\tilde f'_1(z)=-\frac{1}{ (a_j-b_j)}\left(\frac{B_j}{A_j}\right)^{1/(a_j-b_j)}z^{1/(a_j-b_j)-1}+
{\rm O}(z^{(a_jp-a_j+1)/(a_j-b_j)-1}).
\end{equation}

Subtracting (\ref{es44_2}) from (\ref{es44_1}) we obtain
\begin{equation}\label{es44_4}
A_j(\tilde f_1^{a_j}-\tilde f_2^{a_j})+
B_jz(|\tilde f_1|^{b_j}-|\tilde f_2|^{b_j})=
\end{equation}
$$f_1(C_j\tilde f_1^{a_j}+D_j|\tilde f_1|^{b_j}z)-
f_2(C_j\tilde f_1^{a_j}+D_j|\tilde f_1|^{b_j}z)+
f_2(C_j\tilde f_1^{a_j}+D_j|\tilde f_1|^{b_j}z)-
f_2(C_j\tilde f_2^{a_j}+D_j|\tilde f_2|^{b_j}z).$$
Since
$$\tilde f_1^{a_j}-\tilde f_2^{a_j}=
(\tilde f_1-\tilde f_2){\rm O}(z^{(a_j-1)/(a_j-b_j)}),$$
$$z(|\tilde f_1|^{b_j}-|\tilde f_2|^{b_j})=
(\tilde f_1-\tilde f_2){\rm O}(z^{(a_j-1)/(a_j-b_j)}),$$
$$f_1(C_j\tilde f_1^{a_j}+D_j|\tilde f_1|^{b_j}z)-
f_2(C_j\tilde f_1^{a_j}+D_j|\tilde f_1|^{b_j}z)=
{\rm O}(z^{\tilde pa_j/(a_j-b_j)})$$
and
$$f_2(C_j\tilde f_1^{a_j}+D_j|\tilde f_1|^{b_j}z)-
f_2(C_j\tilde f_2^{a_j}+D_j|\tilde f_2|^{b_j}z)=
(\tilde f_1-\tilde f_2){\rm O}(z^{(pa_j-1)/(a_j-b_j)}),$$
(\ref{es44_4}) implies that
$$\tilde f_1(z)-\tilde f_2(z)={\rm O}(z^{(\tilde pa_j-a_j+1)/(a_j-b_j)}).$$

Due to (\ref{es44_3}) there exist $\tilde\epsilon$ and $K$ such that
$$
\left|z+\frac{A_j}{B_j}w^{a_j-b_j}\right| <K|z|^{(a_jp-b_j)/(a_j-b_j)}
\hbox{ for }(w,z)\in Q_w(\tilde f_1(z),\tilde f_2(z),\tilde\epsilon).$$
Suppose that $\epsilon_j$ satisfies
$$
\epsilon_j<\min(\tilde\epsilon,1),~~~
\epsilon_j^q\left|\frac{2B_j}{A_j}\right|^{a_j/(a_j-b_j)}
\left(|C_j|+\left|\frac{A_jD_j}{B_j}\right|+(|B_j|+|D_j|)K\right)<1,
$$
$$
K\epsilon_j^{a_j(p-1)/(a_j-b_j)}<\min\left(
\frac{1}{2}\left|\frac{C_j}{D_j}-\frac{A_j}{B_j}\right|,\left|\frac{A_j}{B_j}\right|\right),
$$
$$
\epsilon_j^{-q} \frac{1}{4}\left|\frac{B_j}{2A_j}\right|^{a_j/(a_j-b_j)}
\left|C_j-\frac{D_jA_j}{B_j}\right|>1~~\mbox{ and }~~
\epsilon_j^{1/(a_j-b_j)-1}<\left(\frac{B_j}{A_j}\right)^{1/(a_j-b_j)}.
$$
Then
\begin{equation}
\begin{split}
|g_j(w,z)|=|w^{a_j}(A_j+B_jw^{b_j-a_j}z,C_j+D_jw^{b_j-a_j}z)|\\
<
\left|\frac{2B_j}{A_j}\right|^{a_j/(a_j-b_j)}|z|^{a_j/(a_j-b_j)}\left(|C_j|+
\left|\frac{A_jD_j}{B_j}\right|+(|B_j|+|D_j|)K\right)\\
<|(w,z)|^{a_j/(a_j-b_j)-q}
\end{split}
\end{equation}
and
\begin{equation}
\begin{split}
|g_j(w,z)|>|w^{a_j}(C_j+D_jw^{b_j-a_j}z)|\\
>|w|^{a_j}\frac{1}{2}\left|C_j-\frac{D_jA_j}{B_j}\right|\\
>2|z|^{a_j/(a_j-b_j)+q}>
|(w,z)|^{a_j/(a_j-b_j)+q}.
\end{split}
\end{equation}
Hence part (a) is proved. The proof for the part (b) is similar.
\qed

\begin{corollary}\label{cor_lem_44}
Lemma~\ref{lem_44} implies that the maps $f_l\to \tilde f_l$ are
monotonic in the following sense:
\begin{itemize}
\item[(a)]
Let $f_0$ be such that $f_1(z)<f_0(z)<f_2(z)$ for all
$|z|<\epsilon_j^{a_j/(a_j-b_j)-q}$.
Define $\tilde f_0(z)$ by analogy with $\tilde f_{1,2}(z)$ in part (a) of that
lemma.
$$
\hbox{If }\tilde f_1(z)>\tilde f_2(z),\hbox{ then }
\tilde f_1(z)>\tilde f_0(z)>\tilde f_2(z),
$$
$$
\hbox{If }\tilde f_1(z)<\tilde f_2(z),\hbox{ then }
\tilde f_1(z)<\tilde f_0(z)<\tilde f_2(z).
$$
for all $|z|<\epsilon_j$.
\item[(b)]
Let $f_0$ be such that $f_1(w)<f_0(w)<f_2(w)$ for all
$|w|<\epsilon_j^{a_j/(a_j-b_j)-q}$.
Define $\tilde f_0(z)$ by analogy with $\tilde f_{1,2}(z)$ in part (b) of that
lemma.
$$
\hbox{If }\tilde f_1(z)>\tilde f_2(z),\hbox{ then }
\tilde f_1(z)>\tilde f_0(z)>\tilde f_2(z),
$$
$$
\hbox{If }\tilde f_1(z)<\tilde f_2(z),\hbox{ then }
\tilde f_1(z)<\tilde f_0(z)<\tilde f_2(z).
$$
for all $|z|<\epsilon_j$.
\end{itemize}
\end{corollary}

\begin{lemma}\label{lem_450}
Suppose $a_j>0$ and $a_j-b_j<0$. Then for any $\epsilon_0>0$ and $r>0$ there exists an
$\epsilon>0$ such that
$$
g_j(w,z)\in Q(r,\epsilon_0)\mbox{ for any }(w,z)\mbox{ with }|(w,z)|<\epsilon.
$$
\end{lemma}

\proof
Let $\epsilon$ satisfy
$$
\epsilon^{b_j-a_j+1}<\min\left(\left|\frac{A_j}{2B_j}\right|,\left|\frac{C_j}{2D_j}\right|\right),
$$
$$
\epsilon^{a_j}|(3A_j/2,3C_j/2)|<\epsilon_0
$$
and
$$
\epsilon^{ra_j}<\min\left(\left|\frac{C_j}{2}\left(\frac{3A_j}{2}\right)^{-1-r}\right|,
\left|\frac{A_j}{2}\left(\frac{3C_j}{2}\right)^{-1-r}\right|\right).
$$
Therefore, if $|(w,z)|<\epsilon$
then
$$
|g_j(w,z)|=|w^{a_j}||(A_j+B_jw^{b_j-a_j}z,C_j+D_jw^{b_j-a_j}z)|<
\epsilon^{a_j}|(3A_j/2,3C_j/2)|<\epsilon_0,
$$
$$
|g_j^w(w,z)|^{1+r}/|g_j^z(w,z)|<|w^{ra_j}| \frac{|3A_j/2|^{1+r}}{|C_j/2|}<1
$$
and similarly
$$
|g_j^z(w,z)|^{1+r}/|g_j^w(w,z)|<1.
$$
\qed

\begin{lemma}\label{lem_46}
For any $\tilde q>0$ there exists an $\epsilon_j>0$ such that
$$
|g_j(w,z)|<|(w,z)|^{\beta_j} \mbox{ where }\beta_j=\min(a_j/2,|b_j|\tilde q/2,|1+b_j|/2),
$$
for all $(w,z)\in B_{\epsilon_j}\setminus\tilde Q_1$,
where
$$
\widetilde Q_1=
\left\{\begin{array}{ll}
\emptyset & \mbox{ if } b_j>0\\
Q_z(-|z|^{-1/b_j-\tilde q},|z|^{-1/b_j-\tilde q},\epsilon_j) & \mbox{ if } b_j<0.
\end{array}\right.
$$
\end{lemma}

\proof
If
$$
(|A_j|+|C_j|)\epsilon_j^{a_j/2}<1/2,\
(|B_j|+|D_j|)\epsilon_j^{|b_j|\tilde q/2}<1/2,\
(|B_j|+|D_j|)\epsilon_j^{|1+b_j|/2}<1/2
$$
then one can verify that
$$
|g_j(w,z)|<(|A_j|+|C_j|)|w|^{a_j}+(|B_j|+|D_j|)|w^{b_j}z|<|(w,z)|^{\beta_j}.
$$
\qed

\begin{lemma}\label{lem_47}
For any $\tilde q>0$, $r>0$ and $\epsilon>0$ there exists an
$\epsilon_j>0$ such that
$$g_j(w,z)\in Q(r,\epsilon)\hbox{ for all }(w,z)\in
B_{\epsilon_j}\setminus(\widetilde Q_1\cup\widetilde Q_2\cup\widetilde Q_3),$$
where $\widetilde Q_1$ is defined in Lemma \ref{lem_46},
$$
\widetilde Q_2=
\left\{
\renewcommand{\arraystretch}{1.2}
\begin{array}{ll}
\emptyset & \mbox{ if } a_j-b_j<0\\
Q_z(-(\frac{B_j}{A_j})^{\alpha_1}z^{\alpha_1}-|z|^{\alpha_2},

-(\frac{B_j}{ A_j})^{\alpha_1}z^{\alpha_1}+|z|^{\alpha_2},\epsilon_j)&
\mbox{ if } 0<a_j-b_j<1\\
Q_w(-\frac{A_j}{ B_j}w^{\alpha_3}-|w|^{\alpha_4},
-\frac{A_j}{ B_j}w^{\alpha_3}+|w|^{\alpha_4},\epsilon_j)&
\mbox{ if } a_j-b_j>1
\end{array}\right.,
$$
$$
\widetilde Q_3=
\left\{
\renewcommand{\arraystretch}{1.2}
\begin{array}{ll}
\emptyset & \mbox{ if } a_j-b_j<0\\
Q_z(-(\frac{D_j}{ C_j})^{\alpha^1}z^{\alpha_1}-|z|^{\alpha_2},
-(\frac{D_j}{ C_j})^{\alpha_1}z^{\alpha_1}+|z|^{\alpha_2},\epsilon_j) &
\mbox{ if } 0<a_j-b_j<1\\
Q_w(\frac{C_j}{ D_j}w^{\alpha_3}-|w|^{\alpha_4},
\frac{C_j}{ D_j}w^{\alpha_3}+|w|^{\alpha_4},\epsilon_j) &
\mbox{ if } a_j-b_j>1
\end{array}\right.,
$$
and
$$
\alpha_1=\frac{1}{ a_j-b_j},\ \alpha_2=\frac{a_j(r+1)-2a_j+2}{ 2(a_j-b_j)},\
\alpha_3=a_j-b_j,\ \alpha_4=\frac{a_j(r+1)}{ 2}-b_j.
$$
\end{lemma}

\proof
We start with the proof of existence of $\epsilon_w$ such that in the case
$a_j-b_j>1$ the condition
$$(w,z)\in B_{\epsilon_w}
\setminus(\widetilde Q_1\cup\widetilde Q_2\cup\widetilde Q_3)$$
implies that
\begin{equation}\label{es47_1}
|g^w_j(w,z)|>|g^z_j(w,z)|^{1+r}.
\end{equation}

If $|A_jw^{a_j}+B_j|w|^{b_j}z|<|A_jw^{a_j}/2|$ then
$$|g^w_j(w,z)|>|B_jw^{a_j(r+1)/2}|
\hbox{ and }
|g^z_j(w,z)|<(|C_j|+|3A_jD_j/B_j|)|w|^{a_j}.$$
For $|w^{a_j(r+1)/2}|<|B_j|(|C_j|+|3A_jD_j/B_j|)^{-r-1}$ the
inequality (\ref{es47_1}) holds true.

If $|A_jw^{a_j}+B_j|w|^{b_j}z|>|A_jw^{a_j}/2|$ then
$$
|g^w_j(w,z)|>\min(|A_j/2|,|B_j/3|)\max(|w^{a_j}|,|w^{b_j}z|)
$$
and
$$
|g^z_j(w,z)|<2\max(|C_j|,|D_j|)\max(|w^{a_j}|,|w^{b_j}z|).
$$
Hence if
$$
|(w,z)|^{r\beta_j}<\min(|A_j/2|,|B_j/3|)(2\max(|C_j|,|D_j|))^{-r-1},
$$
where $\beta_j$ is defined in Lemma~\ref{lem_46},
then (\ref{es47_1}) holds.

The proof of existence of $\epsilon_z$ such that
$|g^z_j(w,z)|>|g^w_j(w,z)|^{1+r}$ is similar.
Denote by $\tilde\epsilon_j$ the $\epsilon_j$ from Lemma \ref{lem_46}.
Then $\epsilon_j=\min(\epsilon_w,\epsilon_z,\tilde\epsilon_j)$, by the
definition of the set $Q(r,\epsilon)$, satisfies the the condition of the lemma. For the case
$0<a_j-b_j<1$ the proof is similar and is not presented, for the case
$a_j-b_j<0$ the statement of the Lemma follows from Lemma~\ref{lem_450}.
\qed

Let the collection of functions $\{h_{l,j}(y)\}$ for $1\le j\le m$, $l\le j$,
be defined as follows\footnote{If an index takes values $1,\ldots,m$, then the index
value modulo $m$ is understood here and below.}:
\begin{eqnarray*}
h_{j,j}(y)&=&y,\\
h_{l,j}(y)&=&
\left\{
\renewcommand{\arraystretch}{1.2}
\begin{array}{rl}
\infty & \mbox{ if } a_l-b_l<0\\
\displaystyle \frac{a_lh_{l+1,j}(y)-a_l+1}{ a_l-b_l} & \mbox{ if } 0<a_l-b_l<1\\
a_lh_{l+1,j}(y)-b_l & \mbox{ if } a_l-b_l>1
\end{array}\right.
\end{eqnarray*}
This collection has the following properties:
\begin{itemize}
\item[(a)]
$$
h_{l,j}(y_0+y_1)=h_{l,j}(y_0)+a_{l,j}y_1,
$$
where $a_{l,j}=\prod_{l\le s<j}\max(a_s,a_s/(a_s-b_s))$
is defined for $l\le j$.
\item[(b)]
If there exist $J$ such that $a_J-b_J<0$, then $h_{l,j}(y)=\infty$
for $l\le J$.
\item[(c)]
If $a_{j-m,j}>1$, then
$h_{l-m,j}(y)>h_{l,j}(y)$ for any $y\ge 1$.
\item[(d)]
If $a_{j-m,j}>1$, then
$\lim_{l\to-\infty} h_{l,j}(y)=\infty$ for any $y\ge 1$.
\end{itemize}

The next theorem gives the main result for Type A cycles,
namely it gives the stability indices $\sigma_j$ for the
collection of maps $g_j$ related to Type A cycles. The coefficients $a_j$ and $b_j$
of the map $g_j$ are related to the eigenvalues of linearisation
of (\ref{eq_ode}) near $\xi_j$ as $a_j=c_j/e_j$ and $b_j=-t_j/e_j$.
Recall, that $c_j>0$ and $e_j>0$ for all $j$ and therefore $a_j>0$.
Following \cite{KruMel95a}, we denote
$$
\rho_j=\min(a_j,1+b_j),
$$
$\rho=\rho_1\cdots\rho_m$, and note that generically the non-degeneracy
conditions (\ref{eq_typea_nondeg}) apply.
%\begin{equation}\label{eq_typea_nondeg}
%a_j\ne 1+b_j,~b_j\ne -1,~\rho\ne 1
%\end{equation}
%are satisfied.

\begin{theorem}\label{th_A_repeat}(reproduces Theorem~\ref{th_A})
For the collection of maps $g_j$ associated with a Type A cycle, the stability indices are:
\begin{itemize}
\item[(a)] If $\rho>1$ and $b_j>0$ for all $j$ then
$\sigma_{j,+}=\infty$ and $\sigma_{j,-}=0$ for any $j$.
\item[(b)] If $\rho>1$, $b_j>-1$ for all $j$ and $b_j<0$ for
$j=J_1,\ldots,J_L$ then $\sigma_{j,-}=0$ and $\sigma_{j,+}$ are:
$$
\sigma_{j,+}=\min_{s=J_1,\ldots,J_L}h_{j,s}\left(-\frac{1}{ b_s}\right)-1.
$$
\item[(c)] If $\rho<1$ or there exists $j$ such that $b_j<-1$ then
$\sigma_{j,+}=0$, $\sigma_{j,-}=\infty$
and the cycle is not an attractor.
\end{itemize}
\end{theorem}

\proof
(a) Since $\rho>1$, there exists a $q>0$ such that
\begin{equation}\label{A_es0}
\prod_{j=1}^m(\rho_j-q)>1.
\end{equation}
By Lemma \ref{lem_411}, for any $j$ there exist $\epsilon_j$ such that
\begin{equation}\label{A_es1}
|g_j(w,z)|<|(w,z)|^{\rho_j-q}\mbox{ for any }(w,z)\mbox{ with }|(w,z)|<\epsilon_j.
\end{equation}
For a given $\delta$, choose an $\epsilon>0$ satisfying
\begin{equation}\label{A_es2}
\epsilon^{\rho_{1,j}}<\min(\delta,\epsilon_j,1)\mbox{ and }
\rho_{1,j}=\prod_{s=1}^j(\rho_s-q),\mbox{ for all }1\le j\le m.
\end{equation}
Consider $(w,z)\in H^{(in)}_1$. If $|(w,z)|<\epsilon$, then (\ref{A_es1}) and
(\ref{A_es2}) imply that
$|g_{j,0}(w,z)|<\delta$. Due to (\ref{A_es0}), $|g(w,z)|<\epsilon$ and hence
$|g_{j,k}(w,z)|<\delta$ for all $0\le j\le m-1$, $k\ge0$. Therefore
$\sigma_{1,+}=\infty$ and $\sigma_{1,-}=0$. The proof for $j>1$ is similar.
\vspace{3mm}

\begin{figure}
\centerline{
\epsfig{file=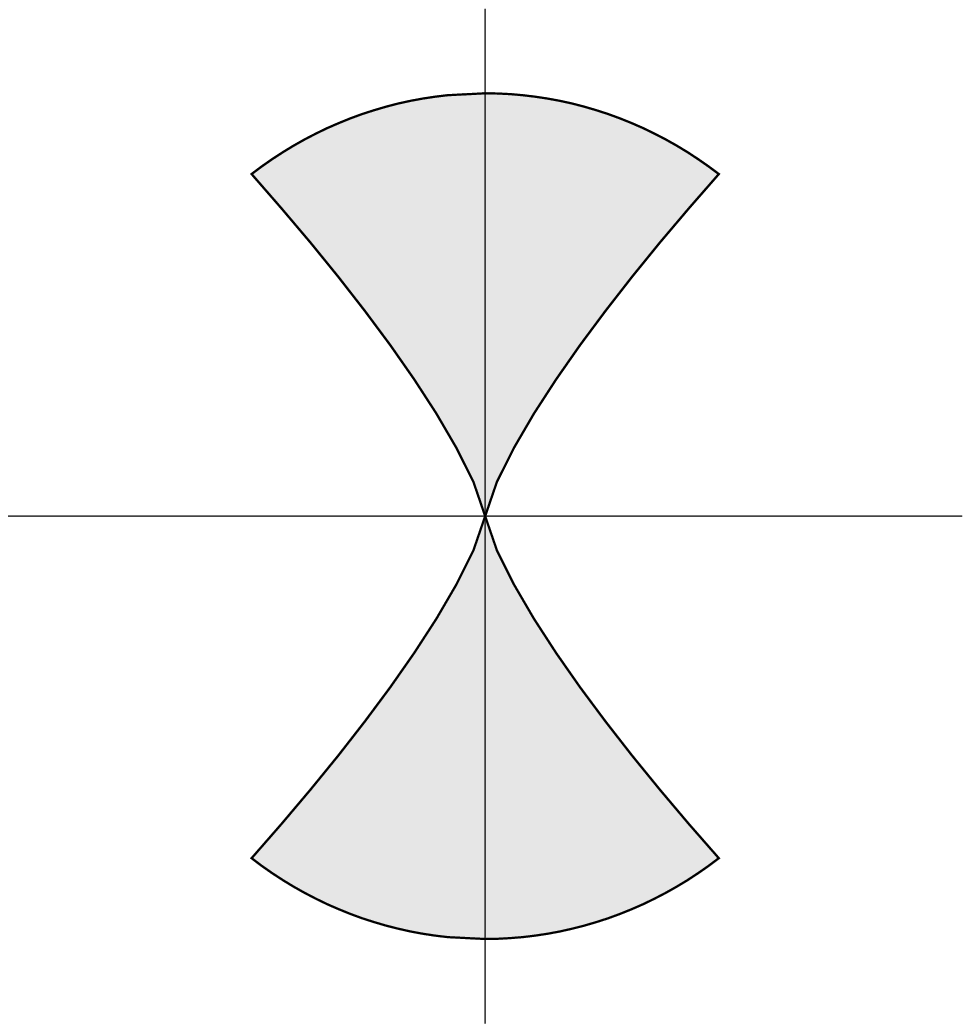,width=6cm}~\hspace{-8mm}
\epsfig{file=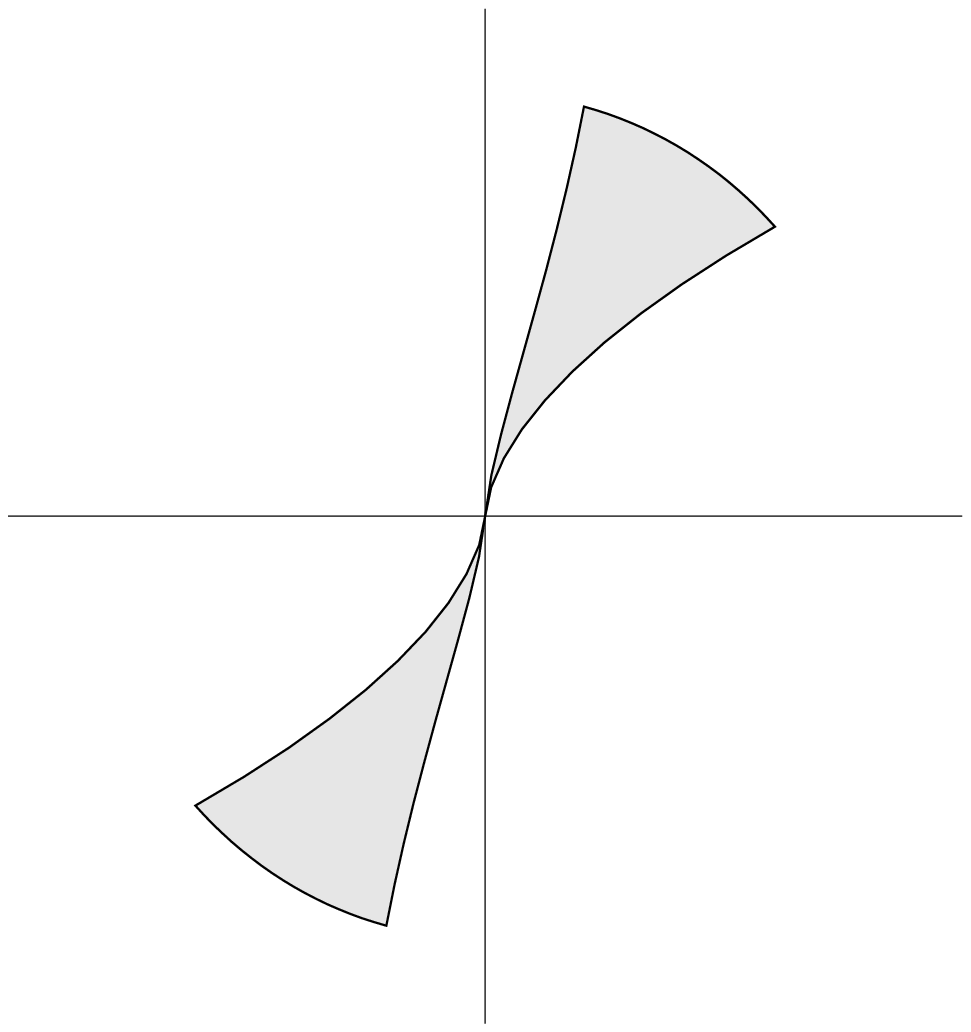,width=6cm}~\hspace{-8mm}
\epsfig{file=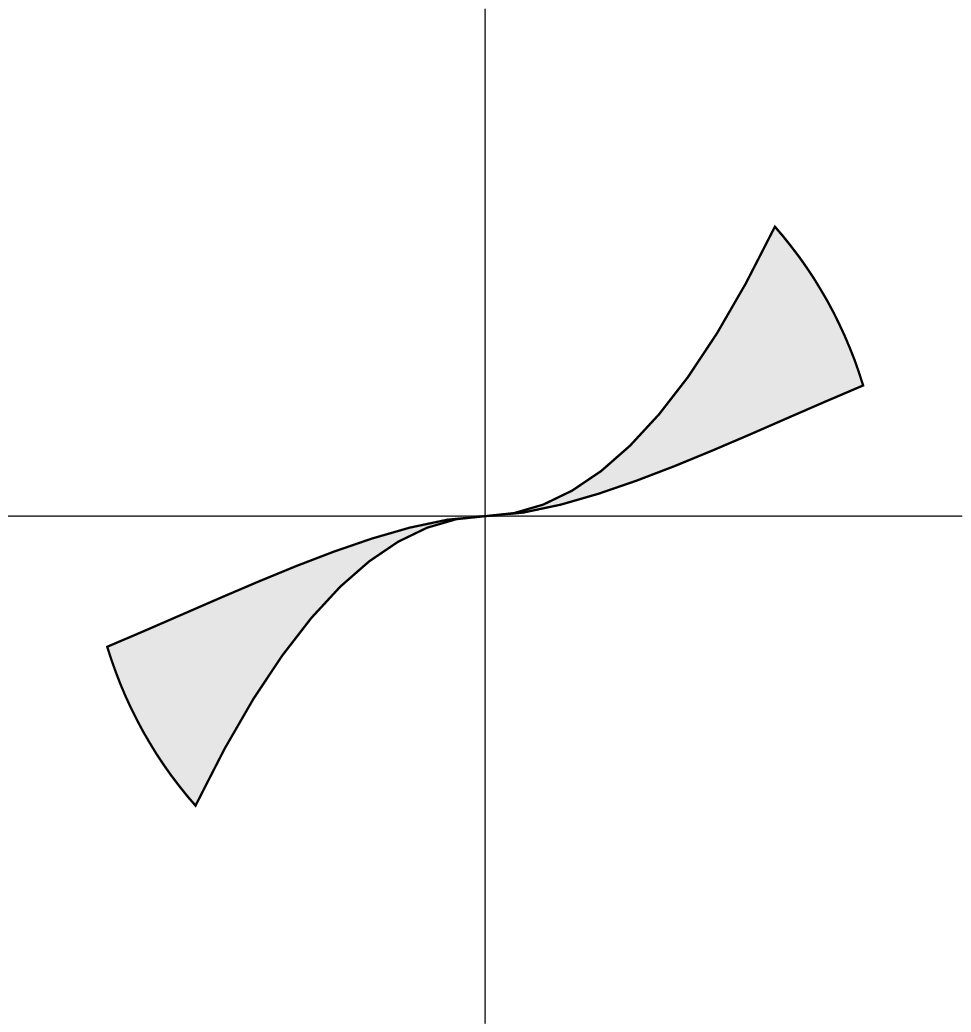,width=6cm}
}

\vspace*{-56mm}
\hspace*{25mm}
{\large $z$}

\vspace*{-.5cm}
\hspace*{80mm}
{\large $z$}

\vspace*{-.5cm}
\hspace*{135mm}
{\large $z$}

\vspace*{1.8cm}
\hspace*{41mm}
{\large $w$}

\vspace*{-.5cm}
\hspace*{96mm}
{\large $w$}

\vspace*{-.5cm}
\hspace*{151mm}
{\large $w$}

\vspace{12mm}
\hspace{40mm}(a)\hspace{50mm}(b)\hspace{50mm}(c)
\vspace{10mm}

\caption{
Examples of the sets (a) $Q^{s,s}(\epsilon_s)$,
(b) $Q^{s-1,s}(\epsilon_{s-1})$ for $C_{s-1}D_{s-1}<0$ and
$0<a_{s-1}-b_{s-1}<1$ and (c) $Q^{s-2,s}(\epsilon_{s-2})$,
for $A_{s-2}B_{s-2}<0$ and $a_{s-2}-b_{s-2}>1$.
}
\label{fig_example2}
\end{figure}

(b) First, let us prove that
\begin{equation}\label{A_ex1}
\sigma_{1,+}< h_{1,s}\left(-\frac{1}{ b_s}\right)-1+\tilde q_1,
\end{equation}
where $\tilde q_1$ is any small number and $s=J_l$ for some $l$. Denote
$q_1=\tilde q_1/a_{1,s}$. Assume that $q$ satisfies (\ref{A_es0}).
Define the sets $Q^{j,s}(\epsilon_j)$ by the following rule:
$$
Q^{s,s}(\epsilon_s)=
Q_z(-|z|^{-1/b_s+q_1},|z|^{-1/b_s+q_1},\epsilon_s).
$$
For a given
$Q^{j+1,s}(\epsilon_{j+1})$ the set $Q^{j,s}(\epsilon_j)$ is
\begin{equation}\label{A_es7}
Q^{j,s}(\epsilon_j)=
\left\{
\renewcommand{\arraystretch}{1.2}
\begin{array}{ll}
\emptyset & \mbox{ if } a_j-b_j<0\\
Q_z(\tilde f_1(z),\tilde f_2(z),\epsilon_j),\
\epsilon_j=\epsilon_{j+1}^{(a_j-b_j)/(a_j-(a_j-b_j)q)} &
\mbox{ if } 0<a_j-b_j<1\\
Q_w(\tilde f_1(w),\tilde f_2(w),\epsilon_j),\
\epsilon_j=\epsilon_{j+1}^{1/(a_j-q)} & \mbox{ if } a_j-b_j>1,
\end{array}\right.
\end{equation}
where $\tilde f_1$ and $\tilde f_2$ are the functions defined in Lemmas
\ref{lem_43} and \ref{lem_44}. Denote by $\tilde\epsilon_j$ the $\epsilon_j$
from Lemmas \ref{lem_43} and \ref{lem_44} and set
$\epsilon_0=\min_j\tilde\epsilon_j^{1/\hat a_{1,j}}$, where
$\hat a_{l,j}=\prod_{l\le s<j}(\max(a_s,a_s/(a_s-b_s))-q)$.
Examples of the sets $Q^{j,s}(\epsilon_s)$
are shown in Figure \ref{fig_example2}.

\vspace{3mm}

For any $\delta>0$ we can find an $\hat\epsilon>0$ such that
$$|g_s(w,z)|>\delta\hbox{ for all }
(w,z)\in Q_z(-|z|^{-1/b_s+q_1},|z|^{-1/b_s+q_1},\hat\epsilon).$$
Hence, if $\epsilon<\min(\epsilon_0,\hat\epsilon^{1/\hat a(1,s)})$ then
$$|g_s\circ\ldots\circ g_1(w,z)|>\delta\hbox{ for all }
(w,z)\in Q^{1,s}(\epsilon).$$
Since
$$\ell(Q^{1,s}(\epsilon))=
{\rm O}(\epsilon^{h_{1,s}(-1/b_s+q_1)+1}),$$
and $h_{1,s}(1/b_s+q_1)-h_{1,s}(1/b_s)=\tilde q_1$
the inequality (\ref{A_ex1}) is proved.

Second, we prove that
\begin{equation}%\label{A_ex1}
\sigma_{1,+}> h_{1,\tilde J}(-1/b_{\tilde J})-1-\tilde q_1,
\end{equation}
where $\tilde q_1$ is
any small number, $q_1=\tilde q_1/a_{1,\tilde J}$, and $\tilde J$ is the value
of $J_l$ where
$$
\min_{1\le l\le L}(h_{1,J_l}(-1/b_{J_l}))
$$
is achieved. Assume that $q$ satisfies (\ref{A_es0}) and $r$ and $q$ satisfy
the conditions of Lemma \ref{lem_41} for all $j$.
Set
$$Q^{s,s}(\epsilon_s)=
Q_z(-|z|^{-1/b_s-q_1},|z|^{-1/b_s-q_1},\epsilon_s)\hbox{ for }s=1,\ldots,J_l,$$
$Q^{1,1}_{l}(\epsilon_1)=\tilde Q_{l}$ where $\tilde Q_{}$,
$l=2,3$, and $\epsilon_1$ are defined in Lemma \ref{lem_47} for $j=1$.
The sets $Q^{j,s}$ and $\widetilde Q^{j,1}_{l}$, $l=2,3$, are defined by
(\ref{A_es7}). Denote
$$y_0=
\left\{\begin{array}{ll}
\infty &\mbox{ if } a_1-b_1<0\\
(a_1(r+1)/2-a_1+1)/(a_1-b_1) &\mbox{ if } 0<a_1-b_1<1\\
a_1(r+1)/2-b_1 & \mbox{ if } a_1-b_1>1.
\end{array}\right.
$$
Since $y_0>1$, by property (d) of the functions $h_{l,j}(y)$, there
exists $k>0$ such that
$h_{1-mk,1}(y_0)>h_{1,\tilde J}(-1/b_{\tilde J})$. Hence by Lemmas
\ref{lem_46} and \ref{lem_47},
$$\hbox{ if }(w,z)\in B_{\epsilon}\setminus
((\cup_{0\le s\le k,\ 1\le l\le L}Q^{1-sm,J_l})\cup
\widetilde Q_2^{1-km,1}\cup\widetilde Q_3^{1-km,1})
\hbox{ then }g^k(w,z)\in Q(r,\tilde\epsilon),$$
where $\epsilon^{k\beta}<\tilde\epsilon$, $\beta=\prod_{1\le j\le m}\beta_j$.
Since
$$\ell(Q^{1,\tilde J}(\epsilon))=
{\rm O}(\epsilon^{h_{1,\tilde J}(-1/b_{\tilde J}-q_1)+1}),$$
$$\ell(\widetilde Q^{1-km,1}_{2,3}(\epsilon))=
{\rm O}(\epsilon^{h_{1-mk,1}(y_0)+1})=
{\rm o}(\epsilon^{h_{1,\tilde J}(-1/b_{\tilde J}-q_1)+1})$$
and
$$\ell(Q^{1-sm,K}(\epsilon))=
{\rm O}(\epsilon^{h_{1-sm,K}(-1/b_K-q_1)+1})=
{\rm o}(\epsilon^{h_{1,\tilde J}(-1/b_{\tilde J}-q_1)+1})\hbox{ if }s\ne0,\ K\ne\tilde J$$
by Corollary~\ref{cor_lem_41}, part (b) is proved.

\vspace{3mm}

(c) Let $r$ and $q_3$ be such that the
conditions of Lemma \ref{lem_41} are satisfied for all $j$ where $b_j>-1$.
For all $j$ where $b_j<-1$ assume that $r$ also satisfies
\begin{equation}\label{A_ex7}
b_j+r<-1.
\end{equation}
Let $\widetilde Q_2$ and $\widetilde Q_3$ be defined as in Lemma \ref{lem_47} for $j=1$.
By the same arguments employed in that lemma, for sufficiently small $\delta$,
if $(w,z)\in B_{\delta}\setminus(\widetilde Q_2\cup\widetilde Q_3)$ then
either $g_1(w,z)>\delta$ or $g_1(w,z)\in Q(r,\delta)$. In the latter case,
if all $b_j>-1$ then $g_{l,k}(w,z)>\delta$ (for small enough $\delta$) for some $l$ and $k$
due to Corollary~\ref{cor_lem_41}(b). If there exist $b_j<-1$ then
$g_{j,1}(w,z)>\delta$ for small $\delta$ due to (\ref{A_ex7}).
Therefore, $g_{l,k}(w,z)>\delta$ for some $l$ and $k$
for all $(w,z)$, such that
$(w,z)\in B_{\delta}\setminus(\widetilde Q_2\cup\widetilde Q_3)$.

We now consider
$(w,z)\notin B_{\delta}\setminus(\widetilde Q_2\cup\widetilde Q_3)$.
There are two (generically) mutually exclusive cases:
\begin{itemize}
\item
Suppose that at least one of the inequalities,
$\prod_{1\le j\le m}a_j>1$ or $a_t-b_t<0$ for some $t$, is satisfied.
Denote $Q^{1,1}_{2,3}=\widetilde Q_{2,3}$
and define the sets $Q^{1-l,1}_{2,3}$ and $y_0$ in the same way as in the proof of the
part (b). Since $\lim_{k\to\infty}h_{1-km,1}(y_0)=\infty$,
$\sigma_{1,-}$ is arbitrary large.
\item
Suppose that $\prod_{1\le j\le m}a_j<1$ and $a_t-b_t>0$ for all $t$.
There exist $q_2>0$ such that
$\prod_{1\le j\le m}(a_j+q_2)<1$. By Corollaries~\ref{cor_lem_43} and \ref{cor_lem_44}
there exist limits as $k\to\infty$ of $f_l$
bounding $Q^{1-km+j,1}_{2,3}(\epsilon)$ for some $\epsilon<\tilde\epsilon$.
Hence, finite values of $\epsilon_j$ can be found such that
Lemmas \ref{lem_43} and \ref{lem_44} hold true for $Q^{1-km+j,1}_{2,3}$
for any $k>0$. For any $\delta<\min_j(\epsilon_j)$,
by Lemmas \ref{lem_43} and \ref{lem_44} the following is satisfied
\begin{equation}\label{A_ex8}
|g_j(w,z)|>(w,z)^{a_j+q_2}\hbox{ for all }(w,z)\in Q^{j-km,1}_{2,3}(\delta).
\end{equation}
Take any $(w_0,z_0)$, $|(w_0,z_0)|=\alpha<\delta$. There exists $k>0$ such that
$\alpha^{k \hat a_{1,m}}>\delta$, where $\hat a_{l,j}=\prod_{l\le s<j}(a_s+q_2)$.
Hence, due to (\ref{A_ex8}), if $|g_{l,s}(w_0,z_0)|<\delta$ for all
$1\le l\le m$ and $s<k$, and $|g^k(w_0,z_0)|<\delta$ then
$g^k(w_0,z_0)\in B_{\delta}\setminus(\widetilde Q_2\cup\widetilde Q_3)$.
\end{itemize}
\qed

\section{Types B and C cycles}

In this section we present proofs of Theorems and Lemmas employed for
calculation of stability indices of Types B and C cycles. To leading order,
the maps $g_j:H^{(in)}_j\to H^{(in)}_{j+1}$ associated
with the cycles of Types B and C reduce to $g_j(w,z)=(Ew^{a_j},Fw^{b_j}z)$ and
$g_j(w,z)=(Ew^{b_j}z,Fw^{a_j})$, respectively. As noted in Section
\ref{typeBC}, it suffices to consider only positive values of $w$ and $z$.
In the coordinates $(\zeta,\eta)$, $\zeta=\ln z$ and $\eta=\ln w$, the maps
$g_j$ take the form:
$$g_j(\zeta,\eta)=M_j\left(
\begin{array}{c}
\zeta\\
\eta
\end{array}\right).$$
(In what follows, the constants $E$ and $F$ are ignored, see discussion in
Section \ref{typeBC}.) The {\em transition matrices} of the maps are
$$
M_j=\left(
\begin{array}{cc}
a_j&0\\
b_j&1
\end{array}
\right)\hbox{ and }
M_j=\left(
\begin{array}{cc}
b_j&1\\
a_j&0
\end{array}\right)
$$
for cycles of Types B and C, respectively.
Recall that the coefficients $a_j$ and $b_j$ of the map $g_j$ are related
to the eigenvalues of linearisation of (\ref{eq_ode}) near
$\xi_j$ as $a_j=c_j/e_j$ and $b_j=-t_j/e_j$.
As in Appendix~\ref{typeA}, the stability indices are calculated in terms
of exponents of the maps $g_j$, $a_j$ and $b_j$. For the map
$g=g_m\circ\ldots\circ g_1$ the transition matrix is $M=M(g)=M_m\cdots M_1$.
We introduce the notation: $M_{j,k}$ and $M^{(j)}$ denote transition matrices
for the maps $g_{j,k}$ and $g^{(j)}$, respectively; $M^{(l,j)}=M_l\cdots M_j$;
$\lambda^j_1$, $\lambda^j_2$, ${\bf v}^j_1=(v^j_{11},v^j_{12})$
and ${\bf v}^j_2=(v^j_{21},v^j_{22})$ denote eigenvalues and associated
eigenvectors of the matrix $M^{(j)}$, respectively. If the eigenvalues are real,
$\lambda^j_1\ge\lambda^j_2$ is assumed.

\subsection{The set $U^{-\infty}(M)$}\label{eig}

A necessary condition for $(w,z)$ to belong to $\cB_{\delta}^g$ (see
Subsection \ref{secindmap}) is that $g^k(w,z)$ is bounded for all $k$
by a small $\delta>0$. Since $g^k(w,z)$ is bounded, in the new coordinates
$(\zeta,\eta)=(\ln w,\ln z)$ the iterates $(\zeta_k,\eta_k)=g^k(\zeta,\eta)$
are bounded from above: $\zeta_k<S$ and $\eta_k<S$ for some large in absolute
value negative $S$. Due to linearity of $g$, this generically implies that
$\lim_{k\to\infty}g^k(\zeta,\eta)=(-\infty,-\infty)$.

We denote
$$
U^{-\infty}(M)=
\{(x,y):\ x\le0,\ y\le0,\ \lim_{n\to\infty}M^n(x,y)^t=(-\infty,-\infty)^t\}.
$$

\begin{lemma}\label{lem_Uinf}
The dependence of $U^{-\infty}(M)$ on eigenvalues and
eigenvectors is as follows:
\begin{itemize}
\item[(i)] If the $\lambda_i$ are complex, then $U^{-\infty}(M)=\emptyset$.
\item[(ii)] If the $\lambda_i$ are real and $\lambda_1\le1$ or $|\lambda_2|>\lambda_1$,
then $U^{-\infty}(M)=\emptyset$;
\item[(iii)] If the $\lambda_i$ are real and $v_{11}v_{12}<0$, then
$U^{-\infty}(M)=\emptyset$;
\item[(iv)] If the $\lambda_i$ are real, $\lambda_1>1$, $v_{11}v_{12}>0$ and
$v_{21}v_{22}\le0$, then
$$
U^{-\infty}(M)=\{(x,y):\ x\le0,\ y\le0\};
$$
\item[(v)] If the $\lambda_i$ are real, $\lambda_1>1$, $v_{11}v_{12}>0$
(whereby we assume $v_{11}>0$ and $v_{12}>0$), and $v_{21}v_{22}>0$, then
$$
U^{-\infty}(M)=\{(x,y):\ x\le0,\ y\le0,\
(v_{11}v_{22}-v_{12}v_{21})^{-1}(v_{22}x-v_{21}y)<0\};
$$
\item[(vi)] If the $\lambda_i$ are real, $\lambda_1>1$, $v_{11}v_{12}=0$ and
$\lambda_2\le1$, then $U^{-\infty}(M)=\emptyset$;
\item[(vii)] If the $\lambda_i$ are real, $\lambda_1>1$, $v_{11}v_{12}=0$,
$\lambda_2>1$ and
$v_{21}v_{22}\le0$, then
$U^{-\infty}(M)=\{(x,y):\ x\le0,\ y\le0\}$;
\item[(viii)] If the $\lambda_i$ are real, $\lambda_1>1$, $v_{11}v_{12}=0$
(whereby we assume $v_{11}\ge0$ and $v_{12}\ge0$), $\lambda_2>1$ and
$v_{21}v_{22}>0$, then
$$
U^{-\infty}(M)=\{(x,y):\ x\le0,\ y\le0,\
(v_{11}v_{22}-v_{12}v_{11})^{-1}(v_{22}x-v_{21}y)<0\}.
$$
\end{itemize}
\end{lemma}

\proof
Let the eigenvalues be complex conjugate, $\lambda_{1,2}=s{\rm e}^{\pm i\phi}$.
Then
$$M^n(\alpha{\bf v}_1+\beta{\bf v}_2)=s^n({\bf v}_1(\alpha\cos n\phi-
\beta\sin n\phi)+{\bf v}_2(\alpha\sin n\phi+\beta\cos n\phi))=$$
$$s^n(\cos n\phi(\alpha v_{11}+\beta v_{21})+\sin n\phi(-\beta v_{11}+\alpha v_{21}),
\cos n\phi(\alpha v_{12}+\beta v_{22})+\sin n\phi(-\beta v_{12}+\alpha v_{22})).$$
Because the eigenvalues are complex, $\phi\ne k\pi$. Hence for any $N_0>0$
there exists $N>N_0$ such that
$$x_n\equiv s^n(\cos n\phi(\alpha v_{11}+\beta v_{21})+
\sin n\phi(-\beta v_{11}+\alpha v_{21}))>0.$$
This proves part (i).

If the eigenvalues are real and distinct, the map $M^n$ in the basis comprised of the
eigenvectors ${\bf v}_1$ and ${\bf v}_2$, $(x,y)=h_1{\bf v}_1+h_2{\bf v}_2$,
takes the form
\begin{equation}\label{mn}
M^n(h_1,h_2)\equiv\left(
\begin{array}{c}
h_1^{(n)}\\
h_2^{(n)}
\end{array}
\right)=
\lambda_1^nh_1\left(
\begin{array}{c}
v_{11}\\
v_{12}
\end{array}
\right)+\lambda_2^nh_2\left(
\begin{array}{c}
v_{21}\\
v_{22}
\end{array}
\right).
\end{equation}
If $|\lambda_1|\le 1$ and $|\lambda_2|\le 1$ (recall that
$\lambda_2<\lambda_1$), then for any $(h_1,h_2)$, (\ref{mn}) has a finite limit as $n\to\infty$.
If $|\lambda_2|>\lambda_1$, then $\lambda_2<0$ and hence in
(\ref{mn}) the sign of $h_j^n$ ($j=1,2$) (\ref{mn}) alternates for odd and
even $n$, if $n$ is large enough. Part (ii) is proved.

Assume that $\lambda_1>1$ and $|\lambda_2|<\lambda_1$.
If $v_{11}v_{12}\ne 0$, to leading order $h_1^{(n)}=\lambda_1^nh_1v_{11}$
and $h_2^{(n)}=\lambda_1^nh_1v_{12}$ for $n\to\infty$.
Thus, if the signs of $v_{11}$ and $v_{12}$ are different, then the limits of $h_1^{(n)}$ and
$h_2^{(n)}$ have different signs, and (iii) is proved.
If $v_{11}v_{12}>0$ (and assuming without any loss of generality that they are
positive), the limits of $h_j^{(n)}$, $j=1,2$, are $-\infty$ in the
points $(x,y)$ such that $h_1<0$. $h_1$ is negative for any $x<0$ and $y<0$,
if $v_{21}v_{22}\le0$, which proves part (iv). If $v_{21}v_{22}>0$, then
the set of $(x,y)$ for which $h_1<0$ satisfies the inequality
$(v_{11}v_{22}-v_{12}v_{11})^{-1}(v_{22}x-v_{21}y)<0$, and so part (v) is proved.

Assume that $v_{12}=0$; if $\lambda_2\le1$, in (\ref{mn}) the limit of $h_2^{(n)}$
for $n\to\infty$ is either $+\infty$ or it does not exist. Thus, part (vi) is
proved. The proofs of statements (vii) and (viii) are similar to the
proofs of (iv) and (v) and are omitted.
\qed

Lemma~\ref{lem_Uinf} can be used to give the dependence of $U^{-\infty}(M)$ on the
matrix entries, $M=(a_{ij})$, $i,j=1,2$.

\begin{lemma}\label{lem_eig}
Let $\lambda_1$ and $\lambda_2$ ($\lambda_1>\lambda_2$, if they
are real; generically $\lambda_1\ne\lambda_2$ ) be the eigenvalues of the
matrix $M=(a_{ij})$, $a_{11}>a_{22}$, and ${\bf v}_1$ and ${\bf v}_2$ be the
associated eigenvectors. Then
\begin{itemize}
\item[(a)]
\begin{itemize}
\item[(i)] the eigenvalues are real
if and only if
\begin{equation}
\frac{(a_{11}-a_{22})^2}{ 4}+a_{12}a_{21}\ge 0
\end{equation}
\item[(ii)] $\lambda_1>1$ if and only if
\begin{equation}
\max\left(\frac{a_{11}+a_{22}}{ 2},~a_{11}+a_{22}-a_{11}a_{22}+a_{12}a_{21}\right)>1
\end{equation}
\item[(iii)] $\lambda_1>|\lambda_2|$ if and only if
\begin{equation}
\frac{a_{11}+a_{22}}{ 2}>0
\end{equation}
\item[(iv)] $v_{11}v_{12}>0$ if and only if
\begin{equation}
a_{21}>0.
\end{equation}
\end{itemize}
\item[(b)] If $\lambda_j$, $j=1,2$, are real, then
$a_{12}v_{22}=(\lambda_2-a_{11})v_{21}$ and $\lambda_2-a_{11}<0$.
\end{itemize}
\end{lemma}

\proof
The eigenvalues of the matrix $M$ can be expressed as
$$
\lambda_{1,2}=\frac{a_{11}+a_{22}}{ 2}\pm
\sqrt{ \left(\frac{a_{11}-a_{22}}{ 2}\right)^2+a_{12}a_{21} }.
$$
Statements (a)(i)-(iv) follow from examination of this formula and on noting that the
eigenvector ${\bf v}_1$ satisfies $a_{21}v_{11}+a_{22}v_{12}=\lambda_1v_{12}$. If the
eigenvalues are real, then $\lambda_2-a_{11}<0$ and
${\bf v}_2$ satisfies $a_{11}v_{21}+a_{12}v_{22}=\lambda_2v_{11}$, which proves
statement (b).
\qed

\begin{corollary}\label{cor_lem_eig}
Assume that entries of a matrix $M$ satisfy the conditions
(i)-(iv) of Lemma \ref{lem_eig}. Then
$U^{-\infty}(M)=\{(x,y):\ x\le0,\ y\le0,\ (\lambda_2-a_{11})x-a_{12}y>0\}$
for $a_{12}<0$, and $U^{-\infty}(M)=\{(x,y):\ x\le0,\ y\le0\}$ for $a_{12}\ge0$.
\end{corollary}

\subsection{Maps and neighbourhoods}\label{mapn}

The condition $w^2+z^2<\epsilon$ in $(\zeta,\eta)$ coordinates is equivalent
to the condition\footnote{In other words, instead of $|(w,z)|=(w^2+z^2)^{1/2}$,
an equivalent norm $|(w,z)|=\max(|w|,|z|)$ can be employed.}
$\max(\zeta,\eta)<R$, where $R<0$; small $\epsilon$ corresponds to large $|R|$.
In this subsection we examine how $R$-neigbourhoods of
$(-\infty,-\infty)$ are transformed by linear maps.

Let $M$ be an invertible linear map $M:\R^2\to\R^2$. Define
$$
U_R=\{(\zeta,\eta)~|~\max(\zeta,\eta)<R\}
$$
and
$$
\begin{array}{c}
U_R(\alpha_1,\beta_1,q_1;\alpha_2,\beta_2,q_2)=\\
\{(\zeta,\eta)\in U_R~:~(\alpha_1+q_1)\zeta+(\beta_1+q_1)\eta<0,\
(\alpha_2+q_2)\zeta+(\beta_2+q_2)\eta<0\},
\end{array}
$$
where $R<0$.

\begin{lemma}\label{lem_nei}
For any $S<0$ and $q>0$ there exists $R<0$ such that
\begin{equation}\label{incl}
M(U_R\cap M^{-1}U_0(1,0,-q;0,1,-q))\subset U_S.
\end{equation}
\end{lemma}

\proof We split the neighbourhood in two parts
$$
U_0(1,0,-q;0,1,-q)=V_1\cup V_2,
$$
where $V_1=U_0(1,0,-q;0,1,-q)\cap U_S$ and $V_2=U_0(1,0,-q;0,1,-q)\setminus U_S$
and denote
$$
\tilde R=\min\{\zeta,\eta~:~(\zeta,\eta)\in M^{-1}V_2\}.
$$
$\tilde R$ is finite, since $V_2$ is bounded and $M$ invertible. The inclusion (\ref{incl})
takes place for any $R<\tilde R$, because
$$
M(U_R\cap M^{-1}U_0(1,0,-q;0,1,-q))\cap V_2=\emptyset
$$
due to $R<\tilde R$.
\qed

Note, that if $(\zeta,\eta)\in U_R\setminus M^{-1}U_0$, then
$\max(\tilde\zeta,\tilde\eta)\ge0$, where $(\tilde\zeta,\tilde\eta)=M(\zeta,\eta)$.

\begin{lemma}\label{lemnew1}
Denote $\widetilde U=U_0(\alpha_1,\beta_1,-q_1;\alpha_2,\beta_2,-q_2)$.
Suppose $\widetilde U\subseteq U^{-\infty}(M)\ne\emptyset$ and
$(v_{11},v_{12})\in \widetilde U$.
\begin{itemize}
\item[(i)] If $\lambda_2\ge0$, then
$M^k(\widetilde U)\subset\widetilde U$ for any $k>0$;
\item[(ii)] $M^{2k}(\widetilde U)\subset\widetilde U$ for any $k>0$.
\end{itemize}
\end{lemma}

\proof
(i) If $(x,y)\in\widetilde U$, then $\alpha(x,y)\in\widetilde U$ for any $\alpha>0$.
If $(x,y)\in\widetilde U$ is represented as $(x,y)=\alpha{\bf v}_1+\beta{\bf v}_2$,
then, due to convexity of $\widetilde U$,
$\alpha{\bf v}_1+\tilde\beta{\bf v}_2\in\widetilde U$ for any $|\tilde\beta|$,
such that $|\tilde\beta|\le|\beta|$ and $\tilde\beta\beta\ge 0$.

For any $(x,y)=\alpha{\bf v}_1+\beta{\bf v}_2\in\widetilde U$ the identity
\begin{equation}\label{eq_n1}
M^k(x,y)=\lambda_1^k(\alpha{\bf v}_1+\beta\lambda_2^k/\lambda_1^k{\bf v}_2)
\end{equation}
holds. Since $\lambda_1>\lambda_2>0$, due to the arguments above,
$M^k(x,y)\in\widetilde U$.

(ii) If $k$ is even, the identity (\ref{eq_n1}) implies the statement for
negative $\lambda_2$ as well.
\qed

\begin{lemma}\label{lemnew2}
Consider the set
$$U^{-\infty}(M)=U_0(\alpha_1,\beta_1,0;\alpha_2,\beta_2,0).$$
For any $q_1>0$, $q_2>0$ and $S<0$ there exists an $R<0$ such that
\begin{itemize}
\item[(i)] If $\lambda_2\ge0$, then
$M^kU_R(\alpha_1,\beta_1,-q_1;\alpha_2,\beta_2,-q_2)\subset U_S$ for any $k>0$;
\item[(ii)] $M^{2k}U_R(\alpha_1,\beta_1,-q_1;\alpha_2,\beta_2,-q_2)\subset U_S$ for any $k>0$.
\end{itemize}
\end{lemma}

\proof
(i) Consider a set $\widetilde W$ comprised of two line segments, one segment
being $x=-1$, $y=[-1,0]$ and the other one $y=-1$, $x=[-1,0]$. Denote
$W=\widetilde W\cap\bar U_0(\alpha_1,\beta_1,-q_1;\alpha_2,\beta_2,-q_2)$.
The iterates $(x_k,y_k)=M^k(x_0,y_0)^t$ are bounded from above for any
$(x_0,y_0)\in W$ by some $\widetilde Q<0$, because $W\subset U^{-\infty}(M)$.
Due to compactness of $W$, the bounds for the iterates are uniform for all
$(x_0,y_0)\in W$ by some $Q<0$. Since the maps $M^k$ are linear,
$\max(x_k,y_k)<Q$ for a given $(x_0,y_0)$ implies
$(\tilde x_k,\tilde y_k)<\alpha Q$ for
$(\tilde x_0,\tilde y_0)=\alpha(x_0,y_0)$. Part (i) is thus proved for $R=-S/Q$.

Statement (ii) is a consequence of Lemma \ref{lemnew1} (ii).
\qed

\subsection{Main theorems}\label{calci}

In this subsection we prove two theorems on stability indices for maps
related to heteroclinic cycles of Types B or C, employing the Lemmas proved
in Sections~\ref{eig} and \ref{mapn}.

\begin{theorem}\label{th_BC1_repeat}(reproduces Theorem~\ref{th_BC1})
Let $g$ be a map related to simple heteroclinic cycle of Types B or C
and $M_j$, $1\le j\le m$, its transition matrices. Suppose that for all $j$,
$1\le j\le m$, all entries of the matrices are non-negative. Then:
\begin{itemize}
\item[(a)] If the transition
matrix $M=M_m\cdots M_1$ satisfies condition (a)(ii) of Lemma \ref{lem_eig},
then $\sigma_{j,+}=\infty$ and $\sigma_{j,-}=0$ for all $j$ and therefore
the cycle is asymptotically stable.
\item[(b)] Otherwise, $\sigma_{j,+}=0$ and $\sigma_{j,-}=\infty$ for all $j$ and
the cycle is not an attractor.
\end{itemize}
\end{theorem}

\proof
(a) Suppose that the matrix $M\equiv M^{(1)}$ satisfies condition
(a)(ii) of Lemma \ref{lem_eig}. For a map $M^{(j)}:\ H^{(in)}_j\to H^{(in)}_j$,
the condition can be expressed as
\begin{equation}\label{eq_mapg1}
\max\left(\tr\left(M^{(j)}\right),2\,\tr \left(M^{(j)}\right)-2\det\left(M^{(j)}\right)\right)>2.
\end{equation}
Hence if the condition is satisfied by $M^{(j)}$ for any one value of $j$,
it is satisfied for all $1\le j\le m$. Any $M^{(j)}$ have non-negative entries, as it
is a product of matrices with non-negative entries.
Therefore, $M^{(j)}$ satisfies conditions (i) and (iii) of part (a) the Lemma.
Due to the assumptions $\lambda_1>\lambda_2$ and $a_{11}>a_{22}$, the condition
(iv) is satisfied. Hence $U^{-\infty}(M^{(j)})=\R_-^2$ for all $j$.

Consider the images of the lines $\alpha(-1+q,-q)$ and $\beta(-q,-1+q)$, where
$0<q<1$, $\alpha\in\R_+$ and $\beta\in\R_+$, under the mappings $M^{j,1}$ or
$M^{j,1}M$. Since all entries of the matrices $M^{j,1}$ and $M^{j,1}M$ are
non-negative, the images take the form $\alpha(-r_1,-r_2)$ for $r_i>0$ and
$\alpha\in\R_+$. Thus, for any positive $q<1$ there exists a $q_j>0$, such that
$$M^{j,1}U_0(1,0,-q;0,1,-q)\subset U_0(1,0,-q_j;0,1,-q_j)$$
and
$$M^{j,1}MU_0(1,0,-q;0,1,-q)\subset U_0(1,0,-q_j;0,1,-q_j).$$

Apply Lemma \ref{lem_nei} to mappings $M^{j,1}$ and $M^{j,1}M$, setting any
$S<0$ and $q=q_j$. According to the Lemma, we can find $S_j$, such that
\begin{equation}\label{inclM}
M^{j,1}U_{S_j}(1,0,-q;0,1,-q)\subset U_S\hbox{ and }
M^{j,1}MU_{S_j}(1,0,-q;0,1,-q)\subset U_S.
\end{equation}
Denote
$$\widetilde S=\min_j S_j.$$

By Lemmas \ref{lemnew1} and \ref{lemnew2} (where Lemma \ref{lemnew2} is applied
for $S=\widetilde S$), there exists $R$ such that
$$M^{2k}U_R(1,0,-q;0,1,-q)\subset U_{\widetilde S}(1,0,-q;0,1,-q)\hbox{ for all }k\ge0.$$
Thus, (\ref{inclM}) implies
$$M_{j,k}U_R(1,0,-q;0,1,-q)\subset U_S\hbox{ for all }k\ge0.$$
Hence, $\sigma_{1,+}>1/q-1$ for any $q$, which implies
$\sigma_{1,+}=\infty$ and $\sigma_{1,-}=0$.
The proof holds true for $j>1$ as well, and therefore part (a) is proved.

For part (b), if the matrix $M\equiv M^{(1)}$ does not satisfy the condition
(ii) of the Lemma, by Lemma \ref{lem_Uinf} the set $U^{-\infty}(M)$ is
empty, $\sigma_{1,+}=0$ and $\sigma_{1,-}=\infty$. Since condition
(\ref{eq_mapg1}) is satisfied or not satisfied by all $M^{(j)}$ simultaneously,
$\sigma_{j,+}=0$ and $\sigma_{j,-}=\infty$ for all $1\ge j\ge m$.
\qed

\begin{theorem}\label{th_BC2_repeat}(reproduces Theorem~\ref{th_BC2}).
Let $X$ be a simple heteroclinic cycle of Types B or C
and $M_j$, $1\le j\le m$ the associated transition matrices. We denote by
$j=j_1,\ldots j_L$ the indices, for which some of the entries of $M_j$
are negative; they are all non-negative for all remaining $j$.
\begin{itemize}
\item[(a)] If at least for one of $j=j_l+1$
the matrix $M^{(j)}$ does not satisfy conditions (i)-(iv)
of Lemma \ref{lem_eig}, then the cycle is repelling and
$\sigma_j=-\infty$ for all $j$.
\item[(b)] If the matrices $M^{(j)}$ satisfy conditions (i)-(iv) of Lemma
\ref{lem_eig} for all $j=j_l+1$, then there exist numbers
$(\alpha_1^j,\beta_1^j,\alpha_2^j,\beta_2^j)$, $1\le j\le m$, such that
\begin{itemize}
\item[(i)] $U_0(\alpha_1^j,\beta_1^j,0;\alpha_2^j,\beta_2^j,0)\ne\emptyset$, $1\le j\le m$.
\item[(ii)] For any $S<0$ and $q>0$ there exists $R<0$ such that
$$M^{(l,j)}(M^{(j)})^k
(U_R(\alpha_1^j,\beta_1^j,-q;\alpha_2^j,\beta_2^j,-q))
\subset U_S\ \mbox{ for all }\,l,\ \ 1\le l<m,\ k\ge0.$$
\item[(iii)] $$
\lim_{k\to\infty}(M^{(l,j)}(M^{(j)})^k(\zeta,\eta))=
(-\infty,-\infty),\ \mbox{ for all }\ (\zeta,\eta)\in
U_0(\alpha_1^j,\beta_1^j,0;\alpha_2^j,\beta_2^j,0).
$$
\item[(iv)]
$$
U_0(\alpha_1^j,\beta_1^j,0;\alpha_2^j,\beta_2^j,0)=U^{-\infty}(M^{(j)})
\cap \left(\bigcap_{1\le l\le L}(M^{(j_l,j)})^{-1}U_0\right)
\cap \left(\bigcap_{1\le l\le L}(M^{(j_l+m,j)})^{-1}U_0\right).
$$
\item[(v)] If $\lambda_2\ge0$ then
$$
U_0(\alpha_1^j,\beta_1^j,0;\alpha_2^j,\beta_2^j,0)=U^{-\infty}(M^{(j)})
\cap \left(\bigcap_{1\le l\le L}(M^{(j_l,j)})^{-1}U_0\right).
$$
\end{itemize}
The cycle is a Milnor attractor.
\end{itemize}
\end{theorem}

\proof For (a), as noted in the proof of Theorem~\ref{th_BC1}, the matrices
$M^{(j)}$ will simultaneously satisfy, or not satisfy,
conditions (i)-(iii) of Lemma \ref{lem_eig} for all $j$.
Suppose the condition (iv) is not satisfied for some $j=J$.
For any $s$, the iterates $(M^{(J)})^kM^{(J,s)}(x,y)$
on increasing $k$ become aligned with $(v^J_{11},v^J_{12})$, see (\ref{mn}). Since
$v^J_{11}v^J_{12}\le0$, the iterates escape from $U_S\subset \hat H^{(in)}_J$
for any $S<0$ for a sufficiently large $k$. Hence part (a) is proved.

For (b) suppose that, for $j_l+1$ the matrix $M^{(j_l+1)}$ satisfies
conditions (iv) of Lemma \ref{lem_eig}.
The matrices $M_j$, $j_l+1\le j\le j_{l+1}-1$, have
positive entries, ${\bf v}^j=M_{j-1}\ldots M_{j_l+2}M_{j_l+1}{\bf v}^{j_l+1}$,
therefore $v^{(j_l+1)}_{11}v^{(j_l+1)}_{12}>0$
implies $v^{(j)}_{11}v^{(j)}_{12}>0$ for any $j$, $j_l+2\le j\le j_{l+1}$.
Hence, it suffices to check condition (iv) for $j=j_l+1$, $1\le l\le L$.

Denote
$$
\widetilde U_j=U^{-\infty}(M^{(j)})\cap
\left(\bigcap_{1\le l\le L}(M^{(j_l,j)})^{-1}U_0\right)
\cap\left(\bigcap_{1\le l\le L}(M^{(j_l+m,j)})^{-1}U_0\right)
$$
The set is non-empty, because it includes a neighbourhood of the point
$(v_{11}^j,v_{12}^j)$ on the plane (since this point belongs to all sets in
the intersection). Since all the sets are of the type
$U_0(\alpha_1,\beta_1,0;\alpha_2,\beta_2,0)$, the intersection is also of the
required type $U_0(\alpha_1^j,\beta_1^j,0;\alpha_2^j,\beta_2^j,0)$. Consider
$\widetilde U=\widetilde U_1$.

Due to Lemma~\ref{lemnew1} and definition of the set $\widetilde U$,
$$M_{j,k}\widetilde U\subset U^{-\infty}(M^{(j)})\hbox{ for all }1\le j\le m,\ k\ge0.$$
By the same arguments as employed in the proof of Theorem~\ref{th_BC1}, this
inclusion implies that for any $q>0$ and $S<0$ there exists $R<0$, such that
$$M_{j,k}U_R(\alpha_1^1,\beta_1^1,-q;\alpha_2^1,\beta_2^1,-q)\subset U_S
\hbox{ for all }1\le j\le m,\ k\ge0,$$
and therefore $-\infty<\sigma_1$. The proof for $\sigma_j$ with $j>1$ is
similar. Finally, by Theorem~\ref{thm_milnor}, $X$ is a Milnor attractor,
since the inequality $-\infty<\sigma_j$ is satisfied for all $j$. \qed

\end{document}